%% file: main.tex
\documentclass[a4paper]{report}
\usepackage[T1]{fontenc}
\usepackage[utf8]{inputenc}
\usepackage{helvet}
\usepackage{titlesec}
\usepackage[super]{nth}
\usepackage{amsfonts}
\usepackage{tikz}
\usepackage{hyperref}
\hypersetup{bookmarksopen=true}

\usepackage{bookmark}
\usepackage{amsmath}
\usepackage{xfrac}
\usepackage{bm}
\usepackage{soul}
\usepackage{graphicx}
\usepackage{float}
\usepackage[inline]{asymptote}
\newcommand{\rom}[1]
    {\MakeUppercase{\romannumeral #1}}
\graphicspath{ {images/} }
\DeclareMathOperator*{\argmin}{arg\,min}
\usepackage[top=2cm,bottom=2cm,left=2.5cm,right=2.5cm]{geometry}
\usepackage{biblatex} 
\begin{document}
\begin{titlepage}

    \begin{center}
        \vspace*{1cm}
            
        \Huge
        \textbf{Improved convergence of forward and inverse finite element models}
            
        \vspace{0.5cm}
        \LARGE
        ENG5041P 
            
        \vspace{1.5cm}
            
        \textbf{Preslav Aleksandrov - 2248797A}
           
        \vspace{0.5cm} 
        A project presented for the degree of\\
        MEng Civil Engineering
        \vfill

            
        \includegraphics[width=0.4\textwidth]{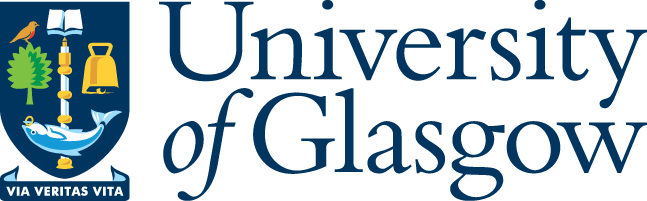}
            
        \Large
        School of Engineering\\
        University of Glasgow\\
        Scotland\\
        2020/2021
            
    \end{center}
\end{titlepage}
\chapter*{Abstract}
Forward and inverse models are used throughout different engineering fields to predict and understand the behaviour of systems and to find parameters from a set of observations. These models use root-finding and minimisation techniques respectively to achieve their goals. This paper introduces improvements to these mathematical methods to then improve the convergence behaviour of the overarching models when used in highly non-linear systems. The performance of the new techniques is examined in detail and compared to that of the standard methods. The improved techniques are also tested with FEM models to show their practical application. Depending on the specific configuration of the problem, the improved models yielded larger convergence basins and/or took fewer steps to converge.
\tableofcontents{}
\listoffigures
\listoftables
\chapter*{Acknowledgements}
Thanks to my advisor Dr. Ankush Aggarwal, whose help, guidance and input enabled me to produce this paper. 
\chapter{Introduction}
\section{Background}
Mathematical modelling is an important pillar of engineering and science. It allows one to define the behaviour of various systems in terms of governing equations, such as the balance of energy, balance of momentum, compatibility and so on. Except in a few special cases, like a statically determined truss, these governing equations cannot be solved analytically, and computational techniques are necessary to find a solution. Mathematical models can be broadly divided into two types: forward models, which one can use to predict the behaviour of a system, and inverse models, which can be used to find certain system properties.
\newline\newline
Constructing a forward model is done by utilising governing equations together with initial and boundary conditions to predict the behaviour of a system. A simple example of the use of forward models is finding the deflection of a beam at a certain location along its length. For this, one must have the beam's material and section properties, its loading and support configuration, a constitutive model (stress-strain relation) and governing equations, such as those provided by the Euler–Bernoulli bending theory. \newline\newline
Inverse models, as the name suggests, do the opposite. They use governing equations together with observations of the system's behaviour to determine some property of a system. Using the beam example, one can create an inverse model to find the modulus of the beam material, given multiple observations of the beam's bending behaviour and the same system of equations used in the forward model example.\newline\newline Even though the difference in the two cases may seem subtle, two different mathematical methods underlay their solutions. These methods are, in the case of forward modelling, root-finding and, in the case of inverse modelling, minimisation. These two methods will be explored in depth through this paper and various solution procedures will be investigated.
\subsection{Available solutions}
There are many ways to solve root-finding and minimisation problems. According to Nocedal and Wright \cite{num_optim}, there are multiple families of methods which can be used. They include Newtonian methods like the BFGS method, Conjugate Gradient methods like the Fletcher–Reeves method, and Trust-Region methods such as the Levenberg–Marquardt method. All of them have particular pros and cons such as differences in computational load, memory usage and so on. The particular focus of this paper will be on two algorithms: the Newton-Raphson (NR) method for root-finding and the Gauss-Newton (GN) method for minimisation.\newline \newline
These methods are popular due to their simplicity and convergence behaviour \cite{GALANTAI200025} \cite{BLASCHKE}. However, there are configurations in which these methods perform sub-optimally. Specifically, both methods undergo performance losses when dealing with highly non-linear systems or when a starting guess, which is required to initialise the methods, is taken too far away from the solution of the system \cite{MEI2018110}. \newline \newline
A particular example of when a forward model, using NR as the root-finding algorithm, undergoes performance losses is when the object being modelled is made from a material which exhibits hyper-elasticity \cite{MEI2018110}. Such a material can be biological soft tissues which usually undergo large strains with small applied stresses \cite{hyper_elastic} or it could be a polyurethane rail pad which exhibits similar behaviour \cite{poly_urethane}. Both of these materials share the same material model, which is described in more detail in \autoref{mooney}.\newline\newline When solving an inverse model, which utilises the same hyper-elastic material, with the GN algorithm, performance loss will also be experienced. This is due to the fact that the direction used by GN to determine the correct parameter value is a low-order approximation of the most optimal one. Improving the performance of GN and NR in those highly non-linear systems will be the topic of this paper.
\subsection{Literature survey}
This paper builds on the work done by Aggarwal \& Pant \cite{exnewton}, which shows how the non-linearity of a function can be reduced to improve the convergence behaviour of root-finding algorithms. Here, an alternate multivariate derivation of their method will be proposed. The paper uses general information about the NR and GN methods from Nocedal \& Wright \cite{num_optim}, and Süli \& Mayers \cite{suli_mayers_2003}. The derivation of non-linear FEM was based on the one given by Belytschko et. al \cite{Lagrangian}. Two data sets were used for testing from Martins et. al \cite{dataset} and Szurgott \& Jarzebski \cite{poly_urethane}.
\section{Aims and Objectives}
The main objective of this paper is to improve the convergence of forward and inverse finite element models through the improvement of their underlying computational solution algorithms. In order to do this, this paper will firstly familiarise the reader with basic concepts of iterative root-finding and minimisation methods by providing definitions and showing some basic examples. Next, the paper will focus on two modifications, one for each method, which can improve the performance of NR and GN when dealing with highly non-linear systems and finally, this paper will show the practical application of these algorithms in forward and inverse finite element models.
\section{Outlines}
This project is broken down into three main parts, prefaced by this introduction and followed by a discussion and conclusion chapter. The three main parts are as follows:
\begin{itemize}
    \item \textbf{Root-Finding} \newline In this section, the paper will present some of the basic concepts of root-finding and provide a simple example in the univariate space. Afterwards, the paper will cover the derivation of the proposed modification and the testing methodology will be shown. It will also explain how the results should be read and show which test functions were used and why. Finally, a comparison will be made between the proposed improvement and the existing method by presenting the results attained from testing.
    \item \textbf{Minimisation} \newline Here, some basic concepts of minimisation will be covered. Then a simple example of linear-least squares minimisation will be shown. Afterwards, the  derivation of the improvement will be presented with explanations of the different steps taken.
    Then the methods of testing will be covered and the set of sample functions will be shown together with their true values used to generate the test data. Finally, a comparison will be made between the proposed improvement and the existing method by presenting the results obtained from testing.
    \item \textbf{Finite Element Methods} \newline A brief description of finite element methods will be given. Then, the derivation of a non-linear finite element system will be shown. Finally, the paper will present a comparison between results obtained by using both conventional and improved methods.
\end{itemize}
\pagebreak
\section{Python and NumPy}
The majority of calculations and experiments presented in this paper were done with Python 3 (3.8.3) and an additional package called NumPy (1.19.1) \cite{harris2020array}. All random numbers were generated using Latin hyper-cube sampling and the LHS-MDU (1.1)
package \cite{sahil_moza_2020_3929531}. The following functions were used to calculate particular elements in the paper:
\begin{itemize}
    \item Moore-Penrose inverse: numpy.linalg.pinv()
    \item Matrix rank: numpy.linalg.matrix\_rank()
    \item Vector norm: numpy.linalg.norm()
    \item Tensor multiplication: numpy.einsum()
    \item Random variables: lhsmdu.sample()
\end{itemize}
\chapter{Root-finding}
\section{Introduction} \label{concepts}
\subsection{Definition}
Root-finding is the process of finding values of $x$ which satisfy the following canonical equation:
\begin{equation}
    f(x) = 0
\end{equation}
where $f(x)$ is any generic function and $x$ is the set of inputs it uses. The values of $x$ which satisfy the above are called zeros or roots. Shown in \eqref{eq: basic} is a generic quadratic equation with its output set to zero, where $a, b$ and $c$ are parameters of the equation and $a\neq0$. 
\begin{equation} \label{eq: basic}
    ax^2+bx+c = 0
\end{equation}
Solving the above equation is by definition root-finding. There are many ways to approach a problem such as this one. However, in this case, the easiest and most common method is to write out the equation's closed-form algebraic expression, a common representation of which is:
\begin{equation}
    x = \dfrac{-b\pm\sqrt{b^2-4ac}}{2a}
\end{equation}
This formula always yields two roots and provides a rigorous method for solving any quadratic equation. However, if the degree of the polynomial is higher than four no closed-form solution is available, as proven by Niels Henrik Abel \cite{abel}. To identify the roots of a function that is not suited for numerical evaluation or that does not have a  closed-form solution, iterative root-finding algorithms can be utilised. These algorithms always produce approximations of the roots and are not guaranteed to find all available roots. They also have various properties discussed below.
\subsection{Convergence behaviour}
The convergence behaviour of an iterative algorithm shows its ability to produce a convergent sequence. A convergent sequence is any infinite sequence $(x_n)$ whose elements become arbitrarily close to each other as the sequence progresses. That is, all but a finite number of elements from the sequence are within a small distance $\epsilon$ from a certain value $x^*$, which is known as the limit of the sequence. The finite number of elements that lay beyond the predetermined distance $\epsilon$ show that there exists another number $\mathcal{N}$ after which the series is considered to have converged. If both $x^*$ and $\mathcal{N}$ exist then the series is said to converge to $x^*$ in $\mathcal{N}$ iterations. This can be mathematically expressed by:
\begin{equation}
    (x_n) \to x^* \Leftrightarrow |x_n-x^*|<\epsilon, n>\mathcal{N}
\end{equation}
\subsection{Tolerance}
The value of $\epsilon$ from the above can be rewritten in terms of tolerances, allowable amounts of variation, which are used by the algorithm as termination criteria. Those will be, firstly, relative tolerance $t_r$, i.e. the tolerance between the last two elements of a sequence and secondly, the absolute tolerance $t_a$ i.e. the tolerance between the latest value of a series and a target value $x^*$.
\begin{equation}
    |x_n-x_{n-1}| \leq t_r
\end{equation}
\begin{equation}
    |x_n-x^*| \leq t_a
\end{equation}
In most cases, setting stricter tolerances ($t_r$ and $t_a$ set to smaller values than another arbitrary set of $t_r$ and $t_a$) will increase the number of iterations necessary for the method to converge.
\subsection{Iterations and Iterative methods}
Iterative methods work because of a phenomenon called a fixed point. This is a point at which an arbitrary function's value is equal to its input. This can be expressed as:
\begin{equation}
    x = g(x)
\end{equation}
The property of functions to have fixed points is described by Brouwer's Fixed Point Theorem \cite{fixed_point}. As any equation of the form $f(x) = 0$ can be rewritten as  $x = g(x)$, by setting $g(x) = x + f(x)$, fixed points can occur in any root-finding problem. Although other theorems such as the Contraction Mapping Theorem \cite{suli_mayers_2003} give further requirements which guarantee convergence, these will not be explored in this paper. 
\newline \newline An iteration is a repetition of a process. Each iteration represents one execution of the entire algorithm and produces an output which is used by the subsequent iteration as a starting point. Iterative methods can continue indefinitely until some end criterion is reached. Particularly, in this paper, the termination criteria were defined in terms of tolerances above. Another factor that impacts the number of iterations is the speed at which the algorithm converges, which can be measured with the rate and order of convergence.
\subsection{Rate and Order of Convergence}
The rate and order of convergence of an algorithm are quantities that represent how quickly it approaches its limit. According to Süli \& Mayers, there are two ways to calculate the speed at which an iterative algorithm converges \cite{suli_mayers_2003}. Both ways are defined with respect to the error $e_n$ of a series which will be defined here as:
\begin{equation}
    e_n = |x_n-x^*|
\end{equation}
Estimating the rate $\mu$ from the above is possible by taking the quotient of two consecutive error terms $e_n$ and $e_{n-1}$ when $n$ is sufficiently large ($n \to \infty $):
\begin{equation} \label{eq: linear rate}
    \lim_{n\to \infty} \dfrac{e_n}{e_{n-1}} = \mu
\end{equation}
Based on the value of $\mu$, the algorithm's rate can be classified as linear if $\mu \in (0,1)$, with a smaller rate being faster than a larger rate. In this case, one can also calculate:
\begin{equation}
    p = -\log_{10}{\mu}
\end{equation}
which is called the asymptotic rate of convergence. If $\mu = 0$, then the rate of the convergence is considered super-linear and if $\mu = 1$, then the rate of the convergence is considered sub-linear.\newline \newline
Another way to calculate the rate of convergence of an algorithm is by first finding the order of convergence, which attempts to describe the polynomial behaviour of the algorithm around the root $x^*$. The base equation for this calculation is similar to \eqref{eq: linear rate} with the addition of the exponent $q$ which in this case represents the order of an algorithm:
\begin{equation} \label{eq: rate order conv}
     \lim_{n\to \infty} \dfrac{e_{n+1}}{e_n^q} = \mu
\end{equation}
If the order of convergence is higher, it will take fewer iterations for a series produced by an iterative method to reach its limit. Methods can be classified by their order, for example, a method is said to have linear convergence if $q = 1$. Other classifications are presented below.
\begin{equation*}
\begin{split}
        q & = 2 \to quadratic\\
        q & = 3 \to cubic
\end{split}
\end{equation*}
The values of $q$ and $\mu$ can be determined both mathematically and numerically from data.
\subsubsection{Estimating rate and order}
Estimating rate and order of convergence starts with the error definition as before. When $n$ becomes sufficiently large from \eqref{eq: rate order conv}, the following is true:
\begin{equation}\label{eq: en+1}
    e_{n+1} = \mu e_{n}^q
\end{equation}
The same relation exists for the previous index pair:
\begin{equation}\label{eq: en}
    e_{n} = \mu e_{n-1}^q
\end{equation}
Taking the quotient of \eqref{eq: en+1} and \eqref{eq: en} gives:
\begin{equation}
    \dfrac{e_{n+1}}{e_{n}} = \dfrac{\mu e_{n}^q}{\mu e_{n-1}^q}
\end{equation}
The factor $\mu$ can be cancelled to arrive at:
\begin{equation}
    \dfrac{e_{n+1}}{e_{n}} =\biggr( \dfrac{e_{n}}{e_{n-1}}\biggr)^q
\end{equation}
Solving for $q$ gives:
\begin{equation}\label{order}
    q = \dfrac{\log e_{n+1}/e_{n}}{\log e_{n}/e_{n-1}}
\end{equation}
From the above, it can be seen that at least three iterations are needed before the order of convergence can be calculated. With $q$ the rate of convergence $\mu$ can be found.
This is done by rearranging \eqref{eq: en+1} to:
\begin{equation}\label{rate}
    \mu = \dfrac{|e_{n+1}|}{|e_{n}|^q}
\end{equation}
In this paper, \eqref{rate} and \eqref{order} will be used for analysis. Using the error definition, another property can be defined.
\subsection{Basin of Attraction}
The basin of attraction is a region of space from which any initial guess will ultimately lead to the same answer, formally known as an attractor $x^*$. In order for the basin of an attractor to have a positive radius $r > 0$, the following has to be true \cite{Basin}:  
\begin{equation}
    \dfrac{e_{n+1}}{e_n } < 1
\end{equation}
Using the above and \eqref{eq: rate order conv}, one can derive an inequality which will define the basin of attraction of any iterative method, as:
\begin{equation}
    \mu e_n^{q-1} < 1
\end{equation}
In this paper, the basin of attraction will be found numerically, by testing a method across a range of values and recording whether it converged. The paper's focus will be on systems of non-linear functions, which produce very discontinuous basins of attraction.
\subsection{Non-linearity}
A linear function, or otherwise known as a linear map, is one which satisfies the following properties:
\begin{equation} \label{supperpos}
    f(a+b) = f(a)+f(b)
\end{equation}
\begin{equation} \label{homo}
     f(aT) = af(T)
\end{equation}
Equation \ref{supperpos} demonstrates the principle of superposition, while equation \ref{homo} shows the principle of homogeneity. If a function does not follow the conditions above, it is considered non-linear. Another definition of a non-linear function is one which possesses a second or higher derivative. One could argue that if a function has a higher-order derivative than another, it is more non-linear. However, as there is no formal definition for the term ``highly non-linear", this paper will consider any function $f(x)$ as highly non-linear if following holds true:
\begin{equation} \label{eq: nonlin1}
    \dfrac{d^2f(x)}{dx^2}\geq\lambda\dfrac{df(x)}{dx}
\end{equation}
where $\lambda$ is sufficiently large.
\section{Existing univariate methods}
Two common univariate root-finding methods are shown in this section. The univariate case is used to introduce the main subject of this section which is the Newton-Raphson method.
\subsection{Newton-Raphson}
The Newton-Raphson (NR) method, more commonly known as the Newton method, is an iterative root-finding algorithm. Its derivation begins by writing out the Taylor's series expansion of a function $f(x)$ at $x^*$ where $f(x^*) = 0$. The Taylor's series expansion is as follows: 
\begin{equation}
    f(x^*) = f(x_n) + f'(x_n)(x_n-x^*)+\dfrac{1}{2}f''(x_n)(x_n-x^*)^2+...
\end{equation}
where $'$ represent derivation with respect to the independent variable, $x_n$ shows the current value of $x$, and $(.)_n$ is the index of the iteration number. Ignoring all terms after the second one, the following equation is left:
\begin{equation} \label{taylors}
    f(x^*) = f(x_n) + f'(x_n)(x_n-x^*) = 0
\end{equation}
One can denote the difference between $x_n$ and $x^*$ as:
\begin{equation}
   (x_n-x^*) = \Delta x
\end{equation}
From here, \eqref{taylors} can be rearranged to the NR fixed point iteration:
\begin{equation} \label{eq: NM}
    f'(x_n)\Delta x = -f(x_n)
\end{equation}
The value of $\Delta x$ can be found by dividing $-f(x_n)$ by $f'(x_n)$, and then $\Delta x$ can be used to update the current value of $x$ as:
\begin{equation}
    x_{n+1} =x_{n}+\Delta x
\end{equation}
\subsection{Secant Method} \label{secant}
The secant method (SM) is an iterative root-finding algorithm that uses successive secant lines to a given function to create an approximation of the gradient at the latest point. It can be seen as a finite difference approximation of NR and due to its simplicity, it has been used throughout history to solve equations \cite{Secant}. The derivation of the secant method begins by choosing two potential roots of a function and evaluating the function at the choices. Afterwards, a secant line is drawn between the points and the x-intercept of the line is taken as the subsequent value in the series of $\{x_n\}$.
This is expressed by:
\begin{equation}
    x_{n+2} = x_{n+1}-\dfrac{x_{n+1}-x_{n}}{f(x_{n+1})-f(x_{x})}f(x_{n+1})
\end{equation}
which can be rearranged as:
\begin{equation}
   \dfrac{f(x_{n+1})-f(x_{x})}{x_{n+1}-x_{n}}\Delta x = -f(x_{n+1})
\end{equation}
In the above, the difference between  consecutive elements of $x$ is expressed as $\Delta x = x_{n+2}-x_{n+1}$. The secant method is a unique algorithm when it comes to setting up. In one dimension, $N=1$, SM needs two guess, $N+1$, to initialise, where $N$ represents the number of dimensions. This translates to higher dimensions. Hence, using this method in multivariate space becomes unviable due to the large number of initial guesses which need to be taken. 
\subsection{Example of root-finding}
\input{Newton_example}
\section{Existing multivariate methods}
There are multiple multivariate methods available which perform root-finding. However, the main focus of this paper will be the multivariate version of NR, which is discussed below.
\subsection{Index notation}
When describing tensors, this paper will use index notation. An index subscript, such as $(.)_i$, shows the number of the element in a first order tensor (vector). Multiple indices show higher order tensors, for example $(.)_{ij}$ shows a matrix with rows indexed by $(.)_{i}$ and columns indexed by $(.)_{j}$. Multiplying tensors with the same index indicates summation over that index and a subscript of a $(.)_{,j}$ shows derivation with respect to a variable with an index of $(.)_j$.
\subsection{Newton-Raphson}
NR in multivariate space ($x \in \mathbb{R}^N $) takes a similar form as before. The variable $x$ becomes a vector of size $N$, which will be denoted by $x_i^{(n)}$, where $i$ is the index of $x$, $(.)^n$ shows the iteration number and $N$ is the number of dimensions. The function $f$ also becomes a vector of size $N$ and is denoted by $F_i(x^{(n)})$. The multivariate form of NR is shown below:
\begin{equation} \label{eq: multivar newt}
    F_{i,j}(x^{(n)})\Delta x_i = -F_i(x^{(n)})
\end{equation}
where $\Delta x_i$ is the step used to update the value of $x_i$.
\section{Extended Method}
This paper will propose a modification to NR by applying a non-linear modification aimed to reduce the non-linearity of a function to the base multivariate function $F_i$.
\subsection{Non-linear modification \texorpdfstring{$P$}{P}}
In order to improve the performance of root-finding methods, the non-linearity of the problem can be reduced \cite{exnewton}. The way this is achieved is by using the definition of non-linearity in \eqref{eq: nonlin1} shown here for convenience:
\begin{equation*}
     \dfrac{d^2f(x)}{dx^2}\geq\lambda\dfrac{df(x)}{dx}
\end{equation*}
Rearranging the above gives:
\begin{equation} \label{eq: extended expl}
    \dfrac{\displaystyle\dfrac{d^2f(x)}{dx^2}} {\displaystyle\dfrac{df(x)}{dx}} \geq \lambda
\end{equation}
where $\lambda$ is a factor of non-linearity. To reduce the non-linearity of the problem, $\lambda$ can be set to zero. This can only be true when the numerator of \eqref{eq: extended expl} is zero and that in itself is true only if $f(x)$ is linear. Thus, a modification to the function is needed that will convert $f(x)$ to a linear function. This new converted function, $r(x, c)$, can be expressed through a conversion function \cite{exnewton}, $P(x, c)$, as:
\begin{equation}
    r(x,c) = mx-c = P(x,c)f(x)
\end{equation}
From here $P(x,c)$ can be found as:
\begin{equation}
    P(x,c) = \dfrac{mx-c}{f(x)}
\end{equation}
The function $P(x,c)$ introduces new roots and eliminates the desired root. Hence, extra modifications are needed. As demonstrated by Aggarwal \& Pant \cite{exnewton}, the final modified form of the function is:
\begin{equation}
    P(x,c) = \dfrac{x-c}{f(x)-f(c)}
\end{equation}
The conversion eliminates the linear factor $m$ by setting it to $1$ to maintain generality and introduces $-f(c)$ in the denominator to eliminate the newly introduced root of $x = c$. Subsequently, the final expression of $r(x,c)$ becomes:
\begin{equation} \label{eq: ENR fin}
   r(x,c) = P(x,c)f(x) = \dfrac{x-c}{f(x)-f(c)}f(x)
\end{equation}
\subsubsection{Multivariate expansion}
The above-described method was defined in the univariate space. This paper will show the development of the method in the multivariate space.\newline\newline The base equation \eqref{eq: ENR fin} needs to be modified. This is done by changing the variables and functions into vectors:
\begin{equation}
\begin{array}{c}
    x\longrightarrow x_i\\
    c\longrightarrow c_i\\
    f(x) \longrightarrow F_i(x)\\
    r(x,c) \longrightarrow q_{ij}(x,c)
\end{array}
\end{equation}
and then rearranging \eqref{eq: ENR fin} to:
\begin{equation}
   q_{ij}(x, c) = (x_i-c_i)\biggr(\dfrac{F_j(x)}{F_j(x)-F_j(c)}\biggr)
\end{equation}
where $i, j \in \{1,2,3,...,N\}$ and $N$ shows the dimension of the problem. Here a special rule is set where no summation over the index $(.)_j$ will be performed to maintain the proper shape of the answer. 
\subsection{Extended Newton-Raphson (ENR) method}
With the multivariate form of the linear modification, ENR can be derived from \eqref{eq: multivar newt} as:
\begin{equation}\label{eq: ENR der}
    q_{ij,k}(x^{(n)}, c)\Delta x_k^{(n)} = -q_{ij}(x^{(n)},c)
\end{equation}
To solve \eqref{eq: ENR der}, the inverse of $q_{ij,k}$ needs to be found. However, obtaining the inverse cannot be naturally realised because $q_{ij,k}$ is a third-order tensor. A third-order tensor cannot act as a linear map from a vector space to itself and, thus, no such map can have an inverse. In order to to get the inverse, $q_{ij,k}$ needs to be modified to a second order tensor. One can set a matrix $w_{Lk}$ to be equal to:
\begin{equation}
    w_{Lk} = q_{ij,k}
\end{equation}
where $ L = N \times (i-1) + j$, which leads to $L\in\{1,2,3,...,N^2\}$; hence, the shape of $w_{Lk}$ is $[N^2, N]$. This is a second-order tensor, so an inverse is possible. However, the shape of the tensor is guaranteed to be rectangular, so a regular inverse cannot be taken. For that reason, a generalised inverse, better known as a Moore-Penrose inverse \cite{penrose_1955}, has to be taken. This inverse is constructed using singular value decomposition (SVD) and is denoted by the superscript $(.)^+$. According to the generalised inverse theory of matrices \cite{Wang2018}, the matrix $w_{Lk}$ will have a unique Moore-Penrose inverse.
Once the inverse of $w_{Lk}$ has been found, the function matrix $p_{ij}$ has to be transformed to the modified function vector $p_L$, using the same $L = N \times (i-1) + j$ method. This will yield the final equation as:
\begin{equation} \label{eq: ENR dx}
    \Delta x_k = w^+_{Lk}p_{L}
\end{equation}
The construction of $w_{Lk}$ is critical to the functioning of the algorithm. Hence, the derivation will be presented. For the equation to be solvable, the rank of the matrix has to be equal to the number of elements of $x$. In order to find what rank the final matrix will be, Sylvester’s rank inequality will be used \cite{sylvester}. It states that if $a_{ij}$ is a matrix of shape $[m\times n]$ and $b_{jk}$ is a matrix of shape $[n\times p]$ then:
\begin{equation}
    rank(a_{ij})+rank(b_{jk})-n\leq rank(a_{ij}b_{jk})
\end{equation}
Getting the derivative of $q_{ik}$ begins by writing out its expression as:
\begin{equation}
    w_{Lk} =\dfrac{\partial}{\partial x_k}\biggr[(x_i-c_i)\biggr(\dfrac{F_j(x)}{F_j(x)-F_j(c)}\biggr)\biggr]
\end{equation}
Using the product rule, the equation is transformed into:
\begin{equation} \label{eq: ENR fexpand}
    w_{Lk} =\dfrac{\partial(x_i-c_i)}{\partial x_k}\biggr(\dfrac{F_j(x)}{F_j(x)-F_j(c)}\biggr)+(x_i-c_i)\dfrac{\partial}{\partial x_k}\biggr(\dfrac{F_j(x)}{F_j(x)-F_j(c)}\biggr)
\end{equation}
The fraction in the brackets will be set to equal a new tensor:
\begin{equation}
    G_j = \biggr(\dfrac{F_j(x)}{F_j(x)-F_j(c)}\biggr)
\end{equation}
Consequently, \eqref{eq: ENR fexpand} can be rewritten as:
\begin{equation}
     w_{Lk} =\dfrac{\partial(x_i-c_i)}{\partial x_k}G_j+(x_i-c_i)\dfrac{\partial G_j}{\partial x_k}
\end{equation}
The expression of the partial derivative of $G_j$ can then be written out as:
\begin{equation}\label{eq: ENR 2}
    \dfrac{\partial G_j}{\partial x_k} =\left[\dfrac{F_{j,k} \biggr(F_j(x)-F_j(c)\biggr) - \dfrac{\partial}{\partial x_k}\biggr(F_j(x)-F_j(c)\biggr)F_j(x) }{\biggr(F_j(x)-F_j(c)\biggr)^2}\right]
\end{equation}
The derivative of $F_j(x)-F_j(c)$ becomes equal to $F_{j,k}$ as the derivative of $F_j(c)$ with respect to $x_k$ is equal to zero. This yields the final form of \eqref{eq: ENR 2} as:
\begin{equation}\label{eq: ENR 2 sim}
    \dfrac{\partial G_j}{\partial x_k} = F_{j,k}\left( \frac{-F_j(c)}{(F_j(x)-F_j(c))^2}\right)
\end{equation}
Here, the fraction in the brackets will be set to a new tensor:
\begin{equation}
    H_j = \frac{-F_j(c)}{(F_j(x)-F_j(c))^2}
\end{equation}
This gives the final form of \eqref{eq: ENR fexpand} as:
\begin{equation}
    w_{Lk} = \dfrac{\partial(x_i-c_i)}{\partial x_k}G_j+(x_i-c_i)F_{j,k}H_j
\end{equation}
Solving the above yields:
\begin{equation}
    w_{Lk} = \delta_{ik}G_j+(x_i-c_i)F_{j,k}H_j
\end{equation}
where $\delta_{ik}$ is the identity matrix. Not summing over the index $(.)_j$ allows for this equation to preserve its rank. By applying Sylvester's inequality, the maximum rank of each tensor along the $(.)_k$ dimension is found to be $1$. This means that once the $ L = N \times (i-1) + j$ conversion is applied, the total rank of the new tensor $w_{Lk}$ will be equal to the sum of the matrix ranks along $(.)_k$. This will coincide with the dimension of the problem, making the system determined. Using this derivation of the problem, the ENR method is fully formalised. As ENR applies a modification to the base function, the convergence properties of the method are identical to that of NR. Hence, this method retains super-linear convergence near the root. The results of its testing are presented below. 
\section{Results}
In order to produce reliable results, a testing methodology was established. 
\subsection{Methods}
\begin{itemize}
\item To create robust results, a uniform random sample of data was used for the initial choice. This is achieved through Latin hyper-cube sampling (LHS) \cite{sahil_moza_2020_3929531} \cite{DEUTSCH2012763}.
\item Tolerances $t_r$ for $\Delta x$ and $t_a$ for $F(x)$ were set to be equal to $10^{-6}$ and $0.001414$ respectively. If each element of the result is $\pm0.001$ away from the actual result it is considered successful. Taking the norm of such a vector results in $\sqrt{0.001^2+0.001^2} = 0.001414$.
\item All functions were tested in the range of $[-50,50]$.
\item All plotted graphs have a resolution of $[512 \times 512]$ pixels.
\item The iteration limit was set at 100 iterations.
\item White areas represent starting points from which the algorithm did not converge within the predetermined steps (un-converged).
\item Black areas represent starting points from which the algorithm diverged or encountered other problems such as numerical overflow in exponents or zero division. 
\item For graphs with multiple roots, each colour represents a convergence to a separate root and is denoted by $r_i$, where $(.)_i$ shows the index of the root. The opacity of the colour relates to the number of steps taken to converge. 
\item The horizontal axis of each graph shows the range of $x_0$. The vertical axis of each graph shows the range of $x_1$. 
\item The performance of the methods will be gauged based on the percentage of the area in which the algorithm converged. This percentage is later referred to as coverage.
\item A set of equations was used to produce the results. The equations and their roots are as follows:
\begin{equation} \label{rf: fn:5}
\begin{array}{rcl} 
f_{1}(x) & = & x_0^3-3x_0x_1^2-1\\
f_{2}(x) & = & 3x_0^2x_1-x_1^3\\
x^* &= &(-0.500, 0.866),\\
    &&(1.000, 0.000),\\
    &&(-0.500, -0.866)

\end{array}
\end{equation}
\begin{equation} \label{eq: enm1}
\begin{array}{rcl} 
f_{1}(x) & = & e^{x_0}-x_1\\
f_{2}(x) & = & x_0\times x_1-e^{x_0}\\
x^* &=& (1.000, 2.718)\\
\end{array}
\end{equation}
\begin{equation}
\begin{array}{rcl} \label{eq: enm5}
f_{1}(x) & = & x_0^2-\dfrac{1}{x_0}+x_1\\
f_{2}(x) & = & \dfrac{1}{x_1}+x_0\\
x^* &=& (1.260, -0.794)\\
\end{array}
\end{equation}
\end{itemize}
\subsection{Choice of modification parameter \texorpdfstring{$c$}{c}}
Choosing a suitable value for the modification parameter $c$ is vital to the improved convergence characteristics of ENR. A particular choice of $c$, whether it is a constant value or a function of $x$, will be referred to as a configuration of $c$. If $c$ is picked close to $x^*$, then the convergence regions, which are shown below, are greatly improved. However, this requires prior knowledge of the roots of the system. If such knowledge is available then that configuration will outperform most others. However, to maintain generality and to help the method tackle systems whose roots are completely unknown, $c$ will primarily be taken as a function of $x$. In this situation, knowledge of the system is necessary to pick a good relation between $c$ and $x$, but that is more readily available as the general type of equation is known when setting up the system, i.e. exponential, high-order polynomial, etc. The sections below are going to compare different configurations of $c$ used by ENR. This will show their effect on the convergence area.
\subsection{Validation} \label{validation}
\begin{figure}[tb] 
    \centering
    \includegraphics[width=0.9\textwidth]{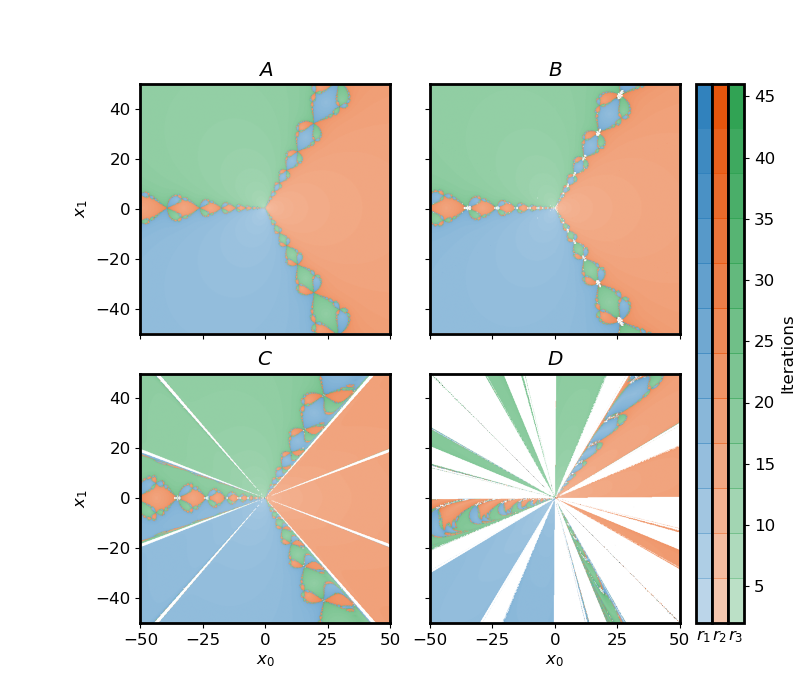}
    \caption[Root-finding: Validation graph \eqref{rf: fn:5}]{Figure showing results of \eqref{rf: fn:5}. Panel A shows the system solved with NR. Panels B-D show the system solved by ENR with different configurations. Panel B uses $c_i = 2x_i$. Panel C uses $[c_0, c_1] = [2x_0,3x_1]$. Panel D uses $c_i = x_i+10^{-5}$. Panel B generates a graph which most closely resembles the original NR graph. The other two modification introduce grater distortions.}
    \label{fig: rf5}
\end{figure}
In order to validate ENR it had to be tested in a configuration which should not be affected by the modification. One such case is \eqref{rf: fn:5}. For the predetermined range given above, NR provides 100\% convergence coverage. It also creates a unique fractal shape which can be often seen in other papers dealing with NR. Hence, in order for the rest of the result to be valid, ENR has to perform almost identically to NR in \eqref{rf: fn:5}. The performance of ENR is dependant on the choice of $c$. Here, multiple different arrangements of $c$ will be presented together with the results from NR. This will allow for comparison not only between NR and ENR but also between the different choices of $c$. In Fig. \ref{fig: rf5}-A the results for \eqref{rf: fn:5} evaluated by NR are presented. The NR method provides a 100\% coverage of the $[-50,50]$ range. A unique fractal pattern can also be seen. The second panel, Fig. \ref{fig: rf5}-B, sets $c$ as a function of $x$, where: 
\begin{equation}
    c_i = 2x_i
\end{equation}
This yields a $99.75\%$ coverage of the range. The white colour in the graph shows areas where the method exceeded the maximum iteration count. These zones are in the highly fractal border area which may explain the reason the method did not converge. The third panel, \ref{fig: rf5}-C, shows an alternate $c$ function:
\begin{equation}
    [c_0, c_1] = [2x_0,3x_1]
\end{equation}
ENR with the above function only yielded a $97.20\%$ coverage of the range. It also produced eight lines which radiate outwards from the centre point. These lines section off four slices which lay close to the diagonals of the graph. Their exact positions depend on the ratio between $c_0$ and $c_1$. If their ratio is one, then the sections lay exactly on the diagonals and produce a symmetric shape. If $c_0<c_1$, then the sections rotate towards the horizontal axis. This can be seen in \ref{fig: rf5}-C. Finally, if $c_0>c_1$, the section rotates towards the vertical axis. The final panel, \ref{fig: rf5}-D, shows the results when $c$ was set to a small perturbation of $x$ as:
\begin{equation}
    c_i = x_i+10^{-5}
\end{equation}
This yielded a coverage of $49.80\%$. The \rom{2} and \rom{4} quadrants are sparsely covered. Using the small perturbation may have caused the new modified function $r$ to have a very small value, as the modified function is proportional to the difference between $x$ and $c$. Having the value of the modified function be a small number reduces the size of the step taken, which will in turn increase the number of iterations. This can be tested by setting the value of:
\begin{equation}
    c_i = x_i+1
\end{equation}
In this case, the plot of the results looks similar to \ref{fig: rf5}-D. However, the coverage is up to $57.68\%$. This is due to the expansion of $r_2$ and $r_3$ convergence zones in the \rom{2} and \rom{4} quadrants. This was further tested by adding $2$ to the value of $x$ and the converge grew to $59.42\%$. This shows that selecting a value far away from the initial guess is beneficial to the convergence properties of ENR. However, this does have diminishing returns as the distance increases.\newline \newline Examining cases \ref{fig: rf5}-A and \ref{fig: rf5}-B, it can be seen that ENR produces the same fractal structure as NR and finds the same number of roots over the same domain. This validates that the method is functioning correctly.
\subsection{Comparison} \label{comp}
\subsubsection{Exponential}
\begin{figure}[htb!] 
    \centering 
    \includegraphics[width=0.9\textwidth]{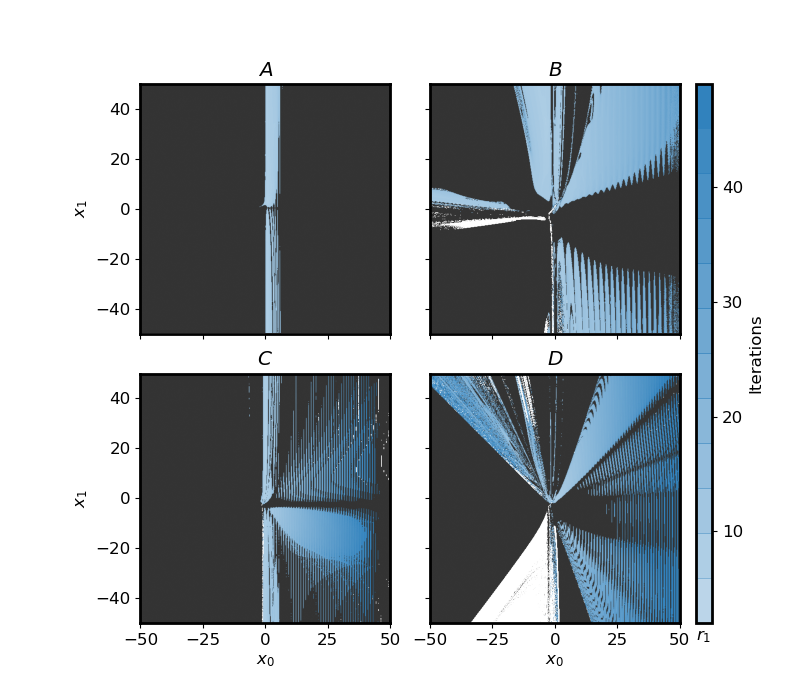}
    \caption[Root-finding comparison for \eqref{eq: enm1}]{The figure shows results of \eqref{eq: enm1} solved by NR ,panel A, and three configurations of ENR, panels B-D. Panel B represents a configuration of $c_i = \frac{1}{2}x_i$. Panel C shows the configuration of $c = [2x_0,3x_1]$ and panel D shows the configuration of $c_i = x_i+1$. All configurations of ENR produced a larger coverage of the range. Sections B and D also created un-converged regions in the \rom{3} quadrant.}
    \label{fig:rtf1}
\end{figure}

\begin{figure}[htb!] 
    \centering 
    \includegraphics[width=0.8\textwidth]{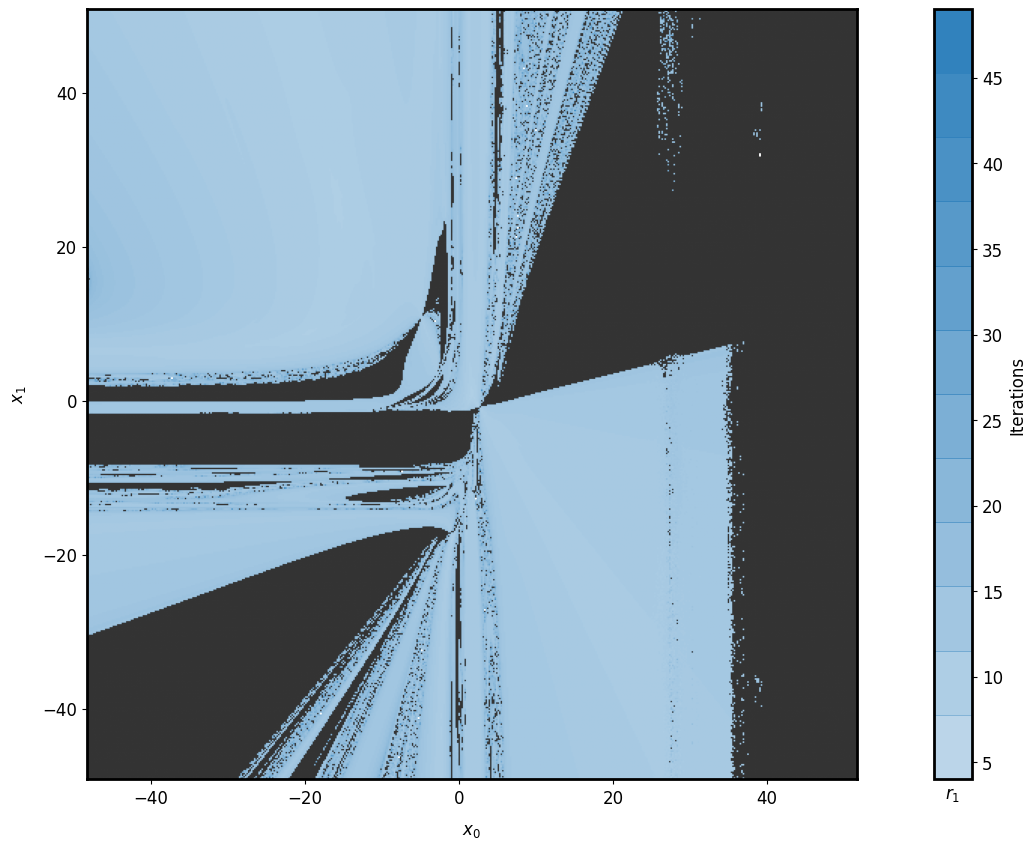}
    \caption[Convergence test range for static parameter (2, 1)]{Figure showing results of \eqref{eq: enm1} solved using ENR and a $c$ configuration of $c = (2,1)$. This yielded two large, smooth converged zones in the \rom{2} and \rom{4} quadrants. The total coverage is $57.24\%$.}
    \label{fig:rtf1_ext}
\end{figure}

The system \eqref{eq: enm1} is one where NR fails to converge for a large area of the testing range. The variable $x_0$ determines the magnitude of the exponential terms, so \eqref{eq: enm1} is exponential in $x_0$. However, the $x_1$ term is linear. It is expected for NR to have a large, smooth convergence range in $x_1$ and a narrow one in $x_0$. This is shown in Fig. \ref{fig:rtf1}-A. In the vertical direction, $x_1$, the graph covers the whole range with some discontinuities. In $x_0$, the covered range is a very slim strip close to the root. The total coverage of the range for this configuration by NR is $04.64\%$. The first configuration of ENR, shown in Fig. \ref{fig:rtf1}-B, $c$ was set to:
\begin{equation}
    c_i = \frac{1}{2}x_i 
\end{equation}
This yielded an improved range in the $x_0$ (non-linear) direction. The total coverage was brought up to $31.22\%$ which is an increase of $26.58\%$. The modified shape is not fully continuous. However, there is a large smooth area in the \rom{1} quadrant, which exhibits predictable behaviour i.e. when the initial guess is further away from the root, the number of iterations increases which can be seen by the darker shade around the edge of the range. There are also some white artefacts around the negative portion of the horizontal axis $(x_0<0, \:x_1 = 0)$. This shows that the iteration limit was exceeded in that area. Unlike in the previous example, the case $c_i = 2x_i$ yielded a similar shape to $c_i = \frac{1}{2}x_i$ with a reduced coverage. That is why it is not shown in this graph. One of the possible explanations for why $\frac{1}{2}x_i$ performed better than $2x_i$ is that as this problem is exponential, small changes in $x$ can cause a large difference in the result. This suggests that choosing smaller scalar modifiers can be beneficial. The third panel of the graph, Fig. \ref{fig:rtf1}-C, shows the same configuration as \ref{fig: rf5}-C, where:
\begin{equation}
    [c_0, c_1] = [2x_0,3x_1]
\end{equation}
Previously, this case caused distortions based on the ratio between the elements, but in the exponential system, the distortions are not present. The total coverage of this configuration amounted to $19.45\%$. It created a continuous region in the \rom{4} quadrant and repeating striations extending towards the upper limit of the $x_0$ range. Reducing the value of the second variable, as shown by the previous case, also improves the convergence properties of the method. For example, the total coverage of $c = [3x_0,2x_1]$ is $46.18\%$. It extended the region of convergence in the \rom{2} quadrant and reduced the spacing between the striations in the \rom{1} and \rom{4} quadrants.
The final configuration shown, in \ref{fig: rf5}-D, is:
\begin{equation}
    c_i = x_i+1
\end{equation}
This configuration yielded a total coverage of $31.36\%$. Comparing panels B and D, it can be observed that both create a smooth region in the \rom{1} quadrant. Both have an area in the \rom{2} quadrant which is discontinuous and striations in the \rom{4} quadrant. The difference in the shapes seems to mimic the order of the $c$ expression, i.e. case B appears to have an area generated by some quadratic relations and case D appears to have an area generated by a linear relation. In panel D the relation between $x$ and $c$ is a constant number which appears to have scaled to a linearly varying shape, whereas in panel B the linear relation between $x$ and $c$ appears to have scaled to a quadratically varying shape.\newline \newline The results from this section show that in the exponential case the ENR method consistently produces a larger coverage of the test range. Different choices of $c$ affect the shape but some general conclusions can be made, i.e. using smaller values when multiplying the exponential term yields better results. Generally, the convergence is improved if the choice of c is close to the root $x^*$. This can be seen in Fig. \ref{fig:rtf1_ext}. By setting $c = (2,1)$, a convergence coverage of $57.24\%$ can be attained with two large, smooth converged zones in the \rom{2} and \rom{4} quadrants.
\begin{table}
\centering
\begin{tabular}{||l l p{0.6\linewidth}||} 
 \hline
 $c$ & Coverage & Remarks
\\ [0.5ex]
 \hline\hline
 $None$ & $04.64\%$ & Narrow convergence on $x_0$ axis\\
 $x_i-10^{-5}$ & $26.18\%$ & Striations and un-converged region\\ 
 $x_i+10^{-5}$ & $26.14\%$ & Striations and un-converged region\\
 $[2x_0,3x_1]$ & $19.45\%$ & Heavy striations, no coverage of $-x_0$ direction\\ 
 $[3x_0,2x_1]$ & $46.18\%$ & Large continuous regions along $x_0$ axis\\ 
 $0.5x_i$ & $31.22\%$ & Continuous regions and small un-converged region\\ 
 $2x_i$ & $11.44\%$ & Small expansion in $+x_0$ direction\\ 
 $[1,2]$ & $73.74\%$ & Large smooth area due to vicinity of $c$ to $x^*$\\ 
 $[10,20]$ & $48.31\%$ & Large smooth area in the \rom{1} and \rom{4} quadrants, repeating stripes in the \rom{2} quadrant\\
 $x^2$ & $05.45\%$ & Wide strip of converged area close to $x_0 = 0$. Large area of un-converged points past $|x_0|>20$\\ 
 [1ex] 
 \hline
\end{tabular}

\caption[Exponential function tabulated coverage]{Tabulated results for \eqref{rf: fn:5} with different $c$ configurations and with brief remarks on the shape of the convergence region. The largest coverage is given by $c = [1,2]$.}
\label{exp1}
\end{table}
\subsubsection{Negative exponent}
\begin{figure}[htb!] 
    \centering 
    \includegraphics[width=0.9\textwidth]{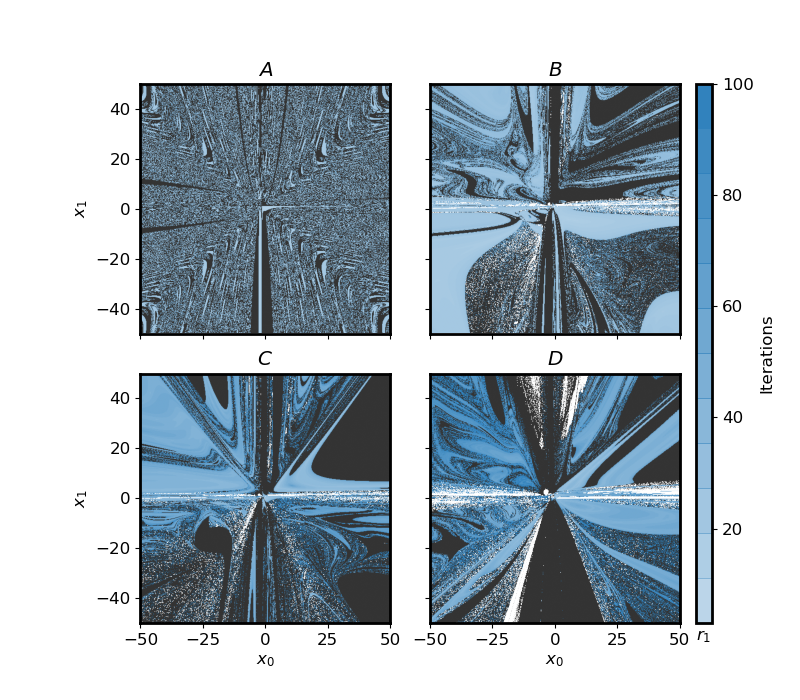}
    \caption[Root-finding comparison for \eqref{eq: enm5}]{Figure showing results of \eqref{eq: enm5} solved by both NR and ENR. NR yielded a discontinuous coverage of the range at $36.63\%$ shown in panel A. Panels B-D show different configuration of ENR. These are: B - $c_i = 2x_i$, C - $[c_0, c_1] = [2x_0,3x_1]$, D - $c_i = x_i+10^{-5}$. All configurations of ENR were able to produce larger coverage than NR. However, their coverage shape was distorted and non-symmetric.}
    \label{fig: rf4}
\end{figure}
Negative exponents in a system, such as \eqref{eq: enm5}, lead to behaviour similar to some non-linear material models. For this system NR has discontinuous coverage of the test range. This can be seen in Fig. \ref{fig: rf4}-A, where NR covers the whole test range with a disjointed pattern which exhibits some symmetries around the vertical and horizontal axes. The total coverage of NR in \eqref{eq: enm5} was $36.63\%$. The second configuration, shown in \ref{fig: rf4}-B, is that of:
\begin{equation}
    c_i = 2x_i
\end{equation}
This was chosen over the previous $c_i = \frac{1}{2}x_i$ because at smaller values a negative exponent function becomes more non-linear. When $c$ was set to half $x$, the total coverage was $44.29\%$, whereas when it was set to double the value of $x$, the total coverage was $49.34\%$. The total coverage of this configuration is larger than that of NR. However, the shape of the convergence region is less well distributed as compared to NR. In \ref{fig: rf4}-B, there are smooth regions close to $x_1 =0$. There is also a large smooth area present in the \rom{3} quadrant. The second configuration:
\begin{equation}
    [c_0, c_1] = [2x_0,3x_1]
\end{equation}
is shown in \ref{fig: rf4}-C. The total coverage was $41.62\%$. This configuration produced a large smooth area in the \rom{2} quadrant, but reduced the density of converged points in the other three quadrants. However, it was still able to produce a larger coverage of the range than NR. The shape of the convergence basin for the given range is affected by the ratio of $c_0$ to $c_1$. If $[c_0, c_1] = [3x_0,2x_1]$, then the coverage increases to $61.07\%$ and smooth regions appear in all 4 quadrants. The third configuration presented in \ref{fig: rf4}-D is:
\begin{equation}
    c_i = x_i+10^{-5}
\end{equation}
This perturbation of $x$ generated a coverage of $46.66\%$. This method changed the pattern of NR drastically but still managed to produce a larger coverage. There are large smooth areas in the \rom{1} and \rom{4} quadrants. The rest of the test range in scattered by converged and un-converged points.
\begin{figure}[htb!] 
    \centering 
    \includegraphics[width=0.8\textwidth]{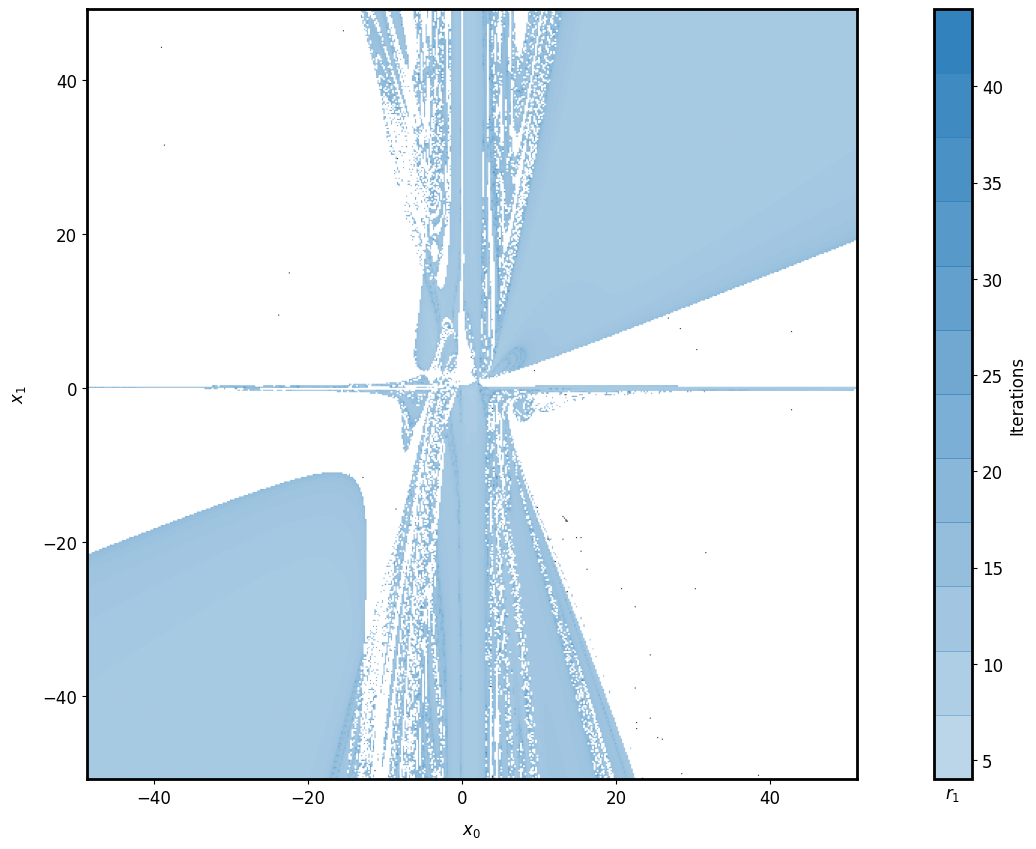}
    \caption[Convergence test range for static parameter (-2, 1)]{Figure showing results of \eqref{eq: enm5} when solved using ENR with $c = [-2, 1]$. This configuration yields two large smooth regions in the \rom{1} and \rom{3} quadrants. It also changes the behaviour of the method across the range. Rather than ENR diverging, like in the previous cases (black area), it exceeds the iteration limit (white area). The total coverage of this configuration is $44.88\%$.}
    \label{fig: rf4_ext}
\end{figure}
These results show the positive effects of the linear modification used by ENR. Even though the convergence area produced by ENR is on average less well distributed than NR in this case, ENR was able to produce a larger coverage overall. The final configuration to be examined is when $c$ was set to equal a constant $(-2, 1)$. This choice of $c$ is relatively close to the true value of $x^* = (1.260, -0.794)$ and it created a very desirable shape with large smooth sections, shown in Fig. \ref{fig: rf4_ext}. It also reduced the number of iterations by almost a half. The total coverage was $44.88\%$. This shows that if a value of $c$ is close to $x^*$, the convergence properties of the method improve drastically. This is also true if the initial guess $x_0$ is close to $x^*$, so using ENR with static values allows for two guesses at initial values. The results from the rest of the tested configurations are presented in Table \ref{nexp1}.
\begin{table}[H]
\centering
\begin{tabular}{||l l l||} 
 \hline
 $c$ & Coverage & Remarks
\\ [0.5ex]
 \hline\hline
 $None$ & $36.63\%$ & Disjoint converged areas across range\\
 $x_i-10^5$ & $46.62\%$ & Large distorted converged area with un-converged artefacts\\ 
 $x_i+10^5$ & $46.66\%$ & Large distorted converged area with un-converged artefacts\\
 $[2x_0,3x_1]$ & $41.62\%$ & Large smooth converged are in the \rom{2} quadrant\\ 
 $[3x_0,2x_1]$ & $61.07\%$ & Large converged areas in the \rom{1}, \rom{3} and \rom{4} quadrants\\ 
 $0.5x_i$ & $44.29\%$ & Large converged areas in the \rom{3} and \rom{4} quadrants\\ 
 $2x_i$ & $49.34\%$ & Large converged areas in the \rom{3} and \rom{4} quadrants\\ 
 $[1,2]$ & $13.58\%$ & Smooth area in the \rom{1} quadrant\\ 
 $[10,20]$ & $53.70\%$ & Large heavily striated area\\ 
 [1ex] 
 \hline
\end{tabular}
\caption[Negative exponent function tabulated coverage]{Tabulated results for the different $c$ configurations of \eqref{eq: enm5} with brief remarks on the shape of the convergence region. The largest coverage is given by $c = [10,20]$.}
\label{nexp1}
\end{table}
\section{Conclusion}
\begin{figure}
    \centering
    \includegraphics[width=0.75\textwidth]{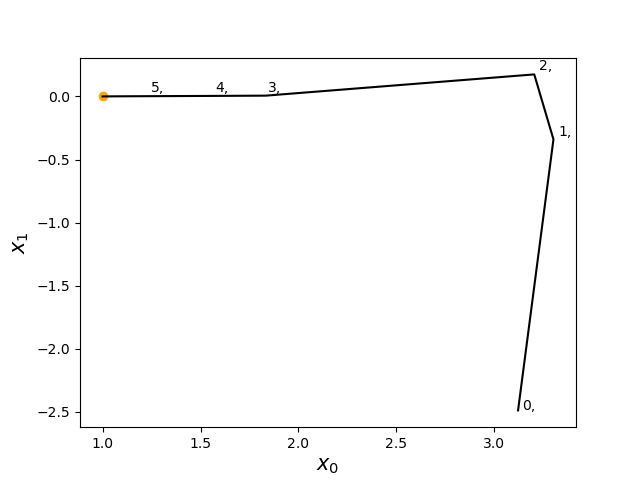}
    \caption[Diagonal secant method steps]{Figure showing steps taken by DS method for \eqref{rf: fn:5}. Numbers indicate the number change in the current approximation of $x^*$ and the orange dot shows the root of the system tested.}
    \label{fig: fs_steps}
\end{figure}
The ENR method provides improved convergence properties when compared to the NR method, as shown in \autoref{comp}. In highly non-linear systems, ENR extended the coverage of the proposed test range and in some cases, depending on the choice of the modification parameter $c$, reduced the number of iterations needed to converge. The modification, however, can also reduce coverage when systems are not highly non-linear, as demonstrated in \autoref{validation}. ENR, as shown by the derivation, is more computationally demanding than NR. However, due to the method's improved properties, it can serve as a useful technique when dealing with highly non-linear systems.
\section{Exploration of Diagonal Secant Method (DS)}
\begin{figure}
    \centering
    \includegraphics[width=0.75\textwidth]{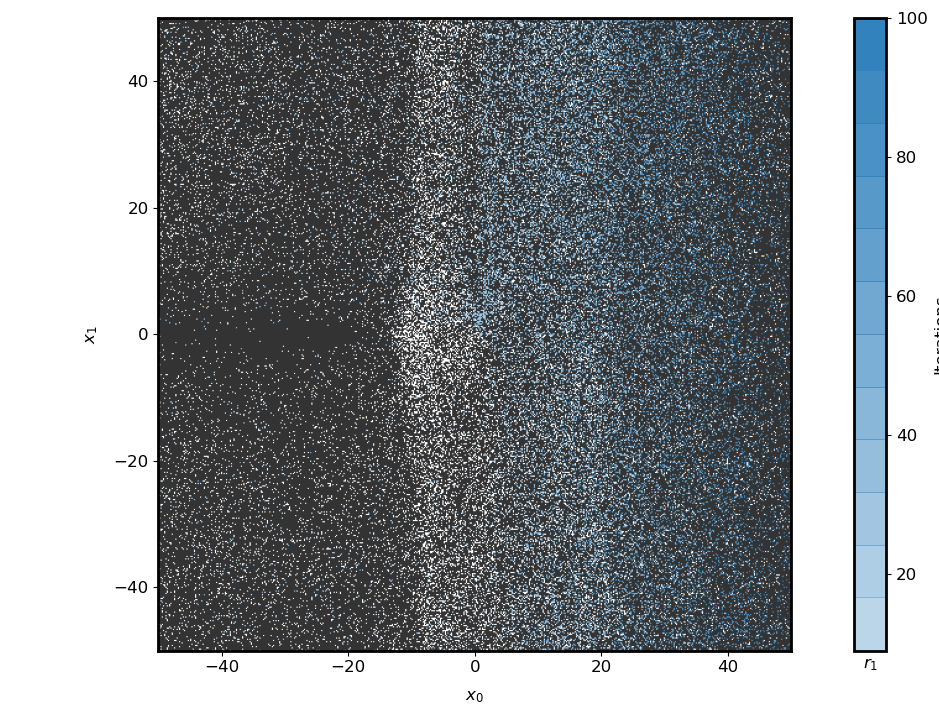}
    \caption[Diagonal secant convergence basin]{Figure showing convergence basin of the DS method.}
    \label{fig: fs_main}
\end{figure}
As mentioned in \autoref{secant}, the secant method with two initial values only exists in one dimension. If the secant method is to be expanded to the multivariate space, one has to increase the number of initial guesses. This makes this method very cumbersome for high dimensional problems. It was attempted to create a method based on the secant approximation which works in multiple dimensions by using only two initial guesses. This was attempted due to the secant method's low computational demand. The derivation of the method began by first taking two guesses denoted by $x^{(0)}_i$ and $x^{(1)}_i$. The system $F_i$ was evaluated at both points and the difference between both evaluations was recorded in a variable:
\begin{equation}
    \Delta H_i = F_i(x^{(1)}) - F_i(x^{(0)})
\end{equation}
The difference in the two guesses was taken as:
\begin{equation}
    \Delta X_i = x^{(n+1)}_i-x^{(n)}_i
\end{equation}
From here, the secant step $\Delta s_{i}$ is calculated as:
\begin{equation}
   \Delta s_{i} = -\frac{-F_i(x^{(n+1)})}{\Delta H_i}\Delta X_i
\end{equation}
where, as in the ENR derivation, no summation over $(.)_i$ was done. This yielded a method which was able to find roots in \eqref{rf: fn:5} and  \eqref{eq: enm1}, but not in \eqref{eq: enm5}. The path which the DS method took to find a root of \eqref{rf: fn:5} with $x^{(0)} = (3,2)$ and $x^{(1)} = (3.5,1)$ is shown in Fig. \ref{fig: fs_steps}. The method finds one element of the variable first and then searches for the other. This behaviour led to a step-like path to be taken to the root. The DS method was also tested across the same test range as ENR which produced Fig. \ref{fig: fs_main}. The second choice $x^{(1)}$ was made by choosing two random values in the range of $[-50,50]$. Hence, the graph shows all initial choices $x^{(0)}$ paired with a random $x^{(1)}$. Fig. \ref{fig: fs_main} shows that the coverage of the range is very poor and very discontinuous. The number of steps taken to converge varied widely and if $x^{0}_i = x^{1}_i$ for any $i$, then the method will fail immediately. All of this combined meant that the method will not be expanded upon further in this paper. 
\chapter{Minimisation}
\section{Introduction}
\subsection{Definition}
\begin{figure}
    \centering
    \includegraphics[width=0.75\textwidth]{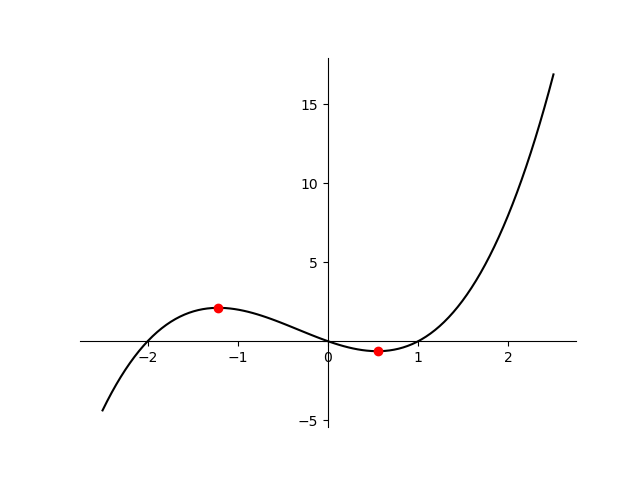}
    \caption[Local extrema]{A graph of the cubic equation $y = (x-2)x(x+1)$ plotted in the range of [-3,3]. Red markers show the positions of local extrema in the range.}
    \label{fig: extremum}
\end{figure}

Minimisation is a mathematical procedure used to locate local or global minima in a function. This involves calculating or estimating multiple derivatives of the function and using them to find a direction leading to a minimum. Minimisation is related to root-finding as it performs root-finding on the derivative of square error of a function rather than the function itself. The error of a function is expressed through the residual, which is the difference between a function's output $f(x, \theta)$, in its current configuration $\theta$, and an observed result. This observed result can be interpreted as the output of the same function $f(x,\theta^*)$ with the observed configuration $\theta^*$ and with some added noise, arising from measurement errors. The goal of minimisation is finding the desired configuration. The process is called minimisation because, as the observed configuration is approached by the current one, the difference between $f(x,\theta)$ and $f(x,\theta^*)$ becomes smaller (minimised).\newline \newline
Graphically this can be seen as finding the location of zero slope of the base function. Using the function from the root-finding example, one can see the local extremums in Fig. \ref{fig: extremum}. Finding the location of zero slope can lead to either of the two extremums in Fig. \ref{fig: extremum}, which is why a minimisation algorithm is constructed in such a way to only find minima. 
\subsection{Error}
\begin{figure}
    \centering
    \includegraphics[width=0.75\textwidth]{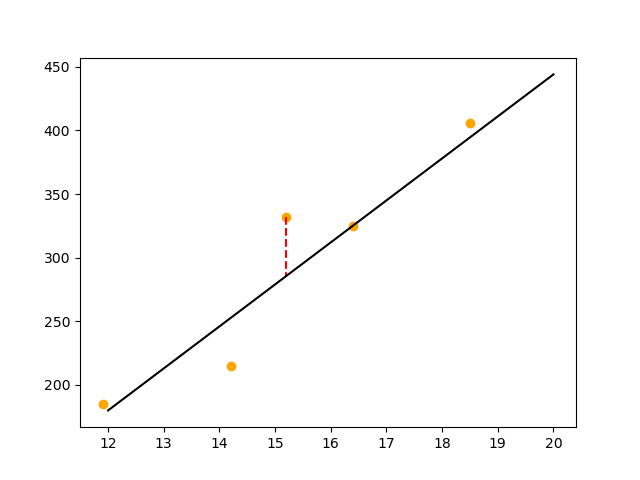}
    \caption[Residual visualisation]{The figure shows a scatter of random data (yellow points) with a linear approximation of that data drawn (black line). One residual is shown on the graph by a dashed red line connecting one of the data points and the approximation line. The length of that line is the magnitude of the residual.}
    \label{fig: residual}
\end{figure}

In minimisation problems, error is the target parameter. The most common way of defining error is as the difference between an observed result and a calculated result. This difference between results is known as a residual and can be defined using $f(x,\theta^*)$, which is the function $f$ evaluated with the desired configuration $\theta^*$. This definition will be used, even though residuals can be defined in terms of empirical observations. All observed results for this paper were artificially generated using $f(x,\theta^*)$ and added noise. The residual can also be shown graphically, which is done in Fig. \ref{fig: residual}. There, one can see that the residual is the distance between the observed points and the proposed model (straight line).
As minimisation tries to reduce the total error in a function, the residual needs to be taken all across the domain of $x$. Hence, the sum of all residuals is taken. To account for different signs which the residuals might have the residuals is are squared. This is known as the square error (SE) and can be calculated as:
\begin{equation}
    SE = r_ir_i
\end{equation}
This allows for an accurate estimation of the system's total error. By using this definition of error and previously explained concepts in \autoref{concepts}, existing minimisation methods can be explored.
\section{Existing methods}
\subsection{Gauss-Newton (GN)}
The Gauss-Newton method is an iterative minimisation algorithm based on the NR method. It performs root-finding on the derivative of the SE function, which locates a minimum. The GN method ignores second order derivatives of the residual to minimise computational costs, but that means that the GN step is an approximation to the most efficient direction. The derivation of GN shows this in more detail. Firstly, the residual is calculated, as:
\begin{equation}
    r_i(\theta) = y_i-f(x_i,\theta)
\end{equation}
Then the function to be minimised $F(\theta)$ is derived by summing the squared residuals and multiplying by a half. The factor is put here in order to eliminate a $2$ which is produced by derivation later on.
\begin{equation}
    F(\theta) = \dfrac{1}{2}r_i(\theta)r_i(\theta)
\end{equation}
The desired value of $\theta$ is $\theta^*$. In order to find $\theta^*$, it is set equal to $\argmin{F(\theta)}$ as:
\begin{equation}
    \theta^* = \argmin{F(\theta)}
\end{equation}
The function $\argmin{}$ represents the value of $\theta_j$ for which the following holds true:
\begin{equation}
    \dfrac{\partial}{\partial\theta_j}\biggr[\dfrac{1}{2}r_i(\theta) r_i(\theta)\biggr] = 0
\end{equation}
The above equation can be simplified to produce the following:
\begin{equation}
    \dfrac{\partial}{\partial\theta_j}\biggr[\dfrac{1}{2}r_i r_i\biggr] = r_i(\theta)\dfrac{\partial r_i}{\partial\theta_j}\biggr\rvert_{\theta_j} = 0
\end{equation}
\begin{equation}
    r_i(\theta)\dfrac{\partial r_i}{\partial\theta_j}\biggr\rvert_{\theta_j} = r_i(\theta) r_{i,j}(\theta) = 0
\end{equation}
where $(.)_{,j}$ shows derivation with respect to index $\theta_j$. Next, the Gauss-Newton iteration has to be derived. This can be done using the Taylor's series expansion of $r_ir_{i,j}$ around an initial guess of $\theta$ called $\theta^0$. The expansion is written out as:
\begin{equation}
r_i(\theta^*)r_{i,j}(\theta^*) = r_i(\theta^0)r_{i,j}(\theta^0) +\dfrac{\partial}{\partial\theta_k} \biggr[r_i(\theta^0)r_{i,j}(\theta^0)\biggr]_{\theta_k^0}(\theta_k^*-\theta_k^0)+...
\end{equation}
Evaluating this expansion up to the second term yields:
\begin{equation}\label{eq: taylor exp}
r_i(\theta^*)r_{i,j}(\theta^*) = r_i(\theta^0)r_{i,j}(\theta^0) + (r_{i,k}(\theta^0)r_{i,j}(\theta^0)+ r_i(\theta^0)r_{i,jk}(\theta^0))\Delta\theta_k
\end{equation}
where $\Delta\theta_k = \theta_k^*-\theta_k^0$. This is the full expansion up to the second term and from here, one can derive GN by ignoring second order derivatives i.e. $(.)_{i,jk}$ and re-arranging to get:
\begin{equation}\label{eq: GN primer}
r_i(\theta^*)r_{i,j}(\theta^*) = r_i(\theta^0)r_{i,j}(\theta^0) + r_{i,k}(\theta^0)r_{i,j}(\theta^0)\Delta\theta_k = 0
\end{equation}
From the above, the problem can be re-arranged again to get the final form of the GN method:
\begin{equation}\label{eq: Full GN}
r_{i,k}(\theta^0)r_{i,j}(\theta^0)\Delta\theta_k  = -r_i(\theta^0)r_{i,j}(\theta^0)
\end{equation}
In the above equation, $\Delta\theta_k$ is the so called GN step. To find its value, LU (lower-upper) decomposition can be used. Practically, this was done using numpy.solve() which internally uses an algorithm called dgesv() that utilises LU decomposition. Once $\Delta\theta_k$ has been calculated it is used to update the initial guess $\theta^0$ and every other consecutive guess until the termination criteria are reached. The update of $\theta$ is done by:
\begin{equation}
    \theta^{(n+1)}_k = \theta^{(n)}_k+\Delta\theta_k
\end{equation}
where the superscript shows the iteration number. The GN step is always in a decent direction. This can be shown by examining the gradient along the step. To find the gradient one can take the inner product of \eqref{eq: Full GN} with $\Delta \theta_j$ on both sides:
\begin{equation}
    \Delta \theta_jr_{i,k}(\theta^0)r_{i,j}(\theta^0)\Delta\theta_k  =-\Delta \theta_jr_i(\theta^0)r_{i,j}(\theta^0)
\end{equation}
If $r_{i,k}(\theta^0)$ is full rank, then the matrix $r_{i,k}(\theta^0)r_{i,j}(\theta^0)$ is positive definite and therefore the left-hand side of the equations is always positive. Thus, on the right-hand side, $\Delta \theta_jr_i(\theta^0)r_{i,j}(\theta^0)$ is always negative. Hence, the GN method always has a step in the decent direction. To further explain how minimisation works, a simple example will be given.
\subsection{Example of minimisation}
\input{GN_example}
\section{Corrected Gauss-Newton (CGN)}
This project proposes a correction to the Gauss-Newton method which utilises the second derivative of the residual, which was omitted during the derivation, between \eqref{eq: taylor exp} and \eqref{eq: GN primer}. This will increase the computational complexity, but it will also improve the convergence behaviour in highly non-linear functions. The proposed improvement is a two step process which uses the original GN step to then calculate a corrected step using the second derivative. The derivation begins by getting the regular GN step, shown in \eqref{eq: Full GN}. This is restated here for convenience.
\begin{equation} \label{eq: cgn og step}
    r_{i,k}(\theta^0)r_{i,j}(\theta^0)\Delta\hat{\theta}_k  = -r_i(\theta^0)r_{i,j}(\theta^0)
\end{equation}
The only modification to this equation is the appearance of $\Delta\hat{\theta}_k$ which will be used to denote the original GN step. Once the GN step is found, the Taylor's series expansion of the residual is written out to the second term as:
\begin{equation}
    r_i(\theta) = r_i(\theta^0) + r_{i,j}(\theta^0)\Delta\theta_j + \dfrac{1}{2}r_{i,jk}(\theta^0)\Delta\theta_j\Delta\theta_k
\end{equation}
Then the corrected GN step $\Delta\theta_j$ can be factored out to produce:
\begin{equation}
    r_i(\theta) = r_i(\theta^0) + \biggr(r_{i,j}(\theta^0) + \dfrac{1}{2}r_{i,jk}(\theta^0)\Delta\hat{\theta}_k\biggr)\Delta\theta_j
\end{equation}
For convenience, the term in the brackets is set to equal $s_{ij}$ and the equation is re-written again as:
\begin{equation}
    r_i(\theta) = r_i(\theta^0) + s_{ij}(\theta^0)\Delta\theta_j
\end{equation}
Substituting the new definition of $r_i(\theta)$ into \eqref{eq: cgn og step} and noting that $r_{i,j}(\theta) \approx s_{ij}(\theta^0)$ and that:
\begin{equation}
    s_{ij}(\theta^0) = \biggr(r_{i,j}(\theta^0) + \dfrac{1}{2}r_{i,jk}(\theta^0)\Delta\hat{\theta}_k\biggr)
\end{equation}
the following can be derived:
\begin{equation} \label{eq: cgn step}
    s_{ik}(\theta^0)s_{ij}(\theta^0)\Delta\bar{\theta}_k  = -r_i(\theta^0)s_{ij}(\theta^0)
\end{equation}
where $\Delta\bar{\theta}_k$ denotes the CGN step. Its value can be found using the same methods as before, namely LU decomposition. If $s_{ik}(\theta^0)$ is full rank, then the the gradient along the step direction is always negative.
\section{Results} \label{GN_Results}
\subsection{Methods}
\begin{itemize}
\item In order to create robust results, a uniform random sample of data is needed to create the initial guess. In this project, this is achieved through Latin hyper-cube sampling (LHS) which provides such uniform data in a given domain \cite{DEUTSCH2012763} \cite{sahil_moza_2020_3929531}.
\item To make the results more general, the average of three samples was taken per specific domain.
\item The distance parameter, which shows how far away the initial guess is from the actual parameters of the equation, is taken on a log scale.
\item A noise parameter was introduced to the data, represented by a Signal to Noise ratio in decibels. The range was adjusted for each experiment.
\item The resulting graphs use distance on the x axis and noise on the y axis. The steps are plotted on a contour plot with different colours representing a different amount of steps. Fractional steps are can be observed due to the averaging of results and due to linear interpolation used between data points. 
\item White space in the graphs represents areas where the method failed by exceeding 100 iterations steps. This region will be referred to as un-converged.
\item A set of equations was used to generate data for testing. The equations, their true values $\theta^*$ and the sampling ranges $s$ are shown:
\begin{equation} \label{gn:1}
\begin{array}{rcl} 
 f(x, \theta) &=& \theta_0x^3 + \theta_1x^2 + \theta_2x + \theta_3 + \theta_4sin(x),\\
 &&\theta^* = \{-0.001, 0.1, 0.1, 2, 15\}\\
 &&s = [1,10]
\end{array}
\end{equation}
\begin{equation} \label{gn:2}
\begin{array}{rcl}
 f(x, \theta) &=& \theta_0^3x^3 + \theta_1^2x^2 + \theta_2^2x + \theta_3^3 + \theta_4sin(x),
 \\
 &&\theta^* = \{-0.001, 0.1, 0.1, 2, 15\}\\
 &&s = [1,10]
\end{array}
\end{equation}
\begin{equation} \label{gn:3}
\begin{array}{rcl}
f(x, \theta) &=& \theta_0x_0^{\theta_1}+\theta_2x_1^{\theta_3},\\ &&\theta^* = \{0.1,4,0.1,2\}\\
&&s = [1,10]
\end{array}
\end{equation}
Note: if a negative variable is raised to a non-integer power only the real part of the result is taken.
\begin{equation} \label{gn:4}
\begin{array}{rcl}
f(x, \theta) &=& \theta_0exp(-x_0/\theta_1)+\theta_2exp(-x_1/\theta_3)\\ &&\theta^* = \{4,2,1,10\}\\
&&s = [0.1,10]
\end{array}
\end{equation}
\end{itemize}
\subsection{Validation (Linear results)} \label{GN validation}
\begin{figure}
    \centering
    \includegraphics[width=\textwidth]{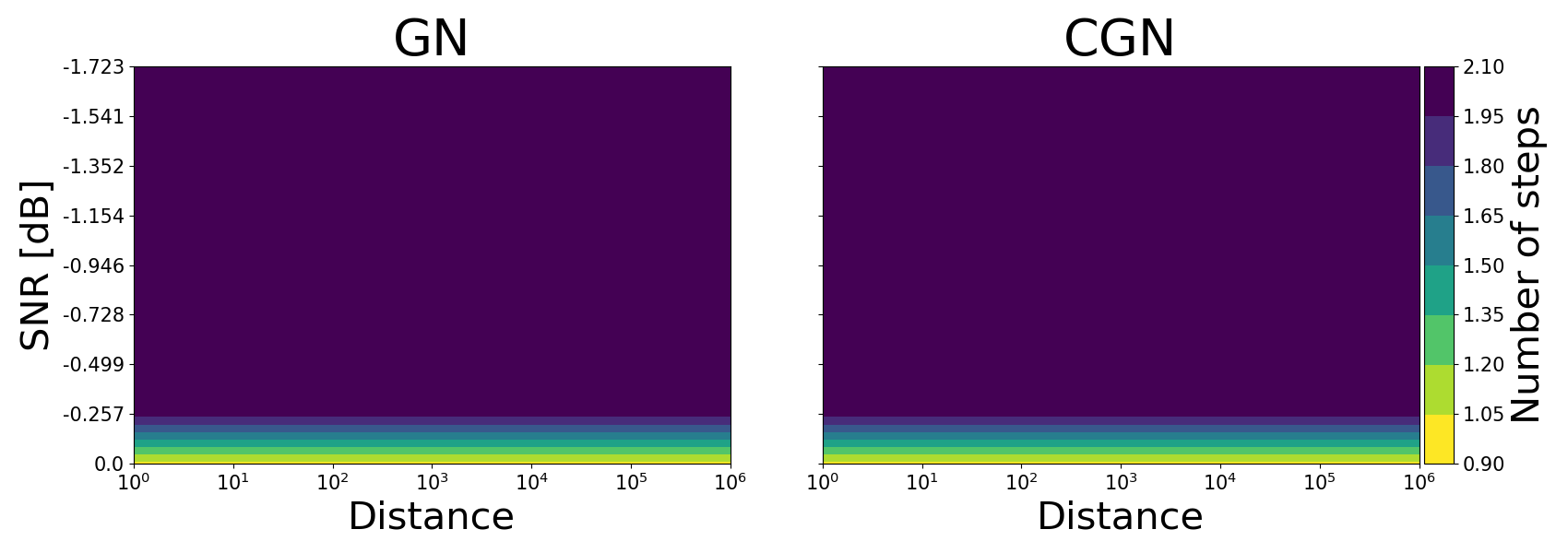}
    \caption[Minimisation \eqref{gn:1} graphs]{The figure shows the results of \eqref{gn:1}. The comparison between GN and CGN was plotted across a range of distances and noise. Both methods produce identical graphs, validating CGN through the linear cases. The information provided by the graphs shows that with the increase of noise both methods need more iterations to pass the tolerance criterion.}
    \label{fig: linear CGN}
\end{figure}

The CGN method should only diverge from the GN method in functions which are non-linear in their parameters. If the parameters are linear, the second derivative of the residual $r_{r,jk} = 0$ and the newly created matrix $s_{ij} = r_{i,j}$. This leads to \eqref{eq: Full GN} being equivalent to \eqref{eq: cgn step}. Hence, comparing result of both methods in such a linear function can be used as a baseline to see if CGN behaves as expected. An equation which is linear in its parameters is \eqref{gn:1}. There it is expected that both GN and CGN will produce the same results. The outcome of this test can be seen in Fig. \ref{fig: linear CGN} which shows two identical graphs. This means that CGN worked as expected and all consequent results can be taken as valid. Fig. \ref{fig: linear CGN} shows graphs where, when there is no noise added, both methods find the answer in one step. This is a known behaviour of the GN method, inherited by CGN. The region of dark blue near the top of both graphs is caused by the increased noise level.
\subsection{Comparison}
In the comparison section, the two methods will be contrasted with each other. The first function which will be examined is \eqref{gn:2}. This function is a modification of \eqref{gn:1}, where powers were added to the parameters of $x$. The function was chosen to represent a classical polynomial case.
\subsubsection{Polynomial}
\begin{figure}[htb!] 
    \centering
    \includegraphics[width=\textwidth]{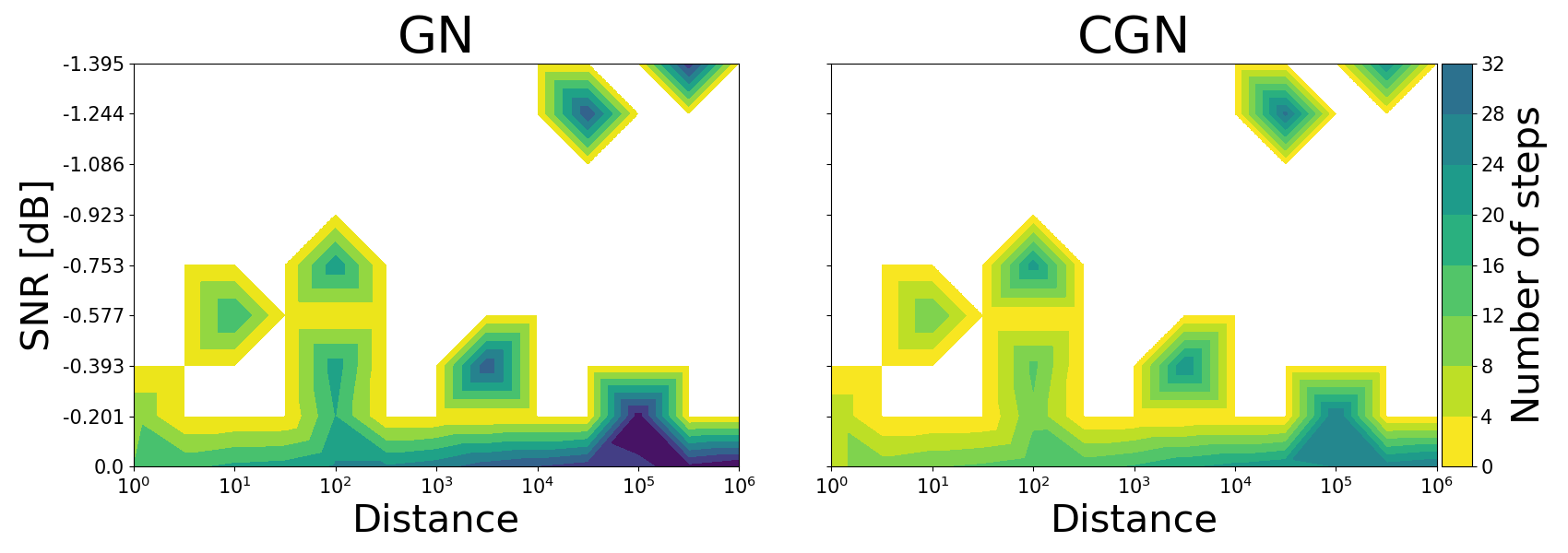}
    \caption[Minimisation \eqref{gn:2} graphs]{The figure shows results of \eqref{gn:2}. The comparison between GN and CGN was plotted across a range of distances and noise. As can be seen, CGN consistently took fewer steps to converge. This is particularly evident at large noise levels, where the CGN converged with almost half of the iteration steps. Both methods share the same un-converged area.}
    \label{fig: CGN_fig2}
\end{figure}

The results of the polynomial case are shown in Fig. \ref{fig: CGN_fig2}. There, CGN consistently requires fewer steps to yield a result. Both methods share the same un-converged areas at large noise levels. They also share the discontinuous convergence area. Near the zero noise level, it can be seen that CGN produces a results with about 8 steps fewer than GN which is equal to a $25\%$ increase in efficiency.
\subsubsection{Exponential parameters}
\begin{figure}[htb!] 
    \centering
    \includegraphics[width=\textwidth]{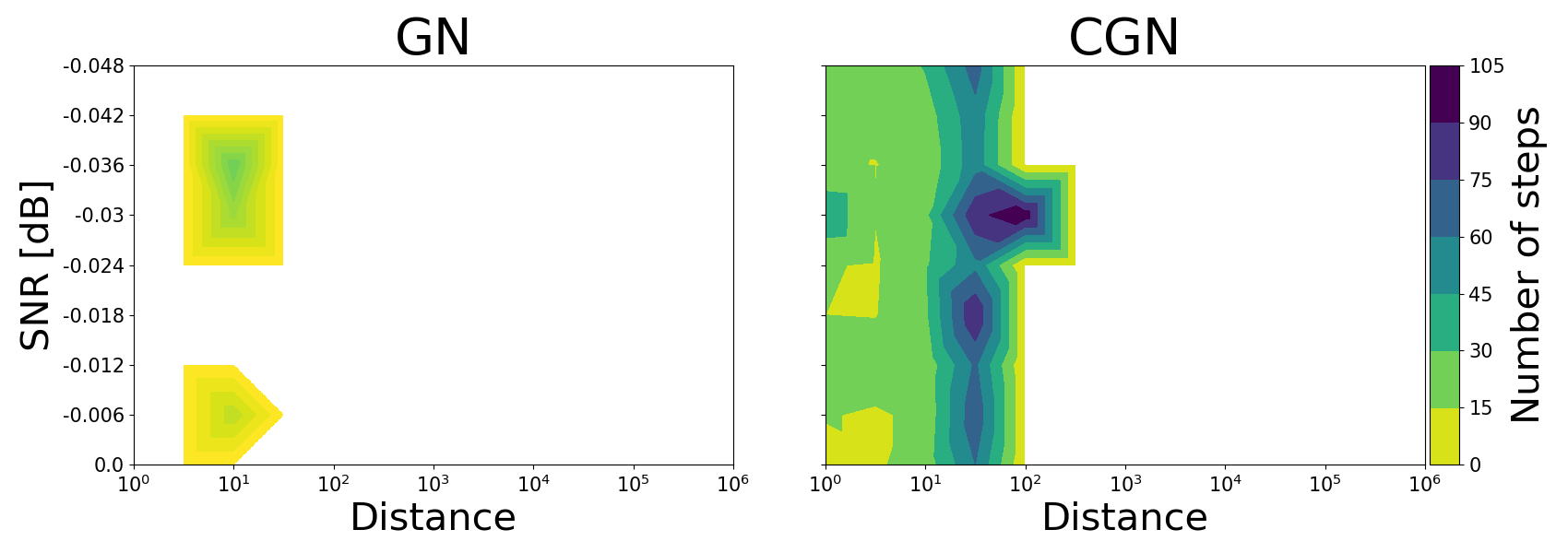}
    \caption[Minimisation \eqref{gn:3} graphs]{This figure shows the system described by \eqref{gn:3}. The system is a representation of a case which has exponential parameters. As can be seen, CGN provides a much larger and more continuous convergence range. As distance increases the number of steps needed increases as well. GN produced only two converged ``islands" whose border areas took fewer steps to converge than the middle sections.}
    \label{fig: CGN_fig3}
\end{figure}

The next system to be examined is \eqref{gn:3}. This function has parameters as exponents and two different independent variables. This system was chosen as the initial guess heavily impacts the systems value and as such an initial guess which is far away from the desired value will take much longer to compute. The results of this test are shown in Fig. \ref{fig: CGN_fig3}. There, it can be seen that CGN provides an expanded convergence range as compared to GN. CGN provides full coverage on the noise range and covers initial distances of up to $10^2$. This is contrasted by GN's coverage of a few discontinuous areas. These areas vary in distance and noise, with a maximum noise level covered of $-0.042dB$.
\subsubsection{Negative exponent in exponential function}
\begin{figure}[htb!] 
    \centering
    \includegraphics[width=\textwidth]{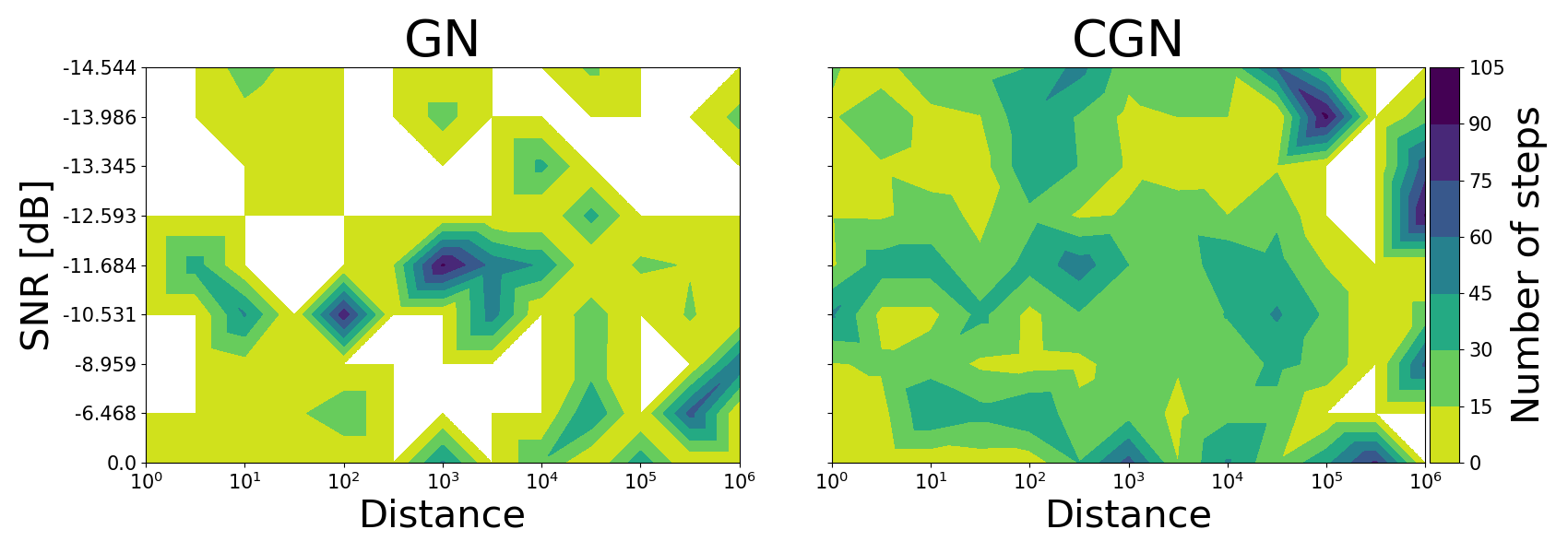}
    \caption[Minimisation \eqref{gn:4} graphs]{This figure shows the results from testing the system described by \eqref{gn:4}. The system represents a case where some of the parameters have negative exponents and are part of an exponential function. This case is similar to ones found in some constitutive models. The graphs show that CGN consistently took fewer steps to converge and provided better coverage. Also, due to large difference in iteration steps, CGN took less computation time on average to produce a result.}
    \label{fig: CGN_fig4}
\end{figure}

The final case to examined here is shown by \eqref{gn:4}. This system has parameters with negative exponents which are used as the argument of an exponential function. In testing this function, it was assumed that all elements of the initial guess are of the same magnitude, which does not show the worst case possible. For comparison purposes this will not affect the results, which  are shown in Fig. \ref{fig: CGN_fig4}. There it can be seen that again CGN takes fewer steps to converge. It also shows that CGN does not have any discontinuities across the test domain. This suggests that CGN can be a valid improvement not only by expanding the convergence basin but also reducing step count.
\subsection{Rate and order of convergence}
\begin{figure}[htb!]
    \centering
    \includegraphics[width=0.75\textwidth]{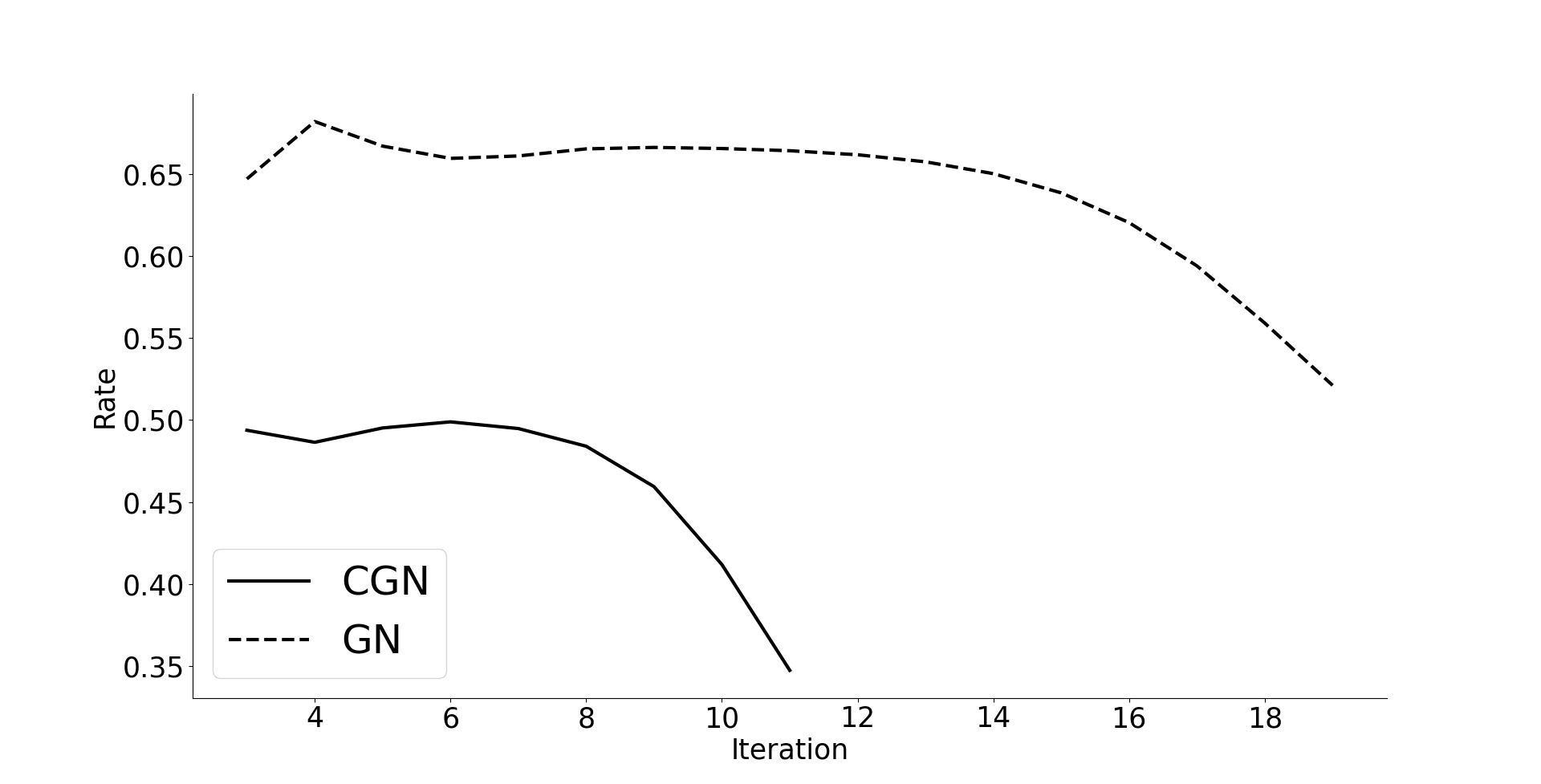}
    \caption[Minimisation Rate comparison]{The figure shows a comparison between the calculated rate of CGN (solid line) and GN (dashed line). The rate of CGN is consistently lower than that of GN which leads to a quicker convergence. The graph shows the rate past the third iteration because, as explained before, three iterations are required to calculate the first instance of rate.}
    \label{fig:rate}
\end{figure}

\begin{figure}[htb!]
    \centering
    \includegraphics[width=0.75\textwidth]{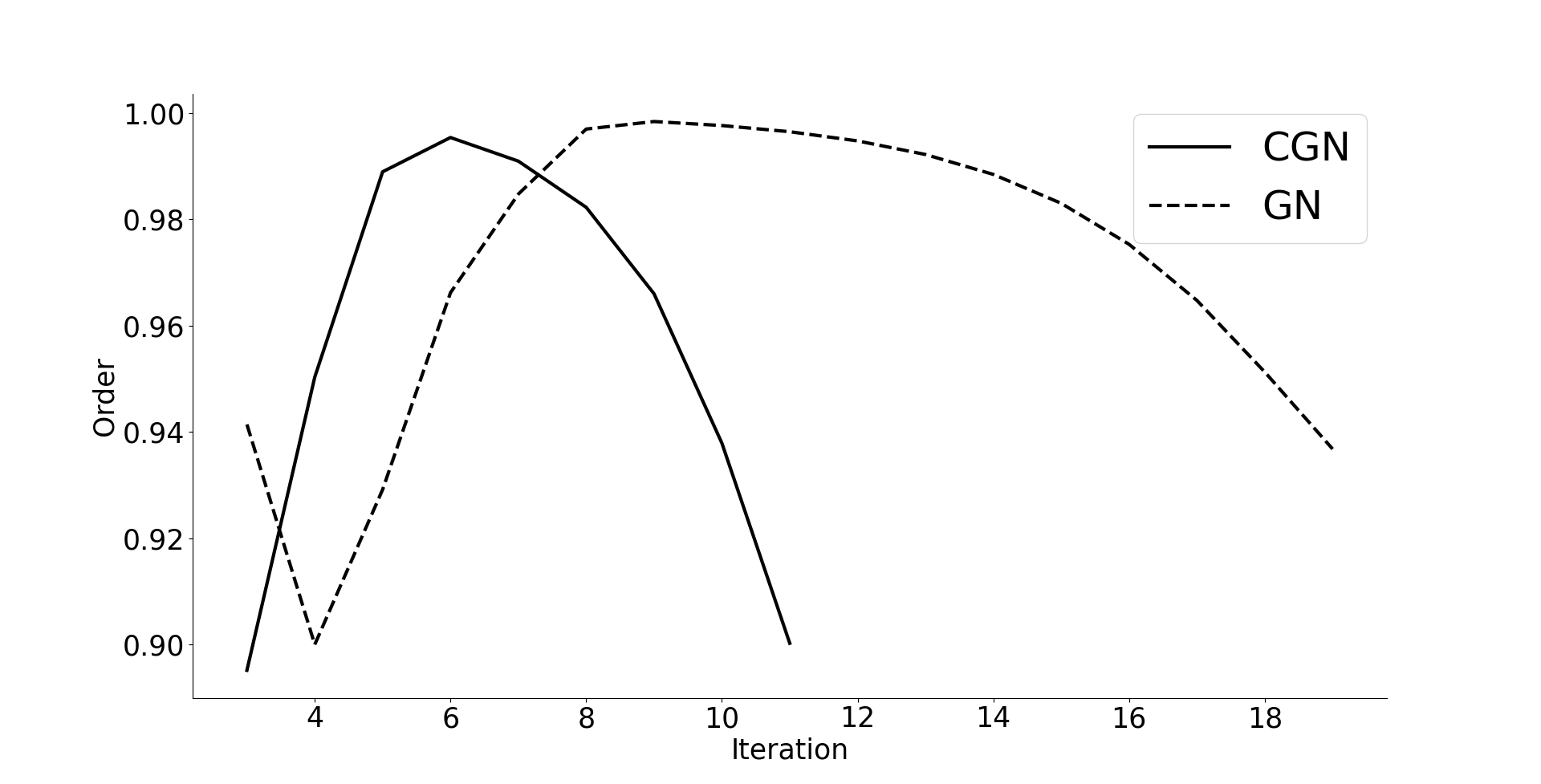}
    \caption[Minimisation Order comparison]{The figure shows a comparison between the calculated order of CGN (solid line) and GN (dashed line). The orders of CGN and GN behave similarly, both laying in the range of $[0.9, 1.0]$ with a similar decrease as the methods converge.The graph shows the rate past the third iteration because, as explained before, three iterations are required to calculate the first instance of order.}
    \label{fig:order}
\end{figure}

In this section, the rate and order of convergence of the CGN and GN methods will be compared. Both methods were tested with \eqref{gn:2}, the noise level was set at $0.0dB$ and an initial guess of the parameters was taken as $[10,10,10,10,10]$. The results of the test are shown in Fig. \ref{fig:rate} and Fig. \ref{fig:order}. The first figure, Fig. \ref{fig:rate}, shows the calculated rate of convergence of both algorithms. As can be seen both of the methods exhibit a similar behaviour where the rate of convergence becomes smaller as the algorithm converges. The CGN method consistently provides a lower rate of convergence compared with GN. The second figure, Fig. \ref{fig:order}, shows the calculated order of convergence of both methods. The order of both varies between $0.9$ and $1.0$. The order of convergence decreases as the algorithm converges. Looking at the behaviour of both algorithms, it can be deduced that both exhibit super-linear convergence. It is known that GN can reach quadratic convergence \cite{bjork}, but only when using an appropriate line-search algorithm on the GN step, which was not used in this experiment. It can be concluded that the CGN method, due to its consistently lower rate of convergence, provides an improvement to the convergence behaviour of the GN method.
\section{Conclusion}
In this chapter, the paper has demonstrated that CGN preforms better than GN in a variety of test cases. In simple linear problems both methods perform identically. However, the added computational load of CGN makes it inefficient to use. In more complex systems, such as the three presented in this paper, CGN provides saving in terms of step count and convergence area. The CGN method also shows a lower rate of convergence and a similar order of convergence to GN, which reinforces the fact that it converges in fewer iteration steps. The computational load of CGN is greater, but this can be offset by the reduced iteration count. In any case, this method can be used to increase the basin of convergence of the GN method and, based on the specific system, yield time savings as well.
\chapter{Finite Element Method (FEM)}
\section{Background}
This paper has so far proposed improvements to NR which resulted in ENR, used for root-finding, and to GN which resulted in CGN, used for minimisation. Now, it will be shown how these new methods can be applied to forward and inverse models through FEM. This paper is concerned with stress analysis of objects with hyper-elastic behaviour. Hence, stretches $\lambda$ and stresses $P$ will be examined. The problem which will be solved in the deformation of a bar shown in Fig. \ref{fig: setup}. Assuming that the material is isotropic in the $y$ and $z$ direction and that it undergoes isochoric deformation (constant volume), which is explained later, one can model the bar as a one-dimensional object. Hence, FEM here will be defined for a one-dimension non-linear system.
\begin{figure}[htb!]
    \centering
    \input{figures/3d_baseconfig}
    \caption{A representation of a bar in three dimensions.}
    \label{fig: setup}
\end{figure}
For the forward modelling case the following procedure will integrate ENR with FEM:
\begin{itemize}
    \item Initially a FEM model will be constructed using the desired object.
    \item An initial guess of the location of each node will be taken. It will, in all cases coincide with the initial configuration, i.e. the starting guess of the deformed shape will be the un-deformed shape.
    \item The FEM model will return information which is equivalent to the previously used function vector $f(x)$ and its derivative.
    \item This information will be used by ENR to produce a better estimate of the locations of all the nodes.
    \item These new locations will be fed back to the FEM model and the cycle will continue until the resulting function vector $f(x)$ is within an absolute tolerance away from zero. In this case, the model has reached a state of static equilibrium.
\end{itemize}
For inverse models a similar procedure will be followed, but rather than varying the locations of the elements, the properties of the object will varied. Testing will be done directly on the material model which is equivalent to a FEM model with one element. The procedure is as follows:
\begin{itemize}
    \item Initially, a FEM model will be constructed using the desired object and its properties.
    \item An initial guess of the material properties will be used by the FEM model to produce a result.
    \item The result will be used together with an observed value to calculate the residual.
    \item A finite difference method will be used to calculate the derivatives of the residual.
    \item The residual and its derivatives will be used to calculate the CGN step which will then be applied to the initial object properties.
    \item The new properties will be passed onto the FEM model and the cycle will continue until the residual becomes less than a predetermined tolerance.
\end{itemize}
\section{Introduction}
Many phenomena in science and engineering can be explained using partial differential equations (PDEs). These equations are constructed using various partial derivatives of a multivariate function together with many independent variables. They are produced naturally and are extensively used in engineering. Examples of where they appear can be seen in the study of groundwater flows or in stress analysis which will be examined in this paper \cite{pdes}. Solving such PDEs by classical analytical methods for arbitrary shapes is almost impossible \cite{FEM}.\newline \newline
That is why in the 1960s the Finite Element Method (FEM) was developed \cite{pdes}. FEM is a numerical method with which PDEs can be solved approximately. This is done by changing the domain of the problem from a continuous to a discretised one. During this process, the body which is being examined is broken down into a finite number of elements, from which the method gets its name, connected by nodes. This newly generated system of local sub-problems can be easily solved and then stitched back together to form the global solution.\newline\newline
If the system which will be analysed is already in its discrete form, i.e. a truss structure, a direct FEM can be applied. However, when that is not the case, a general approach is needed, where an arbitrary domain can be discretised.
\subsection{Derivation}
\begin{figure}[htb!]
    \centering
    \input{figures/LagrVsEul}
    \caption[Lagrangian vs Eulerian meshes]{Figure showing the difference between Lagrangian transformations and Eulerian Transformations.}
    \label{fig: Lagrangian}
\end{figure}
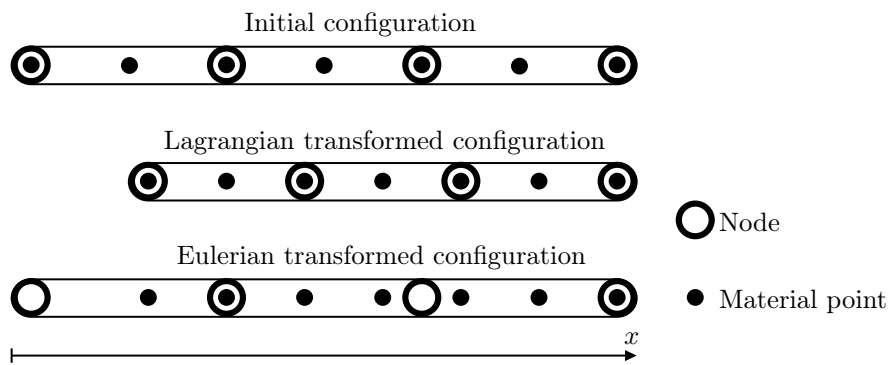

In order to solve a problem of non-linear continua, such as the one at hand, a non-linear formulation of FEM is needed. There are two possible formulations, the Lagrangian and the Eulerian. The difference between the two is in the way the elements, generated by the discretisation of the continuum, behave. Lagrangian elements deform with the material; that is, the location of the nodes is coincident with the material points; whereas Eulerian nodes remain fixed in space, so their location does not change with the material's deformation. The example given in Fig. \ref{fig: Lagrangian} shows this difference graphically. There, it can be seen that in the Lagrangian formulation, because the boundary nodes are coincident with the material points, one has to apply boundary conditions to the nodes, whereas in Eulerian formulation the boundary nodes are not necessarily coincident with the material boundary. Hence, one has to apply boundary conditions at points which are not nodes. Due to this difference, this paper will use the Lagrangian formulation and more specifically the total Lagrangian formulation.\newline \newline As described before, one cannot directly discretise a domain. There are specific steps that need to be taken. Like with linear FEM models, firstly, the strong form of the system needs to derived, which includes the governing equations and the boundary conditions. Then, the weak form has to be derived through integration of the strong form with reduced requirements for continuity. Afterwards, the model can be discretised and solved.
\subsection{Nomenclature}
As the element is expected to experience large strains, a way to describe both the current and initial material locations of the material points is needed. Hence, the initial locations will be denoted by $X$ and the current locations will be denoted by $x$. In both cases, these parameters show a location along the element. The way to convert from one to the other can be done by a function called $\phi(X,t)$ which will act a map between $x$ and $X$. As can be seen, $\phi$ is a function of time as well as a function of the location. This is necessary because the system reacts to the applied loads and deformations, and changes the internal forces. In the case of static equilibrium, which is explored here, $t$ does not need to represent actual time, but can instead just be a monotonically increasing variable. Because an iterative method will be used to solve the non-linear system, $t$ will coincide with the number of iterations. The final relation of $x$, $\phi$ and $X$ can then be written out as:
\begin{equation}
    x = \phi(X,t)
\end{equation}
The displacement of the object will be denoted by $u(X, t)$ which is known as the trial function and is equal to the difference between $x$ and $X$. From this, one can derive the deformation gradient $F$ with regards to the material domain as:
\begin{equation}
    F = \dfrac{\partial \phi}{\partial X}
\end{equation}
Taking the determinant of the deformation gradient gives the Jacobian determinant $J$, which is:
\begin{equation}
    J = det(F)
\end{equation}
and shows the volumetric change of the object. The deformation gradient is a measure of deformation and it can be converted to a parameter known as a stretch, which will be used by the material model to relate stretches to stresses. This is done by using the Polar Decomposition Theorem. The conversion is based on the fact that if the deformation gradient $F$ is non-singular with a positive determinant, it can be expressed as the product of a rigid body rotation $R$ and a deformation $U$ \cite{Stretch} as:
\begin{equation}
    F = RU
\end{equation}
The stretch matrix, used to establish the deformation gradient, is defined as:
\begin{equation}
    U = (F^TF)^{0.5}
\end{equation}
The fractional power of a matrix is used in terms of spectral representation. It is computed by first transforming the matrix into its principal coordinates, where the matrix is diagonal with its eigenvalues on the diagonal. The fractional power is applied to all the diagonal terms, and the matrix is transformed back. 
The rotation $R$ can then be given by:
\begin{equation}
    R = FU^{-1}
\end{equation}
If $F$ shows the deformation gradient in three dimensions then, because $U$ is positive definite and symmetric, it will have three real positive eigenvalues $\lambda_1$, $\lambda_2$ and $\lambda_3$, called the principal stretches, and a corresponding triplet of orthonormal eigenvectors $r_1$, $r_2$ and $r_3$. In the case at hand, no rotation will happen, so:
\begin{equation}
    R = FU^{-1} = I
\end{equation}
From this it can be seen that if no rotation exists, $U = F$. Hence, the components of $F$ in the principal basis are the principal stretches. In the one-dimensional case examined here, the principal stretch of each component can be calculated directly by setting:
\begin{equation}
    \lambda^e_1 = \dfrac{x^e_2-x^e_1}{X^e_2-X^e_1}
\end{equation}
The ratio of the stretches shows what type of deformation the object has undergone. Previously, it was mentioned that the bar object will undergo isochoric (constant-volume) deformation, which now can be defined in terms of stretches as:
\begin{equation}
    J = \lambda_1\lambda_2\lambda_3 = 1
\end{equation}
\subsection{Stress}
Stress in the total Lagrangian formulation is taken as the first Piola-Kirchhoff stress $P$ which is equal to:
\begin{equation}
    P = \frac{\partial W}{\partial F} - pF^{-1}
\end{equation}
where $p$ denotes the Lagrangian multiplier, which represents hydro-static pressure, and $W$ is the strain energy density given by the material model. When a material model is defined, one can solve for the parameter $p$ and use it to find the stress tensor $P$.
\subsection{Governing equations}
There are four governing equations which affect the system at hand \cite{Lagrangian}. These are:
\begin{itemize}
    \item Conservation of mass
    \item Conservation of momentum
    \item A measure of deformation
    \item A constitutive equation which is a material model that relates deformation to stresses
\end{itemize}
There is an additional compatibility requirement, which in this case is that the deformation has to be continuous ($C^1$ in the strong form and $C^0$ in the weak form). For stress analysis, conservation energy can be ignored as deformation as a process can dissipate energy as heat but it will be assumed that the effects of that energy loss are negligible.
\subsubsection{Conservation of mass}
The conservation of mass equation can be written as:
\begin{equation}
    \rho J = \rho_0
\end{equation}
This equation requires that the mass of the original configuration is the same as the mass of the current configuration. The variable $\rho$ shows the current density of the object, whereas $\rho_0$ shows the initial density.
\subsubsection{Conservation of momentum}
Because the area of the element is taken as a constant along the length of the beam, the conservation of momentum formula becomes:
\begin{equation}
    P(X,t)_{,X}+\rho_0 b_i = \rho_0 u_{,tt}
\end{equation}
where:
\begin{equation}
    P(X,t)_{,X} = \frac{\partial P(X,t)}{\partial X}
\end{equation}
In the case of static equilibrium, $\rho_0 u_{,tt}$ vanishes.
\section{One-dimensional case}
In the one-dimensional case, some of the above definitions are modified. The material is assumed to undergo isochoric (constant volume) deformation. Hence, the deformation gradient will become:
\begin{equation}
    F = diag\left[\lambda, \frac{1}{\sqrt{\lambda}}, \frac{1}{\sqrt{\lambda}}\right]
\end{equation}
Also, in the one-dimensional case there are no stresses in the $y$ and $z$ direction, which means that the stress vector $P$ becomes:
\begin{equation}
    P = diag[P^x, 0 ,0]
\end{equation}
This allows the Lagrangian multiplier $p$ to be determined. Moreover, the one-dimensional case modifies the governing equations.
\subsubsection{Final governing equation}
From the above definitions, it can be seen that the equations which govern the behaviour of the object in 1D static equilibrium applications are the conservation of momentum and the material model. The later can be substituted into the former to produce the final governing equation:
\begin{equation}
    P^x(X,t)_{,X}+\rho_0 b = 0
\end{equation}
The above is a boundary value problem and is the target of the FEM model. To finalise the formulation of the problem, the boundary conditions and initial conditions are needed. 
\subsubsection{Boundary and Initial conditions}
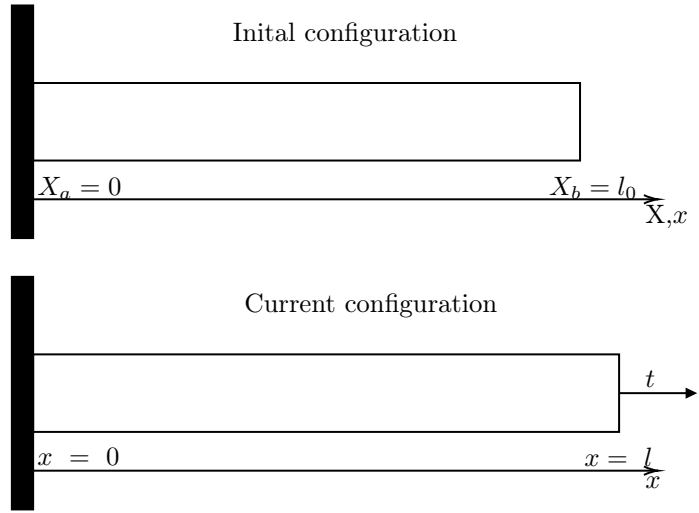
\begin{figure}[htb!]
    \centering
    \input{figures/boundary}
    \caption{Figure showing boundary conditions of a 1D bar.}
    \label{fig: boundary}
\end{figure}

In the one-dimensional case, the boundary consists of two points which are situated at the ends of the domain. In the model, these points are represented by  $X_a$ and $X_b$, shown in Fig. \ref{fig: boundary}. A boundary will be denoted by $\Gamma$. There are two types of boundary, the first is called a displacement boundary and is denoted by $\Gamma_u$. The other boundary is called a traction boundary where the traction force is applied. It is denoted by $\Gamma_t$. Traction and displacement cannot be prescribed at the same boundary. Furthermore, to be able to properly describe the traction boundary, the unit normal $n^0$ has to be defined. It has a value of -1 at $X_a$ and a value of 1 at $X_b$. Values prescribed at the boundaries are denoted by a superposed bar. With these definitions the boundary conditions can be written as:
\begin{equation}
    u = \bar{u} \; on \; \Gamma_u
\end{equation}
\begin{equation}
    n^0P = \bar{t^0} \; on \; \Gamma_t
\end{equation}
The traction force here has a superscript $(.)^0$ to distinguish it from time and to show that it is defined over the original area of the bar. In the one-dimensional bar case, the prescribed boundaries are:
\begin{equation}
    u(X_a,t) = 0 \; on \; \Gamma_u
\end{equation}
\begin{equation}
    n^0P(X_b,t) = P(X_b,t) = \bar{t^0} \; on \; \Gamma_t
\end{equation}
The final set of definitions needed are the initial conditions at $t = 0$. As the body is initially at rest, displacement $u$ and the velocity $\dot{u}$ are zero.
\subsubsection{Continuity}
\begin{figure}[htb!]
    \centering
    \input{figures/continuity}
    \caption[Continuity visualisation]{A graph showing the differences between different continuity levels.}
    \label{fig: cont}
\end{figure}
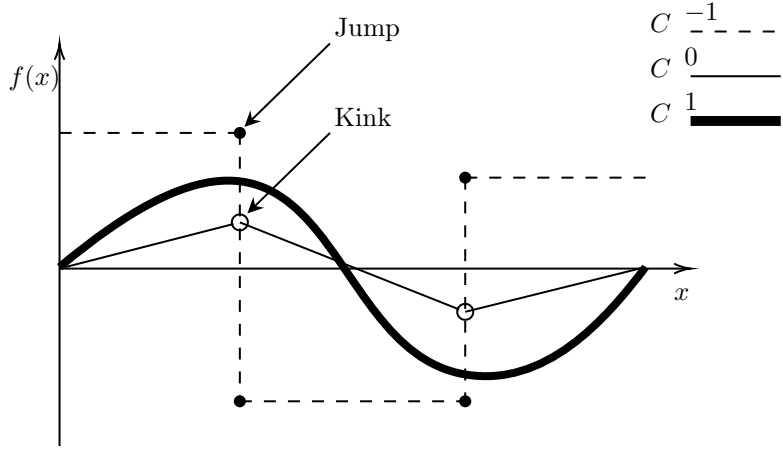

The continuity of the trial solution $u$ and the later shown test function $\delta u$ is vital to the functioning of FEM. That is why the topic of continuity will be examined here.\newline \newline A function is considered to be $C^n$ continuous if its $n$-th derivative exists and is continuous over the entire domain. The derivative of a $C^n$ function will be $C^{n-1}$ continuous. If a function $C^0$ is piece-wise continuously differentiable, its derivative $C^{-1}$ is piece-wise continuous, i.e. $C^{-1}$ is continuous except at specific points called ``jumps". The location of these jumps in the $C^{-1}$ function correspond to the location of "kinks" in the $C^0$ function. A kink is a sharp change is the gradient of the function. Kinks are not exclusive to $C^0$ and they can also exist in $C^{-1}$ functions. A $C^1$ function has no kinks or jumps and is continuously differentiable. The differences between these functions can be seen in Fig. \ref{fig: cont} and are summarised by Table \ref{Continuity}.%

\begin{table}[H]
\centering
\begin{tabular}{||l l l l||} 
 \hline
 Function & Kinks & Jumps & Type
\\ [0.5ex]
 \hline\hline
 $C^{-1}$ & Yes & Yes & piece-wise continuous\\ 
 $C^0$ & Yes & No & piece-wise continuously differentiable\\
 $C^1$ & No & No & continuously differentiable\\ [1ex] 
 \hline
\end{tabular}
\caption{Continuity of functions based on $C^n$}
\label{Continuity}
\end{table}%
The strong form of FEM, which will be formalised below, requires that the trial solution and test function are $C^1$ continuous, whereas the weak formulation only requires that they are $C^0$ continuous. That is why the weak formulation is called ``weak'' because it reduces the continuity requirements for $u$ and $\delta u$.
\subsection{The Strong form}
Using the definition from above, the strong form of the domain can be defined for the one-dimensional bar under static loading as follows:
\begin{equation}
    P^x(X,t)_{,X}+\rho_0 b = 0
\end{equation}
\begin{equation}
    u(X_a,t) = 0 \; on \; \Gamma_u
\end{equation}
\begin{equation}
    n^0P(X_b,t) = P(X_b,t) = \bar{t^0} \; on \; \Gamma_t
\end{equation}
To solve the system the strong from has to be converted to the weak form. This will be shown in the next section.
\subsection{The Weak Form}
The momentum equation cannot be discretised directly by the finite element method. In order to discretise this equation, a weak form, often called the principle of virtual work, is needed \cite{FEM}. It is equivalent to the momentum equation and the traction boundary conditions. To derive the weak form, the strong form has to be multiplied by some arbitrary function $\delta u(X)$, known as the test function. The arbitrary nature of this function makes sure that the weak and strong form are equivalent. The weak form also has to satisfy the previously described continuity requirements and to vanish on $\Gamma_u$. After the multiplication, the integral of both sides can be taken over the domain of the object. This will yield the following:
\begin{equation}
    \int_{X_a}^{X_b} \delta u\left[P_{,X}+\rho_0 b\right]dX= 0
\end{equation}
In the above, the short hand $P$ was used to denote the result of the material model evaluated at $X$ and at time $t$. The first term of the equation can be separated and the whole equations can be rearranged to:
\begin{equation}
    \int_{X_a}^{X_b} \delta u P_{,X}dX + \int_{X_a}^{X_b}\delta u \:\rho_0 b\:dX = 0
\end{equation}
The first term can be expanded using the fundamental theorem of calculus. This will produce:
\begin{equation}
    \delta u n^0P\bigg\rvert_{\Gamma} -\int_{X_a}^{X_b} \delta u_{,X} P dX+ \int_{X_a}^{X_b}\delta u \:\rho_0 b\:dX = 0
\end{equation}
Evaluating the first term across $\Gamma$ and combining the two separate integrals yields the weak form of the momentum equation and the the traction boundary. The test function disappears at $\Gamma_u$ and the nominal stress $P$ at $\Gamma_t$ is equal to the prescribed traction force $\bar{t^0}$. Therefore, the final version of the weak form is:
\begin{equation}
    \delta u \bar{t^0}\bigg\rvert_{\Gamma_t = 0}+ \int_{X_a}^{X_b} \left[\delta u \:\rho_0 b-\delta u_{,X} P\right]dX = 0
\end{equation}
The above equation can be separated into two parts. The first:
\begin{equation}
    \delta W^{ext} = \int_{X_a}^{X_b}\delta u \:\rho_0 b \;dX +\delta u \bar{t^0}\bigg\rvert_{\Gamma_t = 0} 
\end{equation}
shows the external virtual work done on the bar and the second:
\begin{equation}
    \delta W^{int} = \int_{X_a}^{X_b}\delta u_{,X} P \;dX = \int_{X_a}^{X_b}\delta FP \;dX
\end{equation}
which shows the internal virtual work. Using $\delta W^{ext} $ and $\delta W^{int}$, the weak form can be rewritten as:
\begin{equation} \label{eq: virtual work}
    \delta W = \delta W^{ext}-\delta W^{int} = 0
\end{equation}
where $\delta W$ denotes virtual work.
\subsection{Finite Element Discretisation}
With the weak form of the problem, the domain can be discretised. This is done by using the finite element interpolation of the test and trial functions. The domain of the problem, which is contained within $[X_a,\: X_b]$, is broken down into $e$ elements, where $e \in [1, n_e]$ with $n_e+1$ nodes. The nodes are denoted by $X_I$, where $(,)_I$ shows the index of the nodes. The nodes of the generic element will be denoted by $X^e_i$, where $(.)^e$ shows the element and $i \in [1,\: m]$ shows the index of the nodes, and $m$ shows the number of nodes per element. In the one-dimensional bar case, $m = 2$. The total domain is denoted by $\Omega$, whereas the domain of each element, comprising of $[X^e_1,\: X^e_2]$, is denoted by $\Omega_e$.
\subsubsection{Approximation of the trial solution and test function}
For completeness, the approximation of the trial solution for an element $u^e(X,t)$ has to be at least a linear function to meet the completeness condition. This approximation at a node will be called a nodal displacement which is:
\begin{equation} \label{eq: shapef1}
    u^e(X, t) = c^e_1(t)+c^e_2(t)X \equiv \hat{u}^e(t)
\end{equation}
where the $\hat{u}$ denotes the approximation of trial function evaluated at a particular $X$. The time variable $t$ will be omitted from here on. As a one-dimensional element has two nodes and two nodal parameters, there are an equal number of nodes $X_1, X_2$ and nodal parameters $c_1, c_2$. That means that the nodal parameters can be uniquely expressed in terms of nodes. The matrix of nodes can then be described as:
\begin{equation}
    M^e_{ij} = 
    \begin{cases} 
      1 & j = 1 \\
      X^e_j & j = 2
   \end{cases}
\end{equation}
The element parameter vector $c^e$ has to be such that \eqref{eq: shapef1} is true. That means:
\begin{equation}
    c^e = [c^e_1 \; c^e_2]^T = \left[%
    \dfrac{u^e_2X^e_1 - u^e_1X^e_2}{X^e_1-X^e_2}%
    \;%
    \dfrac{u^e_1 - u^e_2}{X^e_1-X^e_2}
    \right]^T
\end{equation}
With $M^e_{ij}$ and $c^e_j$ defined, the next step will be to define the vector of nodal displacements $d^e$. This vector will be equal to the product of $M^e_{ij}$ and $c^e_j$. That is expressed by:
\begin{equation} \label{eq: nodal_dispM}
    d^e_i = M^e_{ij}c^e_j
\end{equation}
Then, a way will be needed to relate the displacement $u^e(X, t)$ to the nodal displacements $d^e_i$. The general equation of displacement \eqref{eq: shapef1} can be rearranged in matrix form to:
\begin{equation} \label{eq: general_dispM}
    u^e(X, t) = p_i(X)c^e_i(t)
\end{equation}
where $p_i(x)$ is equal to 1 when $i=1$ and equal to $X$ when $i=2$. Now, both equations \eqref{eq: nodal_dispM} and \eqref{eq: general_dispM} can be rearranged in terms of $c$. This means that the following system can be created:
\begin{align}
    c^e_i &= d^e_i (M^e_{ij})^{-1}\\
    c^e_i &= u^e(X, t)p^{-1}_i(X)
\end{align}
Setting both right-hand sides equal to each other, the relation between $u^e(X, t)$ and $d^e_i$ can be expressed as:
\begin{equation} \label{eq: shape func der}
    u^e(X, t) = p_i(X)(M^e_{ij})^{-1}d^e_j
\end{equation}
From the above equation, the first two terms can be isolated and set equal to a new vector $N^e$. That vector is known as the shape function. It is equal to:
\begin{equation} \label{eq: shape func}
    N^e_i = p_j(X)(M^e_{ij})^{-1} =
    \begin{bmatrix}
    1 & X
    \end{bmatrix}
    \dfrac{1}{X^e_2-X^e_1}
    \begin{bmatrix}
    X^e_2 & -X^e_1\\
    -1 & 1
    \end{bmatrix} = %
    \dfrac{1}{l^e_0}
    \begin{bmatrix}
    X^e_2-X & X-X^e_1
    \end{bmatrix}
\end{equation}
From here, the final form of the general element displacement can be found as:
\begin{equation} \label{eq: gen_disp}
    u^e(X, t) = N^e_id^e_i =  \dfrac{1}{l^e_0}
    \begin{bmatrix}
    X^e_2-X & X-X^e_1
    \end{bmatrix}
    \begin{bmatrix}
    \hat{u}^e_1 \\
    \hat{u}^e_2
    \end{bmatrix}
\end{equation}
The rate of displacement $u^e_{,X}(X,t)$ can be found by taking the derivative of the above. As $\hat{u}$ is a function of time, only the first terms will be affected by the derivation. Hence, it will take the form of:
\begin{equation}
    u^e_{,X}(X, t) = N^e_{i,X}d^e_i =  \dfrac{1}{l^e_0}
    \begin{bmatrix}
    -1 & 1
    \end{bmatrix}
    \begin{bmatrix}
    \hat{u}^e_1 \\
    \hat{u}^e_2
    \end{bmatrix}
\end{equation}
The same method for deriving the general element trial solution can be applied to the test function. It will yield the following:
\begin{equation}
    \delta u^e(X, t) = N^e_i \delta d^e_i =  \dfrac{1}{l^e_0}
    \begin{bmatrix}
    X^e_2-X & X-X^e_1
    \end{bmatrix}
    \begin{bmatrix}
    \hat{\delta u}^e_1 \\
    \hat{\delta u}^e_2
    \end{bmatrix}
\end{equation}
\begin{equation}
    \delta u^e_{,X}(X, t) = N^e_{i,X}\delta d^e_i =  \dfrac{1}{l^e_0}
    \begin{bmatrix}
    -1 & 1
    \end{bmatrix}
    \begin{bmatrix}
    \hat{\delta u}^e_1 \\
    \hat{\delta u}^e_2
    \end{bmatrix}
\end{equation}
\subsubsection{Global approximation}
Once all local approximations have been derived, they have to be converted back into the global domain. This process is known as scattering. It distributes all local elements $\Omega^e$ for $e = 1,2,...,n_e$ to the global domain $\Omega$. This is done using the connectivity matrix $L_e$. The same matrix is also used in the process called gathering, which relates the global nodal displacements to the local ones. The element nodal displacements are related to global nodal displacements by:
\begin{equation}
    u^e = L^eu
\end{equation}
\begin{equation} \label{eq: local to global}
    u = \sum_{e=1}^{n_e}(L^e)^Tu^e
\end{equation}
The connectivity matrix is a Boolean matrix, which means it only has $0$ and $1$ elements and has a shape of $m \times n_e+1$ in the one-dimensional case. This matrix shows where the position of the local element's nodes $X^e_i$ are in the global node vector $X_I$. This matrix has to be constructed on a per element basis for each specific discretisation of the domain $\Omega$. The global function can also be approximated by the global nodal values by using global shape functions. Hence, the trial solution $u(X,t)$ can be approximated by:
\begin{equation}
    u(X,t) = \sum_{I = 1}^{n_e+1} N(x)_Iu(t)_I
\end{equation}
where the global shape function $N(x)_I$ is equal to:
\begin{equation}
     N(x) = \sum_{e = 1}^{n_e} (L^e)^TN(x)^e_i
\end{equation}
The same procedure can be applied to the test function $\delta u$ which will yield:
\begin{equation}
    \delta u(X,t) = \sum_{I = 1}^{n_e+1} N(x)_I\delta u(t)_I
\end{equation}
\subsubsection{Nodal forces}
The virtual work terms of the weak form can be expressed as the product of a virtual displacement and a virtual force. This will allow for the derivation of the victual force elements:
\begin{equation}
    \delta W = \sum_{I = 1}^{n_e+1}\delta u_I f_I
\end{equation}
where $f_I$ represents the virtual force at the node $I$ and $\delta u_I$ represents the virtual displacement at the same node. The above can be applied to any of the specific work elements to derive their force components. For internal work the equation is:
\begin{equation} \label{eq: virtual force}
    \delta W^{int} = \sum_{I =1}^{n_e+1}\delta u_I f_I^{int} = \int_{X_a}^{X_b}\delta u_{,X} P \;dX 
\end{equation}
The test function $\delta u(X, t)_{,X}$ can be replaced by its equivalent nodal components $N(x)_I$ and $\delta u(t)_I$. As the $\delta u(t)_I$ term is a function of time, it can be taken away from the integral as a constant. This will yield:
\begin{equation}
    \int_{X_a}^{X_b}\delta u_{,X} P \;dX = \sum_{I = 1}^{n_e+1} \delta u(t)_I \int_{X_a}^{X_b} N_{I,X} P \;dX 
\end{equation}
Substituting this back into \eqref{eq: virtual force}, the following can be derived:
\begin{equation}
   \delta W^{int} = \sum_{I =0}^{n_e+1}\delta u_I f_I^{int} = \sum_{I = 0}^{n_e+1} \delta u(t)_I \int_{X_a}^{X_b} N_{I,X} P \;dX 
\end{equation}
Here, because of the arbitrary nature of the test function, it can be deduced that:
\begin{equation}
    f_I^{int} = \int_{X_a}^{X_b} N_{I,X} P \;dX 
\end{equation}
The same procedure is applied to the external virtual work vector $\delta W^{ext}$ and the virtual external force vector $f^{ext}$ can be defined as:
\begin{equation}
    f_I^{ext} = \int_{X_a}^{X_b}N_I\rho_0 b \;dX +N_I \bar{t^0}\bigg\rvert_{\Gamma_t = 0} 
\end{equation}
These global force vectors can be found from the elemental force vectors similarly to what was shown in \eqref{eq: local to global}, as:
\begin{equation}
    f_I = \sum_{e=1}^{n_e}(L^e)^Tf^e_i
\end{equation}
This will allow for the derivation of an equation that will give the local element force. The above will be expressed for a specific force, in this case the virtual internal force:
\begin{equation}
    f^{int}_I = \sum_{e=1}^{n_e}(L^e)^T(f^{int})^e_i = \sum_{e=1}^{n_e}(L^e)^T\int_{X^e_1}^{X^e_2} N^e_{i,X} P \;dX 
\end{equation}
From the above it can be seen that the local element internal virtual force:
\begin{equation}
    (f^{int})^e_i  = \int_{X^e_1}^{X^e_2} N^e_{i,X} P \;dX 
\end{equation}
The same procedure is applied to the external force and this will yield:
\begin{equation}
    (f^{ext})^e_i  = \int_{X^e_1}^{X^e_2}N^e_i\rho_0 b \;dX +N^e_i \bar{t^0}\bigg\rvert_{\Gamma_t = 0}
\end{equation}
Solving the element nodal forces requires the evaluation of an integral. This integral may not be solvable analytically. That is why numerical integration will be used. A popular choice for solving integrals in FEM is Gaussian-Quadrature. It is used because of the accuracy it provides when evaluating the integral of polynomials which are frequent in FEM.
\subsection{Solution procedure}
The weak form was expressed with the help of virtual work terms in \eqref{eq: virtual work}. Later, it was shown that the virtual work terms are equal to the product of a virtual displacement and a virtual force. This will allow for \eqref{eq: virtual work} to be rewritten as:
\begin{equation}
    \sum_{I = 1}^{n_e+1}\delta W_I = \sum_{I = 1}^{n_e+1}\delta u_I f_I^{ext} - \delta u_I f_I^{int} = 0
\end{equation}
The common factor $\delta u$ can be taken out to yield:
\begin{equation}
    \sum_{I = 1}^{n_e+1}\delta u_I( f^{ext}_I -f^{int}_I) = 0
\end{equation}
Now, the displacement boundary conditions can be enforced. They were defined as:
\begin{align}
\begin{split}
    u = \bar{u} \; on \; \Gamma_u\\
    \delta u = 0 \; on \; \Gamma_u
\end{split}
\end{align}
Because the test function $\delta u$ vanishes on $\Gamma_u$ and is zero everywhere else, the following can be deduced:
\begin{equation}
    f^{ext}_I - f^{int}_I = 0, \: I = 2,3,..., n_e+1
\end{equation}
From here the final solution procedure can be constructed. It is as follows:
\begin{enumerate}
    \item Gather element nodal displacements $u^e$
    \item Compute the measure of deformation $F$
    \item Compute stress via the material model
    \item Compute internal nodal forces
    \item Compute external nodal forces
    \item Compute the total force per element: $f^e = f_e^{ext} - f_e^{int}$
    \item Scatter element nodal forces to global matrices.
    \item Check if solution is reached by:
    $ f^{ext}_I - f^{int}_I = 0, \: I = 2,3,..., n_e+1$\newline If yes, then $u$ is the actual displacement vector.\newline Otherwise, go to step 1.
\end{enumerate}
\section{Hyper-Elastic Constitutive models} 
Many different constitutive models can be developed for non-linear elasticity. Also, because many different stress and deformation measures for finite strain are available, the same constitutive equations can be written out in multiple ways. This section will cover the derivation of two constitutive models which will be used by the FEM model to test the proposed improvements. The models are the compressive Mooney-Rivlin (MR) model \cite{mooney} and the Veronda-Westmann (VW) model. MR is a hyper-elastic material model, originally developed to describe the behaviour of rubber, currently used to explain various material behaviours such as the behaviour of polyurethane \cite{poly_urethane} which is extensively used as a bearing material in bridges and under rail tracks \cite{bridge}. VW is a model commonly used for the response of biological tissues \cite{MEI2018110}. All models will be expressed in terms of  the first Piola-Kirchoff stress, denoted by $P$, and principal stretches. 
\subsection{Mooney-Rivlin} \label{mooney}
The compressive Mooney-Rivlin (MR) model defines the strain energy density $W(F)$ as:
\begin{equation}
    W(F) = \frac{\mu}{2}\left[\nu(J^{-2/3}I_1-3)+(1-\nu)(J^{-4/3}I_2-3)\right]+\frac{K}{2}(\ln J)^2
\end{equation}
In the above, $\mu$ is the effective shear modulus, K is the bulk modulus, and $\nu$ is a dimensionless parameter which belongs to $[0,1]$. $I_n$ represents the invariants of the unimodular component of the left Cauchy-Green deformation tensor. $F$ is the deformation gradient and $J = det(F) = \lambda_1 \times \lambda_2 \times \lambda_3 = 1$. This entire expression can be simplified for uniaxial stretch and isochoric deformation for $P$ as:
\begin{equation}
    P(\lambda) = \mu\nu\left(\lambda-\frac{1}{\lambda^2}\right)+\mu(1-\nu)\left(1-\frac{1}{\lambda^3}\right)
\end{equation}
\subsection{Veronda-Westmann}
The Veronda-Westmann (VW) model was chosen for its exponential behaviour. It defines strain energy density as:
\begin{equation}
    W(F) = \frac{A}{B}\left[exp(B(J^{-2/3}I_1-3))-1\right]-\frac{A}{2}\left(J^{-4/3}I_2-3\right)+\frac{K}{2}(\ln J)^2
\end{equation}
In the above, $K$ represents the bulk modulus and $A$ and $B$ are stiffness parameters. The stress equation is:
\begin{equation}
    P(\lambda) = 2A\left(\lambda-\frac{1}{\lambda^2}\right)exp\left(B\left(\lambda^2+\frac{2}{\lambda}-3\right)\right)-A\left(1-\frac{1}{\lambda^3}\right)
\end{equation}
\section{Results}
\subsection{Forward Models}
\subsubsection{Methods}
The FEM model was constructed using 5 linear elements to discretise a one-dimensional bar of length 2m. The system is modelled as one-dimensional as the deformation is assumed to be isochoric and the example material is taken to be isotropic in the $y$ and $z$ axes of the coordinate system. The discretised system is shown in Fig. \ref{fig: Base FEM}. There, each vertical rectangle represents a node of the discretised object, each internal arrow represents a body force, and the final external arrow shows the traction force. They will not necessarily be applied to each configuration and specificities will be given for each test case. All models will have a prescribed displacement of $u(0) = 0$ on $\Gamma_u$
\begin{figure}[htb!]
    \centering
    \input{figures/FEM_base}
    \caption{Base FEM discretisation}
    \label{fig: Base FEM}
\end{figure}
\subsubsection{Validation}
The validation is done by testing NR and ENR on a constitutive model such that both methods successfully converge. One such model can be taken to be the linear elastic model. This model relates stresses and stretches, which will internally be converted to strains, using a constant parameter known as the modulus of elasticity $E$. In this configuration, it is expected that both solving methods will find an answer within the pre-determined iteration limit for sufficiently small applied loadings. The full configuration of the test case is shown in Table \ref{ENR: linear}. In order for ENR to work with FEM, the distance between each pair of elements of $c$ needs to be different, which is why a parameter $\zeta$ is added to each element. $\zeta$ is a small random variable of the order of $10^{-3}$.
\begin{table} [htb]
\centering
\begin{tabular}{||l l||} 
 \hline
 Property & Value\\ [0.5ex]
 \hline\hline
 Constitutive model & Linear Elastic \\
 Elastic modulus $E$ & 100$Pa$\\
 Length & 2$m$\\
 Body loading & 0 \\
 Traction loading & 20Pa \\
 Initial guess $x_0$ & $x_0 = X$ \\
 $c$ value & $c = 2X + \zeta$ \\
 [1ex] 
 \hline
\end{tabular}
\caption[Linear elastic test configuration]{Table showing the test configuration for a linear elastic constitutive model.}
\label{ENR: linear}
\end{table}
Using this configuration, both methods were able to produce a result in 1 iteration. This means that ENR works as expected in the linear case and will produce valid results. The total length of the deformed member was found to be $2.4m$, which can also be verified by hand. Knowing that $E$ relates stresses and strains, the following relation can be written out:
\begin{equation}
    \epsilon = \frac{\sigma}{E} = \lambda-1
\end{equation}
The total length of the new member can then be found using the definition of $\lambda$ as the ratio of deformed length to original length. Hence:
\begin{equation}
   l = l_0\lambda = l_0(\epsilon+1) = 2.0 \times \left(\frac{20}{100}+1\right) = 2.4m
\end{equation}
where $l_0$ represents the original length of the member and $l$ shows the deformed length. The deformed shape of the member has been drawn in Fig. \ref{fig: validation FEM}.
\begin{figure}[htb!]
    \centering
    \input{figures/FEM_linear}
    \caption{Validation FEM discretisation}
    \label{fig: validation FEM}
\end{figure}
\subsubsection{Comparison}
In this section, the performance of ENR and NR will be compared in FEM models using hyper-elastic constitutive models. The first comparison will be using the Veronda-Westmann model, which exhibits non-linear behaviour in tension. That is why only tensile loading (positive) will be applied.
\paragraph{Veronda-Westmann} \label{Veronda}
The VW model has two parameters, $A$ and $B$, which define the behaviour of a material. These parameters are stiffness parameters and will be set to:
\begin{equation*}
    \begin{array}{ll}
     A &= 2.48446MPa \\
     B &= 0.16860 
\end{array}
\end{equation*}
which represent typical values for modelling a silicone rubber material \cite{dataset}. The specific configuration for this test will be shown in Table \ref{ENR: Veronda_1}. With this configuration, ENR converged to a solution in 3 iteration steps, whereas NR took 6 steps to converge. The choice of $c$ affected this result heavily. It was found that if $c$ was taken as $c = 2X + \zeta$, ENR would not converge at all. This lead to further testing of $c$ configurations which revealed that, for this particular case, in order for ENR to converge, $c$ must be taken as at least $c = 3.2X+\zeta$. Past that value, tested up to a value of $c = 10X+\zeta$, ENR converges within 3 iteration steps. The reason for this is probably the exponential nature of VW. When an exponent is taken at a small enough value, it produces an answer which often lays within the predetermined tolerance check. This leads an iterative algorithm to converge to a false root. This behaviour can be seen in Fig. \ref{fig:rtf1}, where both methods with different configurations were unable to converge when an initial guess was taken in the negative $x_0$ range. %
\begin{table} [htb]
\centering
\begin{tabular}{||l l||} 
 \hline
 Property & Value\\ [0.5ex]
 \hline\hline
 Constitutive model & VW \\
 $A$ & 2.48446 MPa\\
 $B$ & 0.16860 \\
 Length & 2$m$\\
 Body load & 5MPa/m \\
 Traction load & 5MPa \\
 Initial guess $x_0$ & $x_0 = X$ \\
 $c$ value & $c = 4X + \zeta$ \\
 [1ex] 
 \hline
\end{tabular}
\caption[Veronda-Westmann test configuration]{Table showing the test configuration used for the Veronda-Westmann example.}
\label{ENR: Veronda_1}
\end{table}
The deformed shape produced by this configuration is shown in Fig. \ref{fig: FEM_VW}. There, the total length of the deformed bar is $4.451m$. \newline \newline In the next step of the testing, the loadings were increased in order to make convergence more difficult to achieve. Hence, the body and traction load was taken as $20MPa$. This caused NR to converge in 34 steps, whereas ENR took 8 steps to converge. In this test, the elongation was $319.16\%$. Further increasing the load to $30MPa$ as a body and traction loading caused NR to converge in 69 steps, whereas ENR took 9 steps to converge. The elongation was at $346.7\%$. These results shows that with a adequate choice of $c$, ENR can outperform NR by up to 7 times the iteration count.
\begin{figure}[htb!]
    \centering
    \input{figures/FEM_VW}
    \caption[Tensile deformed shape]{The figure shows the deformed shape of the test model using the VM material model and the loadings shown in Table \ref{ENR: Veronda_1}.}
    \label{fig: FEM_VW}
\end{figure}
Now, the behaviour of ENR will be examined in a model which behaves in a hyper-elastic manner when under compressive (negative) loads.
\paragraph{Mooney-Rivlin}
The Mooney-Rivlin constitutive model has a wide variety of usage cases, one of which is the modelling of polyurethane. The MR model has two parameters, $\mu$ and $\nu$. Their values for polyurethane are $5.289MPa$ and $0.6417$ respectively \cite{poly_urethane}. Those were used with a compressive body and traction load of $5MPa$ to test the performance of NR and ENR. The $c$ value was set equal to $0.5X+\zeta$. The full configuration of the experiment is presented in Table. \ref{ENR: Mooney_1}.
\begin{table} [htb]
\centering
\begin{tabular}{||l l||} 
 \hline
 Property & Value\\ [0.5ex]
 \hline\hline
 Constitutive model & MR \\
 $mu$ & 5.289MPa\\
 $nu$ & 0.6417 \\
 Length & 2$m$\\
 Body loading & -5MPa/m \\
 Traction loading & -5MPa \\
 Initial guess $x_0$ & $x_0 = X$ \\
 $c$ value & $c = 0.5X + \zeta$ \\
 [1ex] 
 \hline
\end{tabular}
\caption[Mooney-Rivlin test configuration]{Table showing the test configuration used for the Mooney-Rivlin example.}
\label{ENR: Mooney_1}
\end{table}
The test result shows that, for this configuration, ENR produced a result in 4 iteration, whereas NR failed to converge. The deformed shape from the FEM model is presented in Fig. \ref{fig: FEM_MR}.
\begin{figure}[htb!]
    \centering
    \input{figures/FEM_MR}
    \caption[Compressive deformed shape]{The figure shows the deformed shape of the test model using the MR material model and the loadings shown in Table \ref{ENR: Mooney_1}.}
    \label{fig: FEM_MR}
\end{figure}
These results show that the modification, present in ENR, provides an improvement to the convergence. The next step in the testing of MR was to increase the loading and see at which points the ENR method would fail. The next loading configuration was taken as $20MPa$ in compression both as a body and traction load. This caused ENR to fail. The configuration of $c$ was then modified to $c = 0.4X + \epsilon$ which enabled ENR to converge in 6 steps. This result, together with the results from \autoref{Veronda}, can be interpreted as bracketing behaviour of ENR. In \autoref{Veronda} as extension cases were tested, ENR converged in fewer steps than ER when the guess of the modification parameter $c$, if interpreted as geometrical location, laid beyond the deformed object. In this section, the same case arises when testing compressive loads and the MR model.
\paragraph{Modification parameter analysis}
The above results created a need for a method of estimating an adequate value of $c$, which will enable one to use ENR without multiple configuration testing. The two material models were tested independently in an attempt to extrapolate a formula that will provide a suitable choice of $c$. To test MR, the configuration presented in Table \ref{ENR: Mooney_1} was used together with a range of $c$ parameters starting from $c = 0.1X + \zeta$ and finishing at $c = X + \zeta$ with a step change to the variable in front of $X$, called $\phi$, equal to $0.1$. This yielded the graph shown in Fig. \ref{fig: MR C} in the Appendix, where the horizontal axis represents $\phi$ and the vertical shows the number of steps taken. There it can be seen that after $\phi = 0.5$ the method converges with an increasing number of iterations, meaning that in order for ENR to converge, $phi < 0.5$. Interpreted geometrically, this will lead to $c < 1m$. This can be compared with the compressed length of the member which is $0.970m$. The result shows that, in order for ENR to converge, a choice of $c$ which places it beyond the actual deflected configuration is needed.\newline\newline
The same analysis was done for the VW model. In this case, the loading was taken as $20MPa$ in tension both as a body load and traction load and the parameters A and B were taken to be the same as in Table \ref{ENR: Veronda_1}. The initial value for $\phi$ was set to one and a step increase of $0.5$ was used until $\phi = 10$ was reached. The result from this experiment is shown in Fig. \ref{fig: VW C}. There, it can be seen that value of $\phi$ beyond which ENR converges is 4, meaning that $c > 4X+\zeta = 8m$. This can be compared with the deflected length of the member which in this case came out to be $6.383m$. This shows that $c$ must be taken such that it extends beyond the correct answer by $25\%$.\newline \newline From the data gathered from these experiments, it can be seen that setting a $c$ value that lies beyond the deformed length yields consistently good results. A specific equation for this value was not derived as it is application specific. One strategy to always get a result is to over-estimate the deflection and use that as the value of $c$, i.e. in the MR model one can set $c = 0.01X+\zeta$ and guarantee convergence, except when the compressed object's length becomes less $1\%$ of the original. As was shown by Fig. \ref{fig: MR C}, it may not be the most efficient way to choose $c$ but yields consistent results nonetheless.
\subsection{Inverse Models}
\subsubsection{Methods}
Inverse model testing will be based on the previously described CGN and GN methods. Results here will be presented in the same way they were presented in \autoref{GN_Results}. However, rather than using a set of equations, the above-described material models will be adopted as test cases. 
\subsubsection{Validation}
\begin{figure}[htb!]
    \centering
    \includegraphics[width=\textwidth]{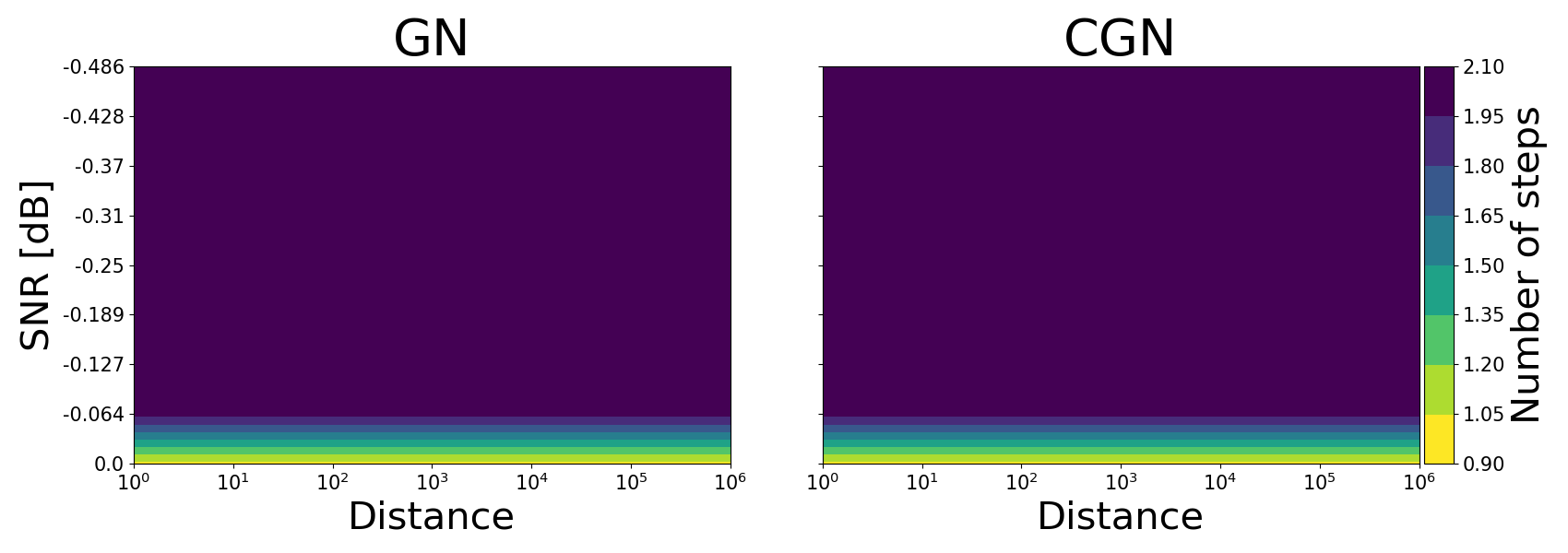}
    \caption[CGN vs. GN on linear elasticity]{Figure showing comparison between CGN and GN when tested on a linear elastic model.}
    \label{fig: GN FEM1}
\end{figure}

The validation of the CGN method will be done by comparing its results to the GN method in the linear case. Here, this will be the linear elasticity equation $\sigma = E\epsilon$, where $\epsilon$ represents the strain which is internally derived from the stretch $\lambda$. The results from this test are presented in Fig. \ref{fig: GN FEM1}. As can be seen, the two methods provided the same results which validate the CGN method in the linear elasticity case. Furthermore, the figure shown here is similar to the one shown in \autoref{GN validation}. 
\subsubsection{Comparison}
\begin{figure}[htb!]
    \centering
    \includegraphics[width=\textwidth]{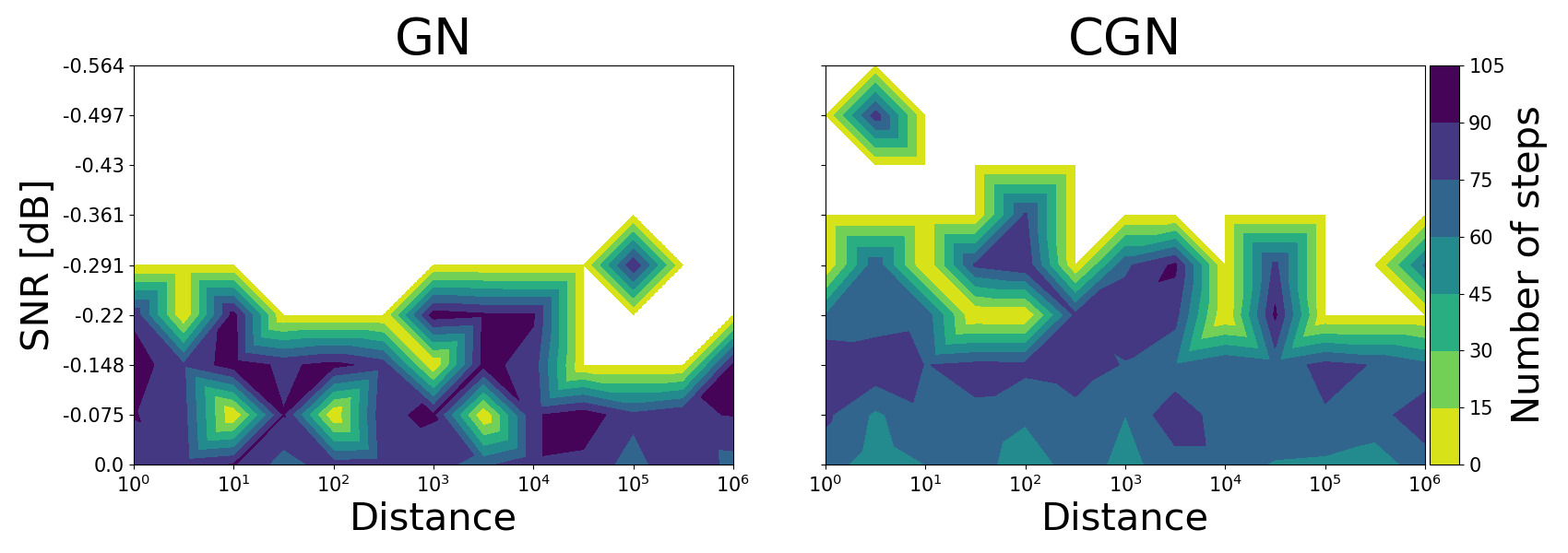}
    \caption[CGN vs. GN on Mooney-Rivlin]{Figure showing comparison between CGN and GN when tested on the Mooney-Rivlin material model. The CGN method produced a larger coverage of the test range and decreased the number of iterations needed to reach a result across the test range.}
    \label{fig: GN FEM3}
\end{figure}

In this section, the performance of CGN and GN will be compared. The first comparison case will be with the VW material model using the parameters shown in \autoref{Veronda}. Ten samples were taken from the test range of $[2,10]$ to generate the test data. All tests were done on the constitutive equations directly rather than on a FEM model which uses them. However, the result shown are still valid when applied to a FEM model.
\paragraph{Veronda-Westmann}
\begin{figure}[htb!]
    \centering
    \includegraphics[width=\textwidth]{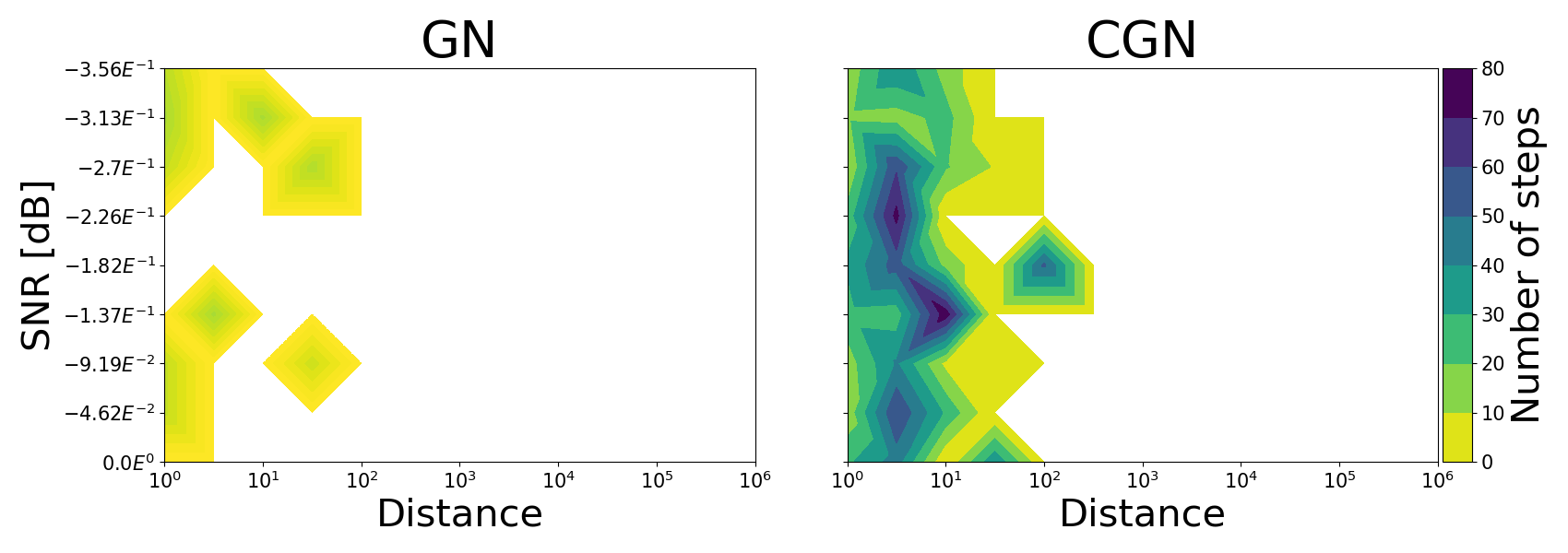}
    \caption[CGN vs. GN on Veronda-Westmann]{Figure showing comparison between CGN and GN when tested on the Veronda-Westmann material model. The CGN method produced a larger coverage of the test range. However, it also increased the iteration count at lower distances.}
    \label{fig: GN FEM2}
\end{figure}
As can be seen in Fig. \ref{fig: GN FEM2}, the CGN reduces the number of steps needed in regions of low noise levels. It also increases the converged area in regions of noise level higher than $-0.22dB$. This shows that CGN improves the convergence properties when applied to the VW material model. 
\paragraph{Mooney-Rivlin}
The second test that will be made is on the MR material model. Here, ten samples were taken from the test range $[0.3 , 0.9]$ to generate the test data. The test range was changed to represent the compressive range of the MR model. The results of the test are shown in Fig. \ref{fig: GN FEM3}.
There it can be seen that CGN provides an increased convergence area. However, the iteration count at low distances has increased. This shows that CGN improves the convergence area when applied to the MR material model. 
\section{Conclusion}
This chapter showed how a non-linear FEM model can be derived and how the two corrected methods can be used to create forward and inverse models. The test result for forwards models revealed that the ENR method improves the convergence behaviour of the FEM model when the materials exhibits hyper-elastic behaviour. This chapter, furthermore, showed the extrapolation of a strategy for selecting an adequate value of the modification parameter $c$, which allows ENR to converge for configurations which cause NR to fail. The CGN was also shown to provide an improvement over the standard GN method in highly non-linear systems in either total coverage of the test range or in integration count or both depending on the system.
\chapter{Discussion \& Conclusion}
\section{Discussion}
The aim of this paper was to propose new methods which can improve the convergence of forward and inverse finite element models. The paper has shown that the two proposed methods possess improved convergence properties when compared to their original counterparts. These improved convergence properties can then translate to improved convergence of the overarching forward and inverse models that this paper began with. When constructing a forward model of an object that is made of a material that is hyper-elastic, the usage of the extended Newton-Raphson method can greatly improve the convergence properties of the model. The degree to which this improvement will be realised is based on the choice of the parameter $c$, for which a strategy has been proposed. The improvement added by ENR can be twofold. It can, firstly, reduce the iteration count needed to find an answer and, secondly, it can expand the convergence basin of the problem allowing for more extreme configurations to be tested.\newline\newline When creating an inverse model, the same improvements can be seen when using the corrected Gauss-Newton method. It was shown in multiple examples that the method yielded a wider convergence range and a reduced iteration count. The CGN method also proved to be more resilient to noise in the data. This can make the CGN a vital tool when creating inverse models using data that has a high degree of noise.\newline \newline
Future development of the methods proposed by this paper can be concentrated on finding alternative formulations of the non-linear modification function. A few different functions were tested in the development of this paper. However, their performance was inadequate due to their insufficient continuity. The non-linear modification can also be applied to other root-finding methods such as BFGS. During the course of this project, an attempt was made to create a multivariate version of the secant method which does not require a large number of starting choices. This was achieved by taking the main diagonal of the approximation of the Jacobian matrix. A similar strategy has been proven to work in quasi-Newtonian methods \cite{diagonal}, so pursuing a simplified multivariate version of the secant method may yield valuable results. A line search algorithm can be applied to the CGN step to further improve the convergence behaviour. It is known that the standard GN method can approach quadratic convergence with an appropriate line search. Hence, if an appropriate line search is used with CGN better than quadratic convergence could be achieved due to CGN's improved rate.
\section{Conclusion}
In conclusion, this paper has covered the improvement of the Newton-Raphson root-finding algorithm through a non-linear modification aimed to reduce the non-linearity of a function. It also covered how the Gauss-Newton step can be corrected to produce improved convergence behaviour. Finally, the paper showed how to derive a non-linear version of FEM and apply the two improved approaches to forward and inverse models.\newline \newline
This paper showed that the ENR method can perform better than NR in a variety of highly non-linear functions. When ENR was applied to a FEM model, the paper showed that for the same initial starting point as NR, ENR was able to converge to a solution in fewer steps. For some configurations in which NR failed to converge, ENR was able to produce an answer from the same starting point. This paper, additionally, showed a bracketing behaviour exhibited by ENR when it comes to the choice of $c$. This led to the development of a strategy for selecting $c$ which produced consistent results.\newline \newline The paper also demonstrated that the CGN method extends the convergence basin of GN in a variety of highly non-linear functions, yet performs identically in the linear case. CGN was also shown to reduce the iteration count needed to reach an answer in some cases. When applied to the two material models examined by this paper, CGN was, again, able to produce a larger convergence basin and a reduced step count. This suggests that it can be used as an alternative to GN when dealing with highly non-linear systems, such as those presented in this paper.\newline \newline
Both improved methods require extra computational resources. However, their convergence properties may make them a suitable choice for the analysis of plastics, soft tissues and other materials which exhibit similar hyper-elastic behaviour.

\printbibliography
\end{document}

%% file: Newton_example.tex
\begin{figure}[tb]
    \centering
    \includegraphics[width=0.75\textwidth]{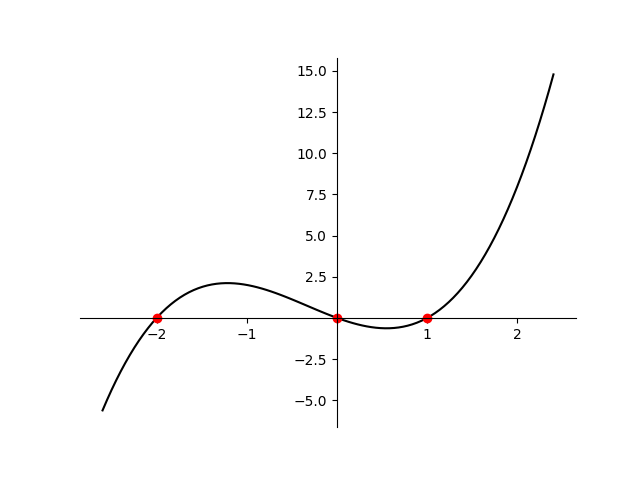}
    \caption[Graph of generic cubic]{A graph of the cubic equation $y = (x-2)x(x+1)$ plotted in the range of [-3,3]. Red markers show the positions of the roots of the equation.}
    \label{fig:cubic 1}
\end{figure}
\begin{figure}[tb]
    \centering
    \includegraphics[width=0.75\textwidth]{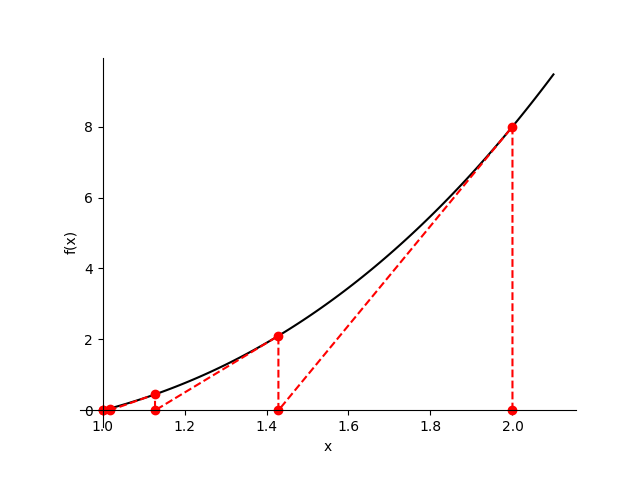}
    \caption[Iteration problem example]{Diagram visualising the NR method iteration. When starting from a initial guess of 2, the method takes 4 steps to converge to the root.}
    \label{fig:cubic 2}
\end{figure}

To demonstrate how root-finding works practically, an example is given. A typical case for root-finding is solving a third order polynomial such as the one shown below:
\begin{equation}
    f(x) = x^3+x^2-2x
\end{equation}
This polynomial yields the curve shown in Fig. \ref{fig:cubic 1}. The function has three roots at $\{-2,0,1\}$.
In order to find one of the equation's roots, the Newton-Raphson iterative method will be used in this example. At each iteration the method produces an estimation of the root denoted by $x_n$, where $(.)_n$ shows the number of iterations. The method also evaluates the function and its derivative at each step. The formulation of the Newton-Raphson method (NM) is:
\begin{equation}
   f'(x_n) \Delta x =-f(x_n)
\end{equation}
where $\Delta x$ shows the difference between the current step $x_n$ and the next step $x_{n+1}$, and $f`$ denotes the derivative of $f$ with respect to $x$. Solving for $\Delta x$ allows one to find the next iteration value through:
\begin{equation}
    x_{n+1} = x_n + \Delta x
\end{equation}
In order to initialise the method, an initial value of $x_0$ needs to be chosen close to the actual value of the solution $x^*$. For the case at hand, an initial choice of $x_0 = 2$ was made. This will be used to evaluate the function and its derivative:
\begin{equation}
\begin{array}{ll}
    f(x_0) = 8\\
    f'(x_0) = 14
\end{array}
\end{equation}
Using these values the next iteration of $x$ can be calculated:
\begin{equation}
    x_1 = x_0 - \dfrac{8}{14} = 2-0.571 = 1.429
\end{equation}
Then the process is repeated until the value of $\Delta x$ becomes less than some previously predetermined tolerance value. Here this was taken as $10^{-3}$ because all values are reported to three decimal places. Table \ref{newt_sample} shows the values at each step of the iteration process. The NR method converged in four iterations to the root closest to the starting choice. The steps taken by the method are graphically show in Fig.\ref{fig:cubic 2}.
\begin{table}[H]
\centering
\begin{tabular}{||c c c c c||} 
 \hline
 iterations & $x_n$ & $f(x_n)$ & $f'(x_n)$ & $\Delta x$
\\ [0.5ex]
 \hline\hline
 0 & 2.000 & 8.000 & 18.000 & 0.571\\ 
 1 & 1.429 & 2.102 & 6.984 & 0.301 \\
 2 & 1.128 & 0.452 & 4.073 & 0.111 \\
 3 & 1.017 & 0.052 & 3.137 & 0.017 \\
 4 & 1.000 & 0.000 & 3.000 & 0.000 \\ [1ex] 
 \hline
\end{tabular}
\caption{Newton-Raphson iterations}
\label{newt_sample}
\end{table}

%% file: GN_example.tex
\begin{figure}
    \centering
    \includegraphics[width=0.75\textwidth]{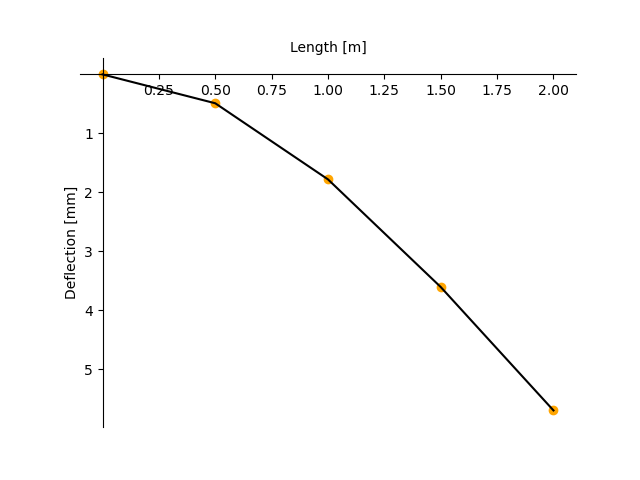}
    \caption[Gauss-Newton fitted curve]{This graphic shows the data points (yellow) used by GN to find $\theta^*$. The black curve shows the estimated value of the deflection function using the value of $\theta$ produced by GN after four iterations.}
    \label{fig: final gn data}
\end{figure}

A typical example of minimisation is finding the best fit of a curve. For this, a set of observations which represents the deflection of a cantilever I-beam will be used. The configuration of the problem is shown in Fig. \ref{fig:beamGN}. The data which will be used is shown in Table \ref{tab:data} and represents the deflection of the beam along its length. This data was created with an $I$ (second moment of area) value of $2340cm^4$ with additional noise added to the result. The known properties of the beam are summarised in Table \ref{tab:prop}. The equation used to calculate the deflection and generate the data is:
\begin{equation}
    y(x) = \frac{Px^2(3L-x)}{6EI}
\end{equation}
where $y$ is the deflection, $P$ is the force, $L$ is the length, $E$ is the modulus and $I$ is the second moment of area. The parameter which will be fitted is the second moment of area $I$. It will be shown in [$cm^4$] and for simplicity, all necessary unit conversions will be done in the background. This is a linear problem. Hence, GN should converge in one step, but as there is noise in the system more steps will be needed.
\begin{table}[htb] 
    \centering
    \begin{tabular}{||ll||}
    \hline
        Modulus $E$ &  $200 GPa$\\
        Length $L$ & $2 m$\\
        Applied Force $P$ & $10 kN$\\ [1ex] 
        \hline
    \end{tabular}
    \caption[Beam properties]{Properties of beam used for Gauss-Newton example.}
    \label{tab:prop}
\end{table}%
\begin{table}[htb]
    \centering
    \begin{tabular}{||c c c c c c||}
    \hline
        $x [m]$ & $0.000$&$0.500$&$1.000$&$1.500$&$2.000$\\
        $y [mm]$&$0.000$&$0.490$&$1.781$&$3.606$&$5.698$\\
        [1ex] 
        \hline
    \end{tabular}
    \caption[Minimisation example data points]{Observed data points used to evaluate the parameter of a deflection equation.}
    \label{tab:data}
\end{table}%
\begin{figure}
    \centering
    \input{figures/beamGN}
    \caption[Problem definition for minimisation]{The figure shows the problem definition for the example solved via the Gauss-Newton method. It shows a cantilever beam of length $L$ with an applied concentrated load $P$ at one side.}
    \label{fig:beamGN}
\end{figure}
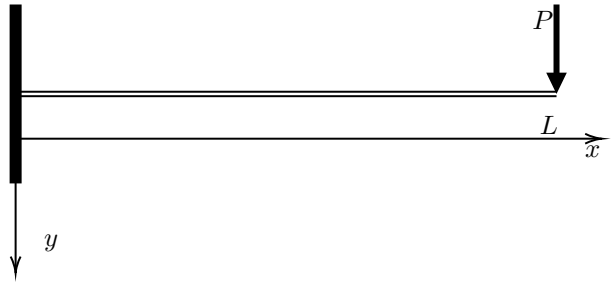\newline%
Firstly, an initial guess of parameter needs to made. For continuity, the second moment of area will be denoted by $\theta$. The above equation can be rewritten in terms of $x$ and $\theta$ as:
\begin{equation}
    f(x, \theta) = \frac{Px^2(3L-x)}{6E\theta}
\end{equation}
For this example, the initial guess $\theta^0$ will be taken to be:
\begin{equation}
    \theta^0 = 2000 cm^4
\end{equation}
From here, the residual can written in terms of $y$ and $f$ as:
\begin{equation}
    r(\theta) = y - f(x, \theta)
\end{equation}
Taking $x$ and $\theta$ to be a vectors the above can be converted to:
\begin{equation}
    r_i(\theta) = y_i - f(x_i, \theta)
\end{equation}
Then the derivative of the residual with respect to $\theta$ can be written out as:
\begin{equation}
   r_{i,j} = \frac{d(y_i(x) - f_i(x_i, \theta_j))}{d\theta_j} = -\frac{df_i(x_i, \theta_j)}{d\theta_j}
\end{equation}
In this case, $r_{i,j}$ can be found analytically, but when this is unavailable a numerical derivative using finite differences can also be used. The closed form of $r_{i,j}$ is:
\begin{equation}
   r_{i,j} = \frac{Px_i^2(3L-x_i)}{6E\theta^2_j}
\end{equation}
Using the residual and its first derivative, one can calculate the GN step using \eqref{eq: Full GN}. This step is used to update the initial guess $\theta^0$ to produce the final answer. All GN steps and values of $\theta$ are presented in Table \ref{tab: GN example}. The termination criterion was set when the step size becomes less than $10^{-3}$ because all values are reported to three decimal places. The result of the process yielded a value for $\theta \equiv I$ of $2339.921cm^4$ which satisfies the system. Using this value, one can see that the test beam used for this example was a UB203x133x25. The final estimated value of $I$ was used to plot the deflection of the beam and the original data points were scattered on top. This can be seen in Fig. \ref{fig: final gn data}.
\begin{table}[htb]
    \centering
    \begin{tabular}{||c c c||}
     \hline
    $Iter.$&$\theta$ & $\Delta\theta$
    \\[0.5ex]
    \hline\hline
    0 & 2000 & 290.541\\
    1 & 2290.541 & 48.339\\
    2 & 2338.880 & 1.0416\\
    3 & 2339.921 & 0.000\\
    [1ex] 
    \hline
    \end{tabular}
    \caption[Gauss-Newton iterations steps]{This table shows the iterations of the GN method. At the fourth iteration the method converges.}
    \label{tab: GN example}
\end{table}

%% file: figures/beamGN.tex
\tikzset{every picture/.style={line width=0.75pt}} 

\begin{tikzpicture}[x=0.75pt,y=0.75pt,yscale=-0.75,xscale=0.75]

\draw [line width=4.5]    (120,60) -- (120,180) ;
\draw    (120,118.5) -- (480,118.5)(120,121.5) -- (480,121.5) ;
\draw [line width=2.25]    (480,60) -- (480,115) ;
\draw [shift={(480,120)}, rotate = 270] [fill={rgb, 255:red, 0; green, 0; blue, 0 }  ][line width=0.08]  [draw opacity=0] (14.29,-6.86) -- (0,0) -- (14.29,6.86) -- cycle    ;
\draw    (120,150) -- (508,150) ;
\draw [shift={(510,150)}, rotate = 180] [color={rgb, 255:red, 0; green, 0; blue, 0 }  ][line width=0.75]    (10.93,-3.29) .. controls (6.95,-1.4) and (3.31,-0.3) .. (0,0) .. controls (3.31,0.3) and (6.95,1.4) .. (10.93,3.29)   ;
\draw    (120,120) -- (120,238) ;
\draw [shift={(120,240)}, rotate = 270] [color={rgb, 255:red, 0; green, 0; blue, 0 }  ][line width=0.75]    (10.93,-3.29) .. controls (6.95,-1.4) and (3.31,-0.3) .. (0,0) .. controls (3.31,0.3) and (6.95,1.4) .. (10.93,3.29)   ;

\draw (497,152.4) node [anchor=north west][inner sep=0.75pt]    {$x$};
\draw (467,132.4) node [anchor=north west][inner sep=0.75pt]    {$L$};
\draw (137,212.4) node [anchor=north west][inner sep=0.75pt]    {$y$};
\draw (462,62.4) node [anchor=north west][inner sep=0.75pt]    {$P$};

\end{tikzpicture}

%% file: figures/3d_baseconfig.tex
\begin{asy}
settings.render=0;

import graph3;
size(14cm,10cm);

currentprojection=perspective(1.5,-2,0.5);

xaxis3("$x$",0,3,grey,OutTicks(3,2));
yaxis3("$y$",-1,1,grey,OutTicks(1,-1));
zaxis3("$z$",-1,1,grey,OutTicks(1,-1));
//draw(O -- 4X, arrow=Arrow3, L=Label("$x$", position=EndPoint,
//align=W));
//draw(O -- 1.5Y, arrow=Arrow3, L=Label("$y$", position=EndPoint));
//draw(O -- 1.5Z, arrow=Arrow3, L=Label("$z$", position=EndPoint));

//real [] x = {0,0.4,0.8,1.2,1.6,2};
real [] l = {1, 1, 1, 1, 1, 1};
real [] l1 = {1,1,1,1,1,1};
real [] b={0,0,0,0,0} ;//{1, 1, 1, 1, 1};
real [] y = {-0.5,0.5};
real [] z = {-0.5,0.5};
real Traction = 0;
int end = 2;
real step = 2;
real x = -step;
for (int i=0; i<end;i+=1)
{
l1[i] = 1/sqrt(l[i]);
}
for (int i = 0; i<end; i+=1)
{
x +=step*l[i];
draw(((x, y[0]*l1[i],z[0]*l1[i])--(x, y[1]*l1[i],z[0]*l1[i])--(x, y[1]*l1[i],z[1]*l1[i])--(x, y[0]*l1[i],z[1]*l1[i])--cycle));
if(i!=end-1)
{
draw((x, y[0]*l1[i],z[0]*l1[i])--(x+step*l[i+1], y[0]*l1[i+1],z[0]*l1[i+1]));
draw((x, y[1]*l1[i],z[0]*l1[i])--(x+step*l[i+1], y[1]*l1[i+1],z[0]*l1[i+1]));
draw((x, y[1]*l1[i],z[1]*l1[i])--(x+step*l[i+1] ,y[1]*l1[i+1],z[1]*l1[i+1]));
draw((x, y[0]*l1[i],z[1]*l1[i])--(x+step*l[i+1] ,y[0]*l1[i+1],z[1]*l1[i+1]));

if(b[i]!=0)
{
draw((x,0,0)--(x+step*l[i+1],0,0),Arrow3);
}
}
}

if (Traction!=0)
{
draw((x,0,0)--(x+step,0,0),Arrow3);
}

\end{asy}

%% file: figures/LagrVsEul.tex
\tikzset{every picture/.style={line width=0.75pt}} 

\begin{tikzpicture}[x=0.75pt,y=0.75pt,yscale=-0.65,xscale=0.65]

\draw   (30,75.5) .. controls (30,67.49) and (36.72,61) .. (45,61) -- (495,61) .. controls (503.28,61) and (510,67.49) .. (510,75.5) .. controls (510,83.51) and (503.28,90) .. (495,90) -- (45,90) .. controls (36.72,90) and (30,83.51) .. (30,75.5) -- cycle ;
\draw   (120,165.5) .. controls (120,157.49) and (126.72,151) .. (135,151) -- (495,151) .. controls (503.28,151) and (510,157.49) .. (510,165.5) .. controls (510,173.51) and (503.28,180) .. (495,180) -- (135,180) .. controls (126.72,180) and (120,173.51) .. (120,165.5) -- cycle ;
\draw   (30,255.5) .. controls (30,247.49) and (36.72,241) .. (45,241) -- (495,241) .. controls (503.28,241) and (510,247.49) .. (510,255.5) .. controls (510,263.51) and (503.28,270) .. (495,270) -- (45,270) .. controls (36.72,270) and (30,263.51) .. (30,255.5) -- cycle ;
\draw  [draw opacity=0][fill={rgb, 255:red, 0; green, 0; blue, 0 }  ,fill opacity=1 ,even odd rule] (34.75,75) .. controls (34.75,69.34) and (39.34,64.75) .. (45,64.75) .. controls (50.66,64.75) and (55.25,69.34) .. (55.25,75) .. controls (55.25,80.66) and (50.66,85.25) .. (45,85.25) .. controls (39.34,85.25) and (34.75,80.66) .. (34.75,75)(30,75) .. controls (30,66.72) and (36.72,60) .. (45,60) .. controls (53.28,60) and (60,66.72) .. (60,75) .. controls (60,83.28) and (53.28,90) .. (45,90) .. controls (36.72,90) and (30,83.28) .. (30,75) ;
\draw  [draw opacity=0][fill={rgb, 255:red, 0; green, 0; blue, 0 }  ,fill opacity=1 ] (38.5,75) .. controls (38.5,71.41) and (41.41,68.5) .. (45,68.5) .. controls (48.59,68.5) and (51.5,71.41) .. (51.5,75) .. controls (51.5,78.59) and (48.59,81.5) .. (45,81.5) .. controls (41.41,81.5) and (38.5,78.59) .. (38.5,75) -- cycle ;

\draw  [draw opacity=0][fill={rgb, 255:red, 0; green, 0; blue, 0 }  ,fill opacity=1 ,even odd rule] (184.75,75) .. controls (184.75,69.34) and (189.34,64.75) .. (195,64.75) .. controls (200.66,64.75) and (205.25,69.34) .. (205.25,75) .. controls (205.25,80.66) and (200.66,85.25) .. (195,85.25) .. controls (189.34,85.25) and (184.75,80.66) .. (184.75,75)(180,75) .. controls (180,66.72) and (186.72,60) .. (195,60) .. controls (203.28,60) and (210,66.72) .. (210,75) .. controls (210,83.28) and (203.28,90) .. (195,90) .. controls (186.72,90) and (180,83.28) .. (180,75) ;
\draw  [draw opacity=0][fill={rgb, 255:red, 0; green, 0; blue, 0 }  ,fill opacity=1 ] (188.5,75) .. controls (188.5,71.41) and (191.41,68.5) .. (195,68.5) .. controls (198.59,68.5) and (201.5,71.41) .. (201.5,75) .. controls (201.5,78.59) and (198.59,81.5) .. (195,81.5) .. controls (191.41,81.5) and (188.5,78.59) .. (188.5,75) -- cycle ;

\draw  [draw opacity=0][fill={rgb, 255:red, 0; green, 0; blue, 0 }  ,fill opacity=1 ,even odd rule] (334.75,75) .. controls (334.75,69.34) and (339.34,64.75) .. (345,64.75) .. controls (350.66,64.75) and (355.25,69.34) .. (355.25,75) .. controls (355.25,80.66) and (350.66,85.25) .. (345,85.25) .. controls (339.34,85.25) and (334.75,80.66) .. (334.75,75)(330,75) .. controls (330,66.72) and (336.72,60) .. (345,60) .. controls (353.28,60) and (360,66.72) .. (360,75) .. controls (360,83.28) and (353.28,90) .. (345,90) .. controls (336.72,90) and (330,83.28) .. (330,75) ;
\draw  [draw opacity=0][fill={rgb, 255:red, 0; green, 0; blue, 0 }  ,fill opacity=1 ] (338.5,75) .. controls (338.5,71.41) and (341.41,68.5) .. (345,68.5) .. controls (348.59,68.5) and (351.5,71.41) .. (351.5,75) .. controls (351.5,78.59) and (348.59,81.5) .. (345,81.5) .. controls (341.41,81.5) and (338.5,78.59) .. (338.5,75) -- cycle ;
\draw  [draw opacity=0][fill={rgb, 255:red, 0; green, 0; blue, 0 }  ,fill opacity=1 ,even odd rule] (484.75,75) .. controls (484.75,69.34) and (489.34,64.75) .. (495,64.75) .. controls (500.66,64.75) and (505.25,69.34) .. (505.25,75) .. controls (505.25,80.66) and (500.66,85.25) .. (495,85.25) .. controls (489.34,85.25) and (484.75,80.66) .. (484.75,75)(480,75) .. controls (480,66.72) and (486.72,60) .. (495,60) .. controls (503.28,60) and (510,66.72) .. (510,75) .. controls (510,83.28) and (503.28,90) .. (495,90) .. controls (486.72,90) and (480,83.28) .. (480,75) ;
\draw  [draw opacity=0][fill={rgb, 255:red, 0; green, 0; blue, 0 }  ,fill opacity=1 ] (488.5,75) .. controls (488.5,71.41) and (491.41,68.5) .. (495,68.5) .. controls (498.59,68.5) and (501.5,71.41) .. (501.5,75) .. controls (501.5,78.59) and (498.59,81.5) .. (495,81.5) .. controls (491.41,81.5) and (488.5,78.59) .. (488.5,75) -- cycle ;

\draw  [draw opacity=0][fill={rgb, 255:red, 0; green, 0; blue, 0 }  ,fill opacity=1 ] (263.5,75.5) .. controls (263.5,71.91) and (266.41,69) .. (270,69) .. controls (273.59,69) and (276.5,71.91) .. (276.5,75.5) .. controls (276.5,79.09) and (273.59,82) .. (270,82) .. controls (266.41,82) and (263.5,79.09) .. (263.5,75.5) -- cycle ;
\draw  [draw opacity=0][fill={rgb, 255:red, 0; green, 0; blue, 0 }  ,fill opacity=1 ] (114,75.5) .. controls (114,71.91) and (116.91,69) .. (120.5,69) .. controls (124.09,69) and (127,71.91) .. (127,75.5) .. controls (127,79.09) and (124.09,82) .. (120.5,82) .. controls (116.91,82) and (114,79.09) .. (114,75.5) -- cycle ;
\draw  [draw opacity=0][fill={rgb, 255:red, 0; green, 0; blue, 0 }  ,fill opacity=1 ] (413.5,76) .. controls (413.5,72.41) and (416.41,69.5) .. (420,69.5) .. controls (423.59,69.5) and (426.5,72.41) .. (426.5,76) .. controls (426.5,79.59) and (423.59,82.5) .. (420,82.5) .. controls (416.41,82.5) and (413.5,79.59) .. (413.5,76) -- cycle ;
\draw  [draw opacity=0][fill={rgb, 255:red, 0; green, 0; blue, 0 }  ,fill opacity=1 ,even odd rule] (34.75,255) .. controls (34.75,249.34) and (39.34,244.75) .. (45,244.75) .. controls (50.66,244.75) and (55.25,249.34) .. (55.25,255) .. controls (55.25,260.66) and (50.66,265.25) .. (45,265.25) .. controls (39.34,265.25) and (34.75,260.66) .. (34.75,255)(30,255) .. controls (30,246.72) and (36.72,240) .. (45,240) .. controls (53.28,240) and (60,246.72) .. (60,255) .. controls (60,263.28) and (53.28,270) .. (45,270) .. controls (36.72,270) and (30,263.28) .. (30,255) ;
\draw  [draw opacity=0][fill={rgb, 255:red, 0; green, 0; blue, 0 }  ,fill opacity=1 ,even odd rule] (544.75,195) .. controls (544.75,189.34) and (549.34,184.75) .. (555,184.75) .. controls (560.66,184.75) and (565.25,189.34) .. (565.25,195) .. controls (565.25,200.66) and (560.66,205.25) .. (555,205.25) .. controls (549.34,205.25) and (544.75,200.66) .. (544.75,195)(540,195) .. controls (540,186.72) and (546.72,180) .. (555,180) .. controls (563.28,180) and (570,186.72) .. (570,195) .. controls (570,203.28) and (563.28,210) .. (555,210) .. controls (546.72,210) and (540,203.28) .. (540,195) ;
\draw  [draw opacity=0][fill={rgb, 255:red, 0; green, 0; blue, 0 }  ,fill opacity=1 ] (308.5,165) .. controls (308.5,161.41) and (311.41,158.5) .. (315,158.5) .. controls (318.59,158.5) and (321.5,161.41) .. (321.5,165) .. controls (321.5,168.59) and (318.59,171.5) .. (315,171.5) .. controls (311.41,171.5) and (308.5,168.59) .. (308.5,165) -- cycle ;
\draw  [draw opacity=0][fill={rgb, 255:red, 0; green, 0; blue, 0 }  ,fill opacity=1 ] (188.5,165) .. controls (188.5,161.41) and (191.41,158.5) .. (195,158.5) .. controls (198.59,158.5) and (201.5,161.41) .. (201.5,165) .. controls (201.5,168.59) and (198.59,171.5) .. (195,171.5) .. controls (191.41,171.5) and (188.5,168.59) .. (188.5,165) -- cycle ;
\draw  [draw opacity=0][fill={rgb, 255:red, 0; green, 0; blue, 0 }  ,fill opacity=1 ] (428.5,165) .. controls (428.5,161.41) and (431.41,158.5) .. (435,158.5) .. controls (438.59,158.5) and (441.5,161.41) .. (441.5,165) .. controls (441.5,168.59) and (438.59,171.5) .. (435,171.5) .. controls (431.41,171.5) and (428.5,168.59) .. (428.5,165) -- cycle ;
\draw  [draw opacity=0][fill={rgb, 255:red, 0; green, 0; blue, 0 }  ,fill opacity=1 ,even odd rule] (124.75,165) .. controls (124.75,159.34) and (129.34,154.75) .. (135,154.75) .. controls (140.66,154.75) and (145.25,159.34) .. (145.25,165) .. controls (145.25,170.66) and (140.66,175.25) .. (135,175.25) .. controls (129.34,175.25) and (124.75,170.66) .. (124.75,165)(120,165) .. controls (120,156.72) and (126.72,150) .. (135,150) .. controls (143.28,150) and (150,156.72) .. (150,165) .. controls (150,173.28) and (143.28,180) .. (135,180) .. controls (126.72,180) and (120,173.28) .. (120,165) ;
\draw  [draw opacity=0][fill={rgb, 255:red, 0; green, 0; blue, 0 }  ,fill opacity=1 ] (128.5,165) .. controls (128.5,161.41) and (131.41,158.5) .. (135,158.5) .. controls (138.59,158.5) and (141.5,161.41) .. (141.5,165) .. controls (141.5,168.59) and (138.59,171.5) .. (135,171.5) .. controls (131.41,171.5) and (128.5,168.59) .. (128.5,165) -- cycle ;

\draw  [draw opacity=0][fill={rgb, 255:red, 0; green, 0; blue, 0 }  ,fill opacity=1 ,even odd rule] (244.75,165) .. controls (244.75,159.34) and (249.34,154.75) .. (255,154.75) .. controls (260.66,154.75) and (265.25,159.34) .. (265.25,165) .. controls (265.25,170.66) and (260.66,175.25) .. (255,175.25) .. controls (249.34,175.25) and (244.75,170.66) .. (244.75,165)(240,165) .. controls (240,156.72) and (246.72,150) .. (255,150) .. controls (263.28,150) and (270,156.72) .. (270,165) .. controls (270,173.28) and (263.28,180) .. (255,180) .. controls (246.72,180) and (240,173.28) .. (240,165) ;
\draw  [draw opacity=0][fill={rgb, 255:red, 0; green, 0; blue, 0 }  ,fill opacity=1 ] (248.5,165) .. controls (248.5,161.41) and (251.41,158.5) .. (255,158.5) .. controls (258.59,158.5) and (261.5,161.41) .. (261.5,165) .. controls (261.5,168.59) and (258.59,171.5) .. (255,171.5) .. controls (251.41,171.5) and (248.5,168.59) .. (248.5,165) -- cycle ;

\draw  [draw opacity=0][fill={rgb, 255:red, 0; green, 0; blue, 0 }  ,fill opacity=1 ,even odd rule] (364.75,165) .. controls (364.75,159.34) and (369.34,154.75) .. (375,154.75) .. controls (380.66,154.75) and (385.25,159.34) .. (385.25,165) .. controls (385.25,170.66) and (380.66,175.25) .. (375,175.25) .. controls (369.34,175.25) and (364.75,170.66) .. (364.75,165)(360,165) .. controls (360,156.72) and (366.72,150) .. (375,150) .. controls (383.28,150) and (390,156.72) .. (390,165) .. controls (390,173.28) and (383.28,180) .. (375,180) .. controls (366.72,180) and (360,173.28) .. (360,165) ;
\draw  [draw opacity=0][fill={rgb, 255:red, 0; green, 0; blue, 0 }  ,fill opacity=1 ] (368.5,165) .. controls (368.5,161.41) and (371.41,158.5) .. (375,158.5) .. controls (378.59,158.5) and (381.5,161.41) .. (381.5,165) .. controls (381.5,168.59) and (378.59,171.5) .. (375,171.5) .. controls (371.41,171.5) and (368.5,168.59) .. (368.5,165) -- cycle ;

\draw  [draw opacity=0][fill={rgb, 255:red, 0; green, 0; blue, 0 }  ,fill opacity=1 ,even odd rule] (484.75,165) .. controls (484.75,159.34) and (489.34,154.75) .. (495,154.75) .. controls (500.66,154.75) and (505.25,159.34) .. (505.25,165) .. controls (505.25,170.66) and (500.66,175.25) .. (495,175.25) .. controls (489.34,175.25) and (484.75,170.66) .. (484.75,165)(480,165) .. controls (480,156.72) and (486.72,150) .. (495,150) .. controls (503.28,150) and (510,156.72) .. (510,165) .. controls (510,173.28) and (503.28,180) .. (495,180) .. controls (486.72,180) and (480,173.28) .. (480,165) ;
\draw  [draw opacity=0][fill={rgb, 255:red, 0; green, 0; blue, 0 }  ,fill opacity=1 ] (488.5,165) .. controls (488.5,161.41) and (491.41,158.5) .. (495,158.5) .. controls (498.59,158.5) and (501.5,161.41) .. (501.5,165) .. controls (501.5,168.59) and (498.59,171.5) .. (495,171.5) .. controls (491.41,171.5) and (488.5,168.59) .. (488.5,165) -- cycle ;

\draw  [draw opacity=0][fill={rgb, 255:red, 0; green, 0; blue, 0 }  ,fill opacity=1 ] (308.5,255) .. controls (308.5,251.41) and (311.41,248.5) .. (315,248.5) .. controls (318.59,248.5) and (321.5,251.41) .. (321.5,255) .. controls (321.5,258.59) and (318.59,261.5) .. (315,261.5) .. controls (311.41,261.5) and (308.5,258.59) .. (308.5,255) -- cycle ;
\draw  [draw opacity=0][fill={rgb, 255:red, 0; green, 0; blue, 0 }  ,fill opacity=1 ,even odd rule] (184.75,255) .. controls (184.75,249.34) and (189.34,244.75) .. (195,244.75) .. controls (200.66,244.75) and (205.25,249.34) .. (205.25,255) .. controls (205.25,260.66) and (200.66,265.25) .. (195,265.25) .. controls (189.34,265.25) and (184.75,260.66) .. (184.75,255)(180,255) .. controls (180,246.72) and (186.72,240) .. (195,240) .. controls (203.28,240) and (210,246.72) .. (210,255) .. controls (210,263.28) and (203.28,270) .. (195,270) .. controls (186.72,270) and (180,263.28) .. (180,255) ;
\draw  [draw opacity=0][fill={rgb, 255:red, 0; green, 0; blue, 0 }  ,fill opacity=1 ] (188.5,255) .. controls (188.5,251.41) and (191.41,248.5) .. (195,248.5) .. controls (198.59,248.5) and (201.5,251.41) .. (201.5,255) .. controls (201.5,258.59) and (198.59,261.5) .. (195,261.5) .. controls (191.41,261.5) and (188.5,258.59) .. (188.5,255) -- cycle ;

\draw  [draw opacity=0][fill={rgb, 255:red, 0; green, 0; blue, 0 }  ,fill opacity=1 ] (428.5,255) .. controls (428.5,251.41) and (431.41,248.5) .. (435,248.5) .. controls (438.59,248.5) and (441.5,251.41) .. (441.5,255) .. controls (441.5,258.59) and (438.59,261.5) .. (435,261.5) .. controls (431.41,261.5) and (428.5,258.59) .. (428.5,255) -- cycle ;
\draw  [draw opacity=0][fill={rgb, 255:red, 0; green, 0; blue, 0 }  ,fill opacity=1 ] (128.5,255) .. controls (128.5,251.41) and (131.41,248.5) .. (135,248.5) .. controls (138.59,248.5) and (141.5,251.41) .. (141.5,255) .. controls (141.5,258.59) and (138.59,261.5) .. (135,261.5) .. controls (131.41,261.5) and (128.5,258.59) .. (128.5,255) -- cycle ;
\draw  [draw opacity=0][fill={rgb, 255:red, 0; green, 0; blue, 0 }  ,fill opacity=1 ] (248.5,255) .. controls (248.5,251.41) and (251.41,248.5) .. (255,248.5) .. controls (258.59,248.5) and (261.5,251.41) .. (261.5,255) .. controls (261.5,258.59) and (258.59,261.5) .. (255,261.5) .. controls (251.41,261.5) and (248.5,258.59) .. (248.5,255) -- cycle ;
\draw  [draw opacity=0][fill={rgb, 255:red, 0; green, 0; blue, 0 }  ,fill opacity=1 ] (368.5,255) .. controls (368.5,251.41) and (371.41,248.5) .. (375,248.5) .. controls (378.59,248.5) and (381.5,251.41) .. (381.5,255) .. controls (381.5,258.59) and (378.59,261.5) .. (375,261.5) .. controls (371.41,261.5) and (368.5,258.59) .. (368.5,255) -- cycle ;
\draw  [draw opacity=0][fill={rgb, 255:red, 0; green, 0; blue, 0 }  ,fill opacity=1 ,even odd rule] (484.75,255) .. controls (484.75,249.34) and (489.34,244.75) .. (495,244.75) .. controls (500.66,244.75) and (505.25,249.34) .. (505.25,255) .. controls (505.25,260.66) and (500.66,265.25) .. (495,265.25) .. controls (489.34,265.25) and (484.75,260.66) .. (484.75,255)(480,255) .. controls (480,246.72) and (486.72,240) .. (495,240) .. controls (503.28,240) and (510,246.72) .. (510,255) .. controls (510,263.28) and (503.28,270) .. (495,270) .. controls (486.72,270) and (480,263.28) .. (480,255) ;
\draw  [draw opacity=0][fill={rgb, 255:red, 0; green, 0; blue, 0 }  ,fill opacity=1 ] (488.5,255) .. controls (488.5,251.41) and (491.41,248.5) .. (495,248.5) .. controls (498.59,248.5) and (501.5,251.41) .. (501.5,255) .. controls (501.5,258.59) and (498.59,261.5) .. (495,261.5) .. controls (491.41,261.5) and (488.5,258.59) .. (488.5,255) -- cycle ;

\draw  [draw opacity=0][fill={rgb, 255:red, 0; green, 0; blue, 0 }  ,fill opacity=1 ,even odd rule] (334.75,255) .. controls (334.75,249.34) and (339.34,244.75) .. (345,244.75) .. controls (350.66,244.75) and (355.25,249.34) .. (355.25,255) .. controls (355.25,260.66) and (350.66,265.25) .. (345,265.25) .. controls (339.34,265.25) and (334.75,260.66) .. (334.75,255)(330,255) .. controls (330,246.72) and (336.72,240) .. (345,240) .. controls (353.28,240) and (360,246.72) .. (360,255) .. controls (360,263.28) and (353.28,270) .. (345,270) .. controls (336.72,270) and (330,263.28) .. (330,255) ;
\draw  [draw opacity=0][fill={rgb, 255:red, 0; green, 0; blue, 0 }  ,fill opacity=1 ] (548.5,255) .. controls (548.5,251.41) and (551.41,248.5) .. (555,248.5) .. controls (558.59,248.5) and (561.5,251.41) .. (561.5,255) .. controls (561.5,258.59) and (558.59,261.5) .. (555,261.5) .. controls (551.41,261.5) and (548.5,258.59) .. (548.5,255) -- cycle ;
\draw    (30,300) -- (507,300) ;
\draw [shift={(510,300)}, rotate = 180] [fill={rgb, 255:red, 0; green, 0; blue, 0 }  ][line width=0.08]  [draw opacity=0] (8.93,-4.29) -- (0,0) -- (8.93,4.29) -- cycle    ;
\draw [shift={(30,300)}, rotate = 180] [color={rgb, 255:red, 0; green, 0; blue, 0 }  ][line width=0.75]    (0,5.59) -- (0,-5.59)   ;

\draw (144.5,123.33) node [anchor=north west][inner sep=0.75pt]   [align=left] {Lagrangian transformed configuration};
\draw (206.5,32.67) node [anchor=north west][inner sep=0.75pt]   [align=left] {Initial configuration};
\draw (154.5,212.67) node [anchor=north west][inner sep=0.75pt]   [align=left] {Eulerian transformed configuration};
\draw (571,187) node [anchor=north west][inner sep=0.75pt]   [align=left] {Node};
\draw (571,247) node [anchor=north west][inner sep=0.75pt]   [align=left] {Material point};
\draw (498,280) node [anchor=north west][inner sep=0.75pt]   [align=left] {$\displaystyle x$};

\end{tikzpicture}

%% file: figures/continuity.tex
\tikzset{every picture/.style={line width=0.75pt}} 

\begin{tikzpicture}[x=0.75pt,y=0.75pt,yscale=-0.75,xscale=0.75]

\draw    (120,241) -- (538,241) ;
\draw [shift={(540,241)}, rotate = 180] [color={rgb, 255:red, 0; green, 0; blue, 0 }  ][line width=0.75]    (10.93,-3.29) .. controls (6.95,-1.4) and (3.31,-0.3) .. (0,0) .. controls (3.31,0.3) and (6.95,1.4) .. (10.93,3.29)   ;
\draw    (120,360) -- (120,92) ;
\draw [shift={(120,90)}, rotate = 450] [color={rgb, 255:red, 0; green, 0; blue, 0 }  ][line width=0.75]    (10.93,-3.29) .. controls (6.95,-1.4) and (3.31,-0.3) .. (0,0) .. controls (3.31,0.3) and (6.95,1.4) .. (10.93,3.29)   ;
\draw    (120,241) -- (235.78,211.09) ;
\draw [shift={(240,210)}, rotate = 345.52] [color={rgb, 255:red, 0; green, 0; blue, 0 }  ][line width=0.75]      (0, 0) circle [x radius= 5.36, y radius= 5.36]   ;
\draw    (240,210) -- (385.95,268.38) ;
\draw [shift={(390,270)}, rotate = 21.8] [color={rgb, 255:red, 0; green, 0; blue, 0 }  ][line width=0.75]      (0, 0) circle [x radius= 5.36, y radius= 5.36]   ;
\draw    (390,270) -- (510,240) ;
\draw  [dash pattern={on 4.5pt off 4.5pt}]  (120,150) -- (240,150) ;
\draw [shift={(240,150)}, rotate = 0] [color={rgb, 255:red, 0; green, 0; blue, 0 }  ][fill={rgb, 255:red, 0; green, 0; blue, 0 }  ][line width=0.75]      (0, 0) circle [x radius= 3.35, y radius= 3.35]   ;
\draw  [dash pattern={on 4.5pt off 4.5pt}]  (240,150) -- (240,330) ;
\draw [shift={(240,330)}, rotate = 90] [color={rgb, 255:red, 0; green, 0; blue, 0 }  ][fill={rgb, 255:red, 0; green, 0; blue, 0 }  ][line width=0.75]      (0, 0) circle [x radius= 3.35, y radius= 3.35]   ;
\draw  [dash pattern={on 4.5pt off 4.5pt}]  (240,330) -- (390,330) ;
\draw [shift={(390,330)}, rotate = 0] [color={rgb, 255:red, 0; green, 0; blue, 0 }  ][fill={rgb, 255:red, 0; green, 0; blue, 0 }  ][line width=0.75]      (0, 0) circle [x radius= 3.35, y radius= 3.35]   ;
\draw  [dash pattern={on 4.5pt off 4.5pt}]  (390,180) -- (390,330) ;
\draw  [dash pattern={on 4.5pt off 4.5pt}]  (390,180) -- (510,180) ;
\draw [shift={(390,180)}, rotate = 0] [color={rgb, 255:red, 0; green, 0; blue, 0 }  ][fill={rgb, 255:red, 0; green, 0; blue, 0 }  ][line width=0.75]      (0, 0) circle [x radius= 3.35, y radius= 3.35]   ;
\draw [line width=3]    (120,240) .. controls (307,82) and (288,295) .. (390,312) ;
\draw [line width=3]    (390,312) .. controls (443,321) and (483,276) .. (510,240) ;
\draw  [dash pattern={on 4.5pt off 4.5pt}]  (540,82) -- (600,82) ;
\draw    (540,112) -- (600,112) ;
\draw [line width=3.75]    (540,142) -- (600,142) ;
\draw    (300,90) -- (245.14,143.9) ;
\draw [shift={(243,146)}, rotate = 315.51] [fill={rgb, 255:red, 0; green, 0; blue, 0 }  ][line width=0.08]  [draw opacity=0] (10.72,-5.15) -- (0,0) -- (10.72,5.15) -- (7.12,0) -- cycle    ;
\draw    (300,150) -- (247.14,201.9) ;
\draw [shift={(245,204)}, rotate = 315.53] [fill={rgb, 255:red, 0; green, 0; blue, 0 }  ][line width=0.08]  [draw opacity=0] (10.72,-5.15) -- (0,0) -- (10.72,5.15) -- (7.12,0) -- cycle    ;

\draw (84,102.4) node [anchor=north west][inner sep=0.75pt]    {$f( x)$};
\draw (527,252.4) node [anchor=north west][inner sep=0.75pt]    {$x$};
\draw (511,57.4) node [anchor=north west][inner sep=0.75pt]    {$C^{ \begin{array}{l}
-1
\end{array}}$};
\draw (511,87.4) node [anchor=north west][inner sep=0.75pt]    {$C^{ \begin{array}{l}
0
\end{array}}$};
\draw (511,117.4) node [anchor=north west][inner sep=0.75pt]    {$C^{ \begin{array}{l}
1
\end{array}}$};
\draw (301,71) node [anchor=north west][inner sep=0.75pt]   [align=left] {Jump};
\draw (301,131) node [anchor=north west][inner sep=0.75pt]   [align=left] {Kink};

\end{tikzpicture}

%% file: figures/boundary.tex
\tikzset{every picture/.style={line width=0.75pt}} 

\begin{tikzpicture}[x=0.75pt,y=0.75pt,yscale=-0.65,xscale=0.65]

\draw   (90,90) -- (510,90) -- (510,150) -- (90,150) -- cycle ;
\draw  [fill={rgb, 255:red, 0; green, 0; blue, 0 }  ,fill opacity=1 ] (73.67,30) -- (90,30) -- (90,210) -- (73.67,210) -- cycle ;
\draw    (90,180) -- (568,180) ;
\draw [shift={(570,180)}, rotate = 180] [color={rgb, 255:red, 0; green, 0; blue, 0 }  ][line width=0.75]    (10.93,-3.29) .. controls (6.95,-1.4) and (3.31,-0.3) .. (0,0) .. controls (3.31,0.3) and (6.95,1.4) .. (10.93,3.29)   ;
\draw   (90,300) -- (540,300) -- (540,360) -- (90,360) -- cycle ;
\draw  [fill={rgb, 255:red, 0; green, 0; blue, 0 }  ,fill opacity=1 ] (73.67,240) -- (90,240) -- (90,420) -- (73.67,420) -- cycle ;
\draw    (540,330) -- (597,330) ;
\draw [shift={(600,330)}, rotate = 180] [fill={rgb, 255:red, 0; green, 0; blue, 0 }  ][line width=0.08]  [draw opacity=0] (8.93,-4.29) -- (0,0) -- (8.93,4.29) -- cycle    ;
\draw    (90,390) -- (568,390) ;
\draw [shift={(570,390)}, rotate = 180] [color={rgb, 255:red, 0; green, 0; blue, 0 }  ][line width=0.75]    (10.93,-3.29) .. controls (6.95,-1.4) and (3.31,-0.3) .. (0,0) .. controls (3.31,0.3) and (6.95,1.4) .. (10.93,3.29)   ;

\draw (558,182) node [anchor=north west][inner sep=0.75pt]   [align=left] {X,$\displaystyle x$};
\draw (91,160) node [anchor=north west][inner sep=0.75pt]   [align=left] {$\displaystyle X_{a} =0$};
\draw (484,160) node [anchor=north west][inner sep=0.75pt]   [align=left] {$\displaystyle X_{b} =l_{0}$};
\draw (558,310) node [anchor=north west][inner sep=0.75pt]   [align=left] {$\displaystyle t$};
\draw (558,392) node [anchor=north west][inner sep=0.75pt]   [align=left] {$\displaystyle x$};
\draw (91,370) node [anchor=north west][inner sep=0.75pt]   [align=left] {$\displaystyle x\ =\ 0$};
\draw (511,370) node [anchor=north west][inner sep=0.75pt]   [align=left] {$\displaystyle x=\ l$};
\draw (241,41) node [anchor=north west][inner sep=0.75pt]   [align=left] {Inital configuration};
\draw (250,251) node [anchor=north west][inner sep=0.75pt]   [align=left] {Current configuration};

\end{tikzpicture}

%% file: figures/FEM_base.tex
\begin{asy}
settings.render=0;
import graph3;
size(10cm,10cm);

currentprojection=perspective(1.5,-2,0.5);

xaxis3("$x$",0,3,grey,OutTicks(3,2));
yaxis3("$y$",-1,1,grey,OutTicks(1,-1));
zaxis3("$z$",-1,1,grey,OutTicks(1,-1));
//draw(O -- 4X, arrow=Arrow3, L=Label("$x$", position=EndPoint,
//align=W));
//draw(O -- 1.5Y, arrow=Arrow3, L=Label("$y$", position=EndPoint));
//draw(O -- 1.5Z, arrow=Arrow3, L=Label("$z$", position=EndPoint));

//real [] x = {0,0.4,0.8,1.2,1.6,2};
real [] l = {1, 1, 1, 1, 1, 1};
real [] l1 = {1,1,1,1,1,1};
real [] b= {1, 1, 1, 1, 1};
real [] y = {-0.5,0.5};
real [] z = {-0.5,0.5};
real Traction = 10;
int end = 6;
real step = 0.4;
real x = -step;
for (int i=0; i<end;i+=1)
{
l1[i] = 1/sqrt(l[i]);
}
for (int i = 0; i<end; i+=1)
{
x +=step*l[i];
draw(((x, y[0]*l1[i],z[0]*l1[i])--(x, y[1]*l1[i],z[0]*l1[i])--(x, y[1]*l1[i],z[1]*l1[i])--(x, y[0]*l1[i],z[1]*l1[i])--cycle));
if(i!=end-1)
{
draw((x, y[0]*l1[i],z[0]*l1[i])--(x+step*l[i+1], y[0]*l1[i+1],z[0]*l1[i+1]));
draw((x, y[1]*l1[i],z[0]*l1[i])--(x+step*l[i+1], y[1]*l1[i+1],z[0]*l1[i+1]));
draw((x, y[1]*l1[i],z[1]*l1[i])--(x+step*l[i+1] ,y[1]*l1[i+1],z[1]*l1[i+1]));
draw((x, y[0]*l1[i],z[1]*l1[i])--(x+step*l[i+1] ,y[0]*l1[i+1],z[1]*l1[i+1]));

if(b[i]!=0)
{
draw((x,0,0)--(x+step*l[i+1],0,0),Arrow3);
}
}
}

if (Traction!=0)
{
draw((x,0,0)--(x+step,0,0),Arrow3);
}
\end{asy}

%% file: figures/FEM_linear.tex
\begin{asy}
settings.render=0;
import graph3;
size(10cm,10cm);

currentprojection=perspective(1.5,-2,0.5);

xaxis3("$x$",0,3,grey,OutTicks(3,2));
yaxis3("$y$",-1,1,grey,OutTicks(1,-1));
zaxis3("$z$",-1,1,grey,OutTicks(1,-1));
//draw(O -- 4X, arrow=Arrow3, L=Label("$x$", position=EndPoint,
//align=W));
//draw(O -- 1.5Y, arrow=Arrow3, L=Label("$y$", position=EndPoint));
//draw(O -- 1.5Z, arrow=Arrow3, L=Label("$z$", position=EndPoint));

real [] l = {1.2, 1.2, 1.2000000000000006, 1.1999999999999993, 1.1999999999999993};
real [] l1 = {1,1,1,1,1};
real [] l2 = l;
real [] y = {-0.5,0.5};
real [] z = {-0.5,0.5};
int end = 5;
real step = 0.4;
real x = 0;
for (int i=0; i<end;i+=1)
{
l1[i] = 1/sqrt(l[i]);
}
for (int i = 0; i<end; i+=1)
{
draw(((x, y[0]*l1[i],z[0]*l1[i])--(x, y[1]*l1[i],z[0]*l1[i])--(x, y[1]*l1[i],z[1]*l1[i])--(x, y[0]*l1[i],z[1]*l1[i])--cycle));
draw((x, y[0]*l1[i],z[0]*l1[i])--(x+step*l[i], y[0]*l1[i],z[0]*l1[i]));
draw((x, y[1]*l1[i],z[0]*l1[i])--(x+step*l[i], y[1]*l1[i],z[0]*l1[i]));
draw((x, y[1]*l1[i],z[1]*l1[i])--(x+step*l[i] ,y[1]*l1[i],z[1]*l1[i]));
draw((x, y[0]*l1[i],z[1]*l1[i])--(x+step*l[i] ,y[0]*l1[i],z[1]*l1[i]));
x +=step*l[i];
draw(((x, y[0]*l1[i],z[0]*l1[i])--(x, y[1]*l1[i],z[0]*l1[i])--(x, y[1]*l1[i],z[1]*l1[i])--(x, y[0]*l1[i],z[1]*l1[i])--cycle));
}

\end{asy}

%% file: figures/FEM_VW.tex
\begin{asy}
settings.render=0;
import graph3;
size(10cm,10cm);

currentprojection=perspective(1.5,-2,0.5);

xaxis3("$x$",0,5,grey,OutTicks(5,2));
yaxis3("$y$",-1,1,grey,OutTicks(1,-1));
zaxis3("$z$",-1,1,grey,OutTicks(1,-1));
//draw(O -- 4X, arrow=Arrow3, L=Label("$x$", position=EndPoint,
//align=W));
//draw(O -- 1.5Y, arrow=Arrow3, L=Label("$y$", position=EndPoint));
//draw(O -- 1.5Z, arrow=Arrow3, L=Label("$z$", position=EndPoint));

real [] l = {2.5754486971760246, 2.439432195351979, 2.27316694937453, 2.061948048699228, 1.7797210216299508};
real [] l1 = {1,1,1,1,1};
real [] l2 = l;
real [] y = {-0.5,0.5};
real [] z = {-0.5,0.5};
int end = 5;
real step = 0.4;
real x = 0;
for (int i=0; i<end;i+=1)
{
l1[i] = 1/sqrt(l[i]);
}
for (int i = 0; i<end; i+=1)
{
draw(((x, y[0]*l1[i],z[0]*l1[i])--(x, y[1]*l1[i],z[0]*l1[i])--(x, y[1]*l1[i],z[1]*l1[i])--(x, y[0]*l1[i],z[1]*l1[i])--cycle));
draw((x, y[0]*l1[i],z[0]*l1[i])--(x+step*l[i], y[0]*l1[i],z[0]*l1[i]));
draw((x, y[1]*l1[i],z[0]*l1[i])--(x+step*l[i], y[1]*l1[i],z[0]*l1[i]));
draw((x, y[1]*l1[i],z[1]*l1[i])--(x+step*l[i] ,y[1]*l1[i],z[1]*l1[i]));
draw((x, y[0]*l1[i],z[1]*l1[i])--(x+step*l[i] ,y[0]*l1[i],z[1]*l1[i]));
x +=step*l[i];
draw(((x, y[0]*l1[i],z[0]*l1[i])--(x, y[1]*l1[i],z[0]*l1[i])--(x, y[1]*l1[i],z[1]*l1[i])--(x, y[0]*l1[i],z[1]*l1[i])--cycle));
}

\end{asy}

%% file: figures/FEM_MR.tex
\begin{asy}
settings.render=0;
import graph3;
size(9cm,8cm);

currentprojection=perspective(1.5,-2,0.5);

xaxis3("$x$",0,2,grey,OutTicks(2,2));
yaxis3("$y$",-1,1,grey,OutTicks(1,-1));
zaxis3("$z$",-1,1,grey,OutTicks(1,-1));
//draw(O -- 4X, arrow=Arrow3, L=Label("$x$", position=EndPoint,
//align=W));
//draw(O -- 1.5Y, arrow=Arrow3, L=Label("$y$", position=EndPoint));
//draw(O -- 1.5Z, arrow=Arrow3, L=Label("$z$", position=EndPoint));

real [] l = {0.6069355841720173, 0.6347441228008357, 0.6690364262694395, 0.7132025153519402, 0.7740849702659658};
real [] l1 = {1,1,1,1,1};
real [] l2 = l;
real [] y = {-0.5,0.5};
real [] z = {-0.5,0.5};
int end = 5;
real step = 0.4;
real x = 0;
for (int i=0; i<end;i+=1)
{
l1[i] = 1/sqrt(l[i]);
}
for (int i = 0; i<end; i+=1)
{
draw(((x, y[0]*l1[i],z[0]*l1[i])--(x, y[1]*l1[i],z[0]*l1[i])--(x, y[1]*l1[i],z[1]*l1[i])--(x, y[0]*l1[i],z[1]*l1[i])--cycle));
draw((x, y[0]*l1[i],z[0]*l1[i])--(x+step*l[i], y[0]*l1[i],z[0]*l1[i]));
draw((x, y[1]*l1[i],z[0]*l1[i])--(x+step*l[i], y[1]*l1[i],z[0]*l1[i]));
draw((x, y[1]*l1[i],z[1]*l1[i])--(x+step*l[i] ,y[1]*l1[i],z[1]*l1[i]));
draw((x, y[0]*l1[i],z[1]*l1[i])--(x+step*l[i] ,y[0]*l1[i],z[1]*l1[i]));
x +=step*l[i];
draw(((x, y[0]*l1[i],z[0]*l1[i])--(x, y[1]*l1[i],z[0]*l1[i])--(x, y[1]*l1[i],z[1]*l1[i])--(x, y[0]*l1[i],z[1]*l1[i])--cycle));
}

\end{asy}